\documentclass[fleqn,usenatbib,useAMS]{mnras}
\usepackage{hhline}
\usepackage{xspace}
\usepackage{natbib}
\usepackage{bigstrut}
\usepackage{lipsum} 
\usepackage[T1]{fontenc}

\usepackage{graphicx}
\usepackage{amsmath}
\usepackage{xcolor}
\usepackage{bm}
\usepackage{times}
\usepackage{amssymb,amsmath}

\newcommand{\RXTE}{\textit{RXTE}\xspace}

\newcommand{\Swift}{\textit{Swift}\xspace}

\newcommand{\Fermi}{\textit{Fermi}\xspace}
\newcommand{\Agile}{\textit{AGILE}\xspace}

\newcommand{\RN}[1]{%
  \textup{\uppercase\expandafter{\romannumeral#1}}%
}

\def\lsim{ \lower .75ex\hbox{$\sim$} \llap{\raise .27ex \hbox{$<$}} }
\def\gsim{ \lower .75ex \hbox{$\sim$} \llap{\raise .27ex \hbox{$>$}} }

\title{Bhjet: a public multi-zone, steady state jet + thermal corona spectral model}
\author[Lucchini et al.]{Lucchini, M.$^{1,2}$, Ceccobello C.$^{3}$, Markoff S.$^{2,4}$, Kini Y.$^{2}$, Chhotray A.$^{2}$, Connors R. M. T.$^{5}$, \newauthor
Crumley P.$^{6}$, Falcke H.$^{7}$, Kantzas D.$^{2,4}$, Maitra D.$^{8}$ \\
$^1$MIT Kavli Institute for Astrophysics and Space Research, MIT, 70 Vassar Street, Cambridge, MA 02139, USA \\
$^2$API -- Anton Pannekoek Institute for Astronomy, University of Amsterdam, Science Park 904, 1098 XH Amsterdam, the Netherlands \\
$^3$Department of Space, Earth and Environment, Chalmers University of Technology, Onsala Space Observatory, 439 92 Onsala, Sweden \\ 
$^4$GRAPPA -- Gravitational and Astroparticle Physics Amsterdam, University of Amsterdam, Science Park 904, 1098 XH Amsterdam, the Netherlands \\
$^5$Cahill Center for Astronomy and Astrophysics, California Institute of Technology, Pasadena, CA 91125, USA\\
$^6$ Osmos.io, Seattle Washington 98122, USA\\
$^7$Department of Astrophysics, Institute for Mathematics, Astrophysics and Particle Physics (IMAPP), Radboud University, P.O. Box9010, 6500 GL Nijmegen, \\ 
The Netherlands\\
$^8$Department of Physics and Astronomy, Wheaton College, Norton, MA 02766, USA
}

\begin{document}

\maketitle

\begin{abstract}
Accreting black holes are sources of major interest in astronomy, particular those launching jets because of their ability to accelerate particles, and dramatically affect their surrounding environment up to very large distances. The spatial, energy and time scales at which a central active black hole radiates and impacts its environment depend on its mass. The implied scale-invariance of accretion/ejection physics between black hole systems of different central masses has been conﬁrmed by several studies. Therefore, designing a self-consistent theoretical model that can describe such systems, regardless of their mass, is of crucial importance to tackle a variety of astrophysical sources. We present here a new and significantly improved version of a scale invariant, steady-state, multi-zone jet model, which we rename BHJet, resulting from the efforts of our group to advance the modelling of black hole systems. We summarise the model assumptions and basic equations, how they have evolved over time, and the additional features that we have recently introduced. These include additional input electron populations, the extension to cyclotron emission in near-relativistic regime, an improved multiple Inverse-Compton scattering method, external photon seed fields typical of AGN and a magnetically-dominated jet dynamical model as opposed to the pressure-driven jet conﬁguration present in older versions. In this paper, we publicly release the code on GitHub and, in order to facilitate the user’s approach to its many possibilities, showcase a few applications as a tutorial.
 
\end{abstract}


\section{Introduction}
Accretion is one of the most efficient mechanism in the Universe for converting rest-mass into energy, and as a result accreting compact objects can have substantial impact on their surroundings \citep{Silk98,Fabian12}. Accreting objects also often launch collimated outflows of plasma called jets; this phenomenon is observed in accreting black holes, both stellar and supermassive \cite[e.g.][]{Fanaroff74,Mirabel94,Bloom11}, neutron stars \citep[e.g.][]{Migliari12,vandenEijnden18}, white dwarfs \citep[e.g.][]{Kellogg01,Sokoloski2008,Koerding2008} and young stellar objects \citep[e.g.][]{Sahai98}. Out of these systems, black holes are particularly exciting targets because they are the only objects that span over nine orders of magnitude in mass/size. Black holes also offer  convenient laboratories to study accretion and ejection physics in the strong gravitational regime without the contamination of a stellar magnetic field and a solid surface.

Coordinated radio/X-ray campaigns have discovered two key properties of accretion/ejection coupling in accreting black holes. First, during black hole X-ray binary (BHXB) hard spectral states, when steady jets are present, the outflowing material is tightly coupled to the accretion flow. This coupling takes the form of a tight correlation \citep{Hannikainen98,Corbel00,Corbel03} between radio luminosity, tracking the power of the jet at large distances ($z\approx 10^{6-8}\,\rm{R_g}$) from the black hole, and the X-ray luminosity, which originates close to the central engine ($z\leq 10^{1-3}\,\rm{R_g}$) and can be thought of as a proxy for the power in the accretion flow. Second, as mentioned above, the accretion/ejection coupling appears to be scale invariant. The existence of scale invariance was first inferred by extending the radio/X-ray correlation to a wide variety of jetted Active Galactic Nuclei (AGN) types  \citep{Merloni03,Falcke04,Koerding2006,Plotkin12}. By including mass, these empirical studies demonstrated that all low-luminosity accreting black holes with jets seem to populate a plane in the three-dimensional space of mass, radio luminosity and X-ray luminosity, known as the fundamental plane of black hole accretion (FP). Such scale invariance means we can study black hole accretion using two different approaches: 1) monitoring the variation of fundamental quantities, such as the accretion rate over time (with constant black hole mass, viewing angle and spin) during the outburst activity of BHXBs, or alternatively 2) focus on AGN, which allows for large-sample studies, with a wide range of viewing angles, black hole masses and spin, but with near constant accretion rate over the length of one or more observations. 

In recent years, global general relativistic magneto-hydrodynamics (GRMHD) simulations have elucidated the long-standing question of which mechanism is responsible for  jet launching. Previously the debate was mostly focused on whether the magnetic field lines driving the outflow are anchored on the disc, using its angular momentum to eject matter via the magneto-centrifugal force (\citealt{Blandford82}, BP from now onward) or on the black hole ergosphere via frame-dragging, using the rotational energy of the compact object itself to launch the jet (\citealt{Blandford77}, BZ from now onward). However, it now appears likely that both mechanisms play a role simultaneously \citep{McKinney06,Monika13,Chatterjee19}. A direct consequence of this composite scenario is the formation of a structured jet, in which a highly magnetised, pair-loaded, BZ-type inner spine results in a highly relativistic, Poynting-flux dominated jet, surrounded by a slower, mass-loaded, BP-type sheath formed at the interface with the accretion disc. This scenario is supported by observational evidence \citep{Mertens16,Giovannini18}. Most likely then, a better question to be asked is not what is the physical mechanism leading to jet launching, but rather which part of the jet dominates the observed emission. A crucial property of non-radiative, ideal GRMHD is that it is inherently scale-free (meaning that to change from code to physical units for a given quantity one simply has to assume a certain black hole mass). A scale-free semi-analytical approach can also explain why the observed properties of accreting black holes appear to be scale invariant \citep{Heinz03}. Consequently, scale-invariant semi-analytical models that can match observational data at a fraction of the computational cost of GRMHD can be used guide simulations by probing the parameter space very quickly. More complex theoretical models can then be invoked to deepen our understanding of the physics at play.

In general, semi-analytical models for jets in BHXBs and AGN are typically used to address different scientific questions, and as a result they tend to be set up differently and use different overall assumptions. Models for AGN jets typically take the ``single-zone'' approach (\citealp{Tavecchio98,Boettcher13}, with a few noticeable exceptions like e.g. \citealt{Potter13a,Potter13b,Potter18,Zacharias22}, whose approach is similar to that described in sec.~\ref{sec:bljet}), in which all the jet emission is assumed to originate from a single, spherical blob of plasma at some location in the jet, usually referred to as the blazar zone. These models focus on addressing the origin of the high energy emission observed in AGN jets (blazars especially), and are tailored to probe how and where particles are accelerated within jets and/or whether cosmic rays and neutrinos could be produced \citep[e.g.][]{Tavecchio98,Ghisellini10,Boettcher13,Reimer19}. Single zone models can account well for the optically thin, high energy fraction of the emission, but when lower frequency emission is included in the data-set, the spectral energy distribution (SED) shows a spectral break where the synchrotron emission becomes optically thick due to synchrotron self-absorption effects; below this break, the spectral slope is nearly flat or inverted ($F(\nu)\propto\nu^{-\alpha}$, with $\alpha\lesssim 0$). Non-thermal synchrotron emission from a single region instead predicts $F(\nu)\propto\nu^{5/2}$, because a flat/inverted spectrum cannot be reproduced with one single emitting zone \citep{Blandford79}. More recent works use a two-zone approach and treat particle acceleration in deeper detail \citep[e.g][]{Baring17,Boettcher19}, but conceptually they are similar to standard one-zone models in that they focus on inferring the detailed particle properties within a relatively localised part of the outflow. While highly successful at reproducing the high-energy spectrum of AGN, these models cannot easily be connected back to, and thus cannot help constrain, the larger scale plasma dynamics of the accretion/ejection coupling. Models for BHXBs, on the other hand, tend to focus on coupling the broad properties of entire the outflow. These models often try to couple the properties of the jet to its launch conditions, in the form of spectral \citep[e.g.][]{Markoff01,Markoff05} or timing information \citep[e.g.][]{Kylafis08,Malzac13,Drappeau17,Peault19}, and/or try to account for the detailed evolution of the plasma as it moves downstream in the jet \citep[e.g.][]{Peer09,Zdziarski14}. They are often based on a multi-zone approach inspired by the ``standard'' compact jet model proposed by \cite{Blandford79} and \cite{Hjellming88}. Furthermore, they rarely include the contribution of accelerated protons, with a few exceptions, e.g., \citealt{Pepe15,Kantzas20}. A stratified, multi-zone approach is favoured over the standard AGN-type single zone model for several reasons. First, unlike AGN, among BHXRBs only a few have confirmed $\gamma$-ray detections. Cygnus X-1  has a \Fermi detection associated with the jet \citep{Zanin16}. Cygnus X-3 hase been observed by both \Agile \citep{Tavani09} and \Fermi \citep{Abdo09b}. In a recent \Fermi survey of high-mass BHXRBs, other sources have shown emission in such band as well \citep{Harvey22}. An excess towards V404 Cygni observed by \Agile has been reported by \citealt{Piano17}, but no significant detection is present in the \Fermi data, \citealt{Harvey21}. Second, while the compact radio emission clearly originates in the jet at distance $z\approx 10^{6-8}\,\rm{r_g}$ away from the black hole \citep[e.g.][]{Fender99,vanderHorst13,Russell14b}, in the optical and infra-red disentangling the optically-thin jet spectrum (which we expect to originate around $z\approx 10^{2-4}\,\rm{R_g}$, e.g., \citealt{Gandhi08,Gandhi11,Russell14a}) from direct or reprocessed emission of the accretion flow \citep[e.g.][]{Tetarenko20} or even the companion star \citep[e.g.][]{AlfonsoGarzon18} can be very challenging. These two issues, namely the lack of constraints on the optically thin, high energy emission and the intertwined contribution of the jet and the disc, and potentially companion star, in the optical and infra-red bands, leave a standard single zone model essentially unconstrained when applied to a typical BHXB spectral energy distribution. 

In this work, we present the first public release of the \texttt{BHJet} code\footnote{https://github.com/matteolucchini1/BHJet/}, which is a semi-analytical, steady-state, multi-zone jet model designed to reproduce the SED of accreting black hole jets for all central masses (in this sense, the model can be thought as being scale-invariant). Broadly speaking, the model calculates the time-averaged emission produced by a bipolar jet. If desired, users can also include the contribution of an accretion disk, described by a simple multicoloured black body component similar to \texttt{diskbb}. A population of thermal electrons injected into the jet base can scatter both external and local cyclo-synchrotron photons; this jet base is typically ignored in the works discussed previously, but in our model it takes the role of the X-ray emitting corona ubiquitously detected in accreting black holes. In the outer regions of the jet, the electrons are continuously accelerated into a non-thermal distribution, leading to the typical synchrotron (and inverse Compton) emission observed in jets. These outer jet segments are opaque to synchrotron radiation and produce the typical synchrotron self-absorbed flat/inverted radio spectrum. In this way, \texttt{BHJet} can link the spectral and dynamical properties of the jet acceleration and collimation zone near the black hole, to those of the outer outflow.

The origins of the model can be traced back to the work of \cite{Falcke95}, who extended the work of \citep{Blandford79} to present the first semi-analytical dynamical jet model in which the disc and jet form a coupled system, enforced by setting the total jet power to be linearly proportional to $\dot{M}c^2$, the accretion power from the disc. This model was successfully applied to explain the broad radio and dynamical properties of radio loud and radio quiet quasars \citep{Falcke95b} and low luminosity AGN and the first 'microquasar' GRS~1915+105 \citep{Falcke99}. \cite{Markoff01} extended the dynamical treatment to a full multi-zone, multiwavelength model incorporating particle distributions, synchrotron and single-scattering inverse Compton radiation, called \texttt{agnjet}, first applied to the Galactic centre supermassive black hole Sgr A* \citep{Falcke00,Markoff01b} in order to model the quiescent and flaring SEDs. \citet{Markoff01} then showed that the same model could be scaled down to reproduce the full hard-state SED of the BHXB XTE\,J1118$-$480. This paper was the first to demonstrate that the X-ray emitting corona, which is a ubiquitous sign of black hole accretion, may in fact be located in the innermost regions of the jet. Subsequently \cite{Markoff03} and \cite{Markoff05} showed that this spectral component can also reproduce the X-ray spectra of BHXBs, strengthening the suggestion that the base of the jet may be associated with the corona. Crucially, this finding also introduced a significant degeneracy to the model, as the two radiative mechanisms (thermal Comptonisation and non-thermal, optically thin synchrotron) can sometimes reproduce the data equally well while requiring very different physical conditions in the outflow (\citealt{Markoff08,Nowak11,Markoff15,Connors17}, although see \citealt{Zdziarski03,Yuan07}). One potential way to break this degeneracy, and isolate the radiative mechanism responsible for the high-energy emission, is to identify the reflection features in the X-ray spectrum of a given source: non-thermal synchrotron from accelerated particles downstream ($z\gtrsim10^{2}\,\rm{R_g}$) in the jet should result in weak, non-relativistic reflection spectra, while inverse Compton near the jet base ($z\approx10\,\rm{R_g}$) predicts that reflection should be more prominent and relativistic \citep{Markoff04}. These different radiative scenarios will also lead to different predicted lags between the various bands; however in this paper we will only cover SED modelling, see, \citep[e.g][]{Gandhi11,Kara19,Wang21} for spectral-timing studies. 

The model has undergone several major changes recently. First, the dynamical treatment of the jet has been overhauled to be more self-consistent and versatile. The main drawback of the approach of \cite{Falcke95} is that the outflow can only accelerate up to mildly relativistic Lorentz factors ($\gamma\approx2-3$). While low Lorentz factors are consistent with observations of both BHXBs and LLAGN \citep[e.g.][]{Fender04a,King16}, this assumption is inconsistent with observations of powerful AGN, particularly blazars \citep[e.g.][]{Cohen71,Whitney71,Aharonian06,Pushkarev09}. Furthermore, in response to criticism from \cite{Zdziarski16}, \cite{Crumley17} pointed out that the original model does not account for the energy required to accelerate the leptons in the jet and power the observed emission, meaning that the original \texttt{agnjet} model violates energy conservation by a factor of $\approx 2$. These issues were addressed in \cite{Lucchini19a}, who introduced an improved dynamical model ``flavour'', called \texttt{bljet}, which treats the overall jet energy budget and magnetic content of the outflow more self-consistently via a Bernoulli approach, allowing for arbitrarily large bulk Lorentz factors. Second, \cite{Connors19} highlighted an issue raised by assuming that the radiating leptons are all relativistic throughout the jet, but particularly in the base. This scenario prevents a smooth power-law from inverse Compton scattering because the orders are visibly separated from each other, requiring a finely-tuned combination of synchrotron, inverse Compton, and reflection, in order to successfully model the spectra of BHXBs. However, these fine tuned models \citep[e.g.][]{Markoff05} are in tension with measurements of low frequency X-ray lags in BHXBs \citep[e.g.][]{Kotov01,Arevalo06}. This issue has been overcome in \cite{Lucchini21}, who updated the treatment of the lepton distribution in the model to allow for non-relativistic temperatures and found, not surprisingly, that in this regime the base of the jet produces spectra that are effectively identical to standard corona ``lamp-post'' models \citep[e.g.][]{Matt91,Beloborodov99,Dauser13,Mastroserio18}. We recently improved the inverse Compton calculation, which now allows the code to transition from single to multiple scattering regimes and, therefore, is able to handle a larger range of optical depths and electron temperatures. A preliminary version of this function was already used in \cite{Connors19} and discussed in \cite{Lucchini21}, but here we present for the first time an updated version that has a more refined treatment of the radiative transfer. We benchmarked it against the widely known \texttt{CompPS} code to define the range of applicability and verify the output spectral shapes across such range.  All of these changes have made the latest version of the model (and source code) much more versatile, and significantly different, compared to the original work it is based on. 

The main goal of this paper is to present the details of the newest features while providing a unified documentation of the newest version of the model to support its associated public release. We refer to this model as \texttt{BHJet}, which joins the \texttt{agnjet} and \texttt{bljet} model flavours. Finally, we describe the general characteristics of the model when it is applied to several types of accreting black hole systems. For completeness, we also discuss legacy features presented in older work, and compare these features to more recent updates. 

In section \ref{sec:overview} we present a brief overview of \texttt{BHJet} and its code structure; in sections \ref{sec:particles} and \ref{sec:radiation} we discuss a new library of \textsc{C++} classes the code is based on, and in section \ref{sec:BHJet} we detail both the model flavours available. In section \ref{sec:GRMHD} we qualitatively compare the physics of the jet in our code with the results of global GRMHD simulations. In section \ref{sec:applications} we apply the model to the SEDs of bright hard state BHXBs, LLAGN, and powerful flat spectrum radio quasars (FSRQs). At last, in section \ref{sec:conclusion}, we draw our conclusions.

\section{Code overview}
\label{sec:overview}

The \texttt{BHJet} family of models presented in this paper is built on a small library of \textsc{C++} classes called \texttt{Kariba}, which are presented here for the first time and can be found in the \texttt{BHJet} GitHub repository. \texttt{Kariba} is designed to account for several standard spectral components observed in accreting black holes, as well as the underlying particle distributions responsible for the multiwavelength emission. \texttt{Kariba} was designed with the goal of being simple and versatile enough to treat many different systems and spectral components, while also being as computationally efficient as possible. The \texttt{BHJet} family of models describes several flavours of black hole jets, detailed below, and calls objects from \texttt{Kariba} to compute the final SED. Beyond standard \textsc{C/C++} dependencies, the only additional library required is the GNU Scientific Library\footnote{https://www.gnu.org/software/gsl/}, which we use for numerical interpolation, integration and derivation.

Along with the \texttt{Kariba} library of classes and the main function, we provide two ways to run the model. We include a \textsc{C++} wrapper, which reads a file with the necessary input parameters, runs the code calculating the emission on an appropriate frequency grid, and runs a Python plotting script to show the output. We also include a \textsc{SLIRP} file and a \textsc{S-Lang} wrapper (similar to the \textsc{C++} wrapper, but without plotting functions), so that users can import the model into the spectral fitting package \textsc{ISIS} \citep{Houck00}. We also note that the format of the array returned by the model is also compatible with \textsc{XSPEC} \citep{Arnaud96}, and thus in principle should be easily used in this package as well. We note however that users should never use \texttt{BHJet} to fit exclusively X-ray spectra, since the model is designed for multiwavelength (radio-to-$\gamma$-ray) emission. As a result, fitting a single part of the spectrum is very likely to result in best fit parameters that are almost entirely unconstrained and/or non-physical.

\begin{figure*}
    \centering
    \includegraphics[width=\textwidth]{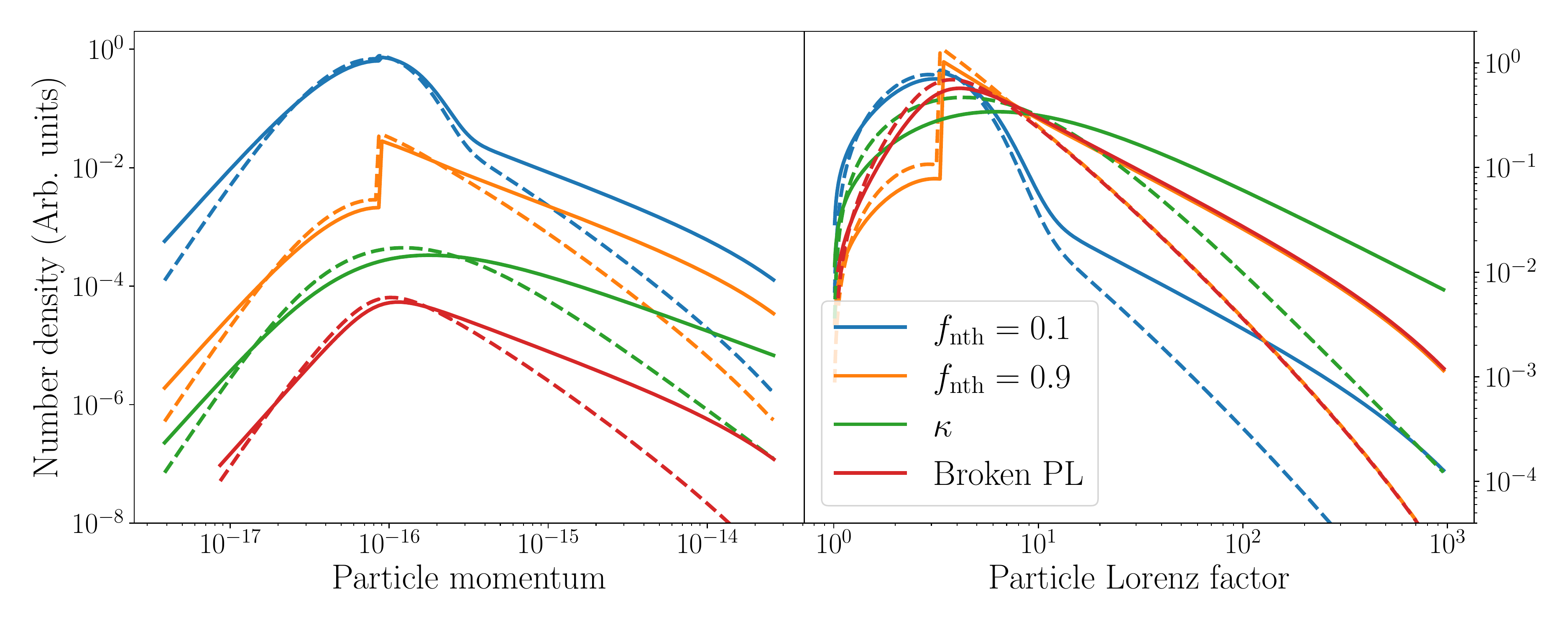}
    \caption{Different hybrid distributions for the same input parameters: emitting region of radius 20 $R_{\rm g}$ for a 10 $M_{\sun}$ black hole, corresponding to $3\times 10^{7}\rm{cm}$, temperature $511\,\rm{keV}$, maximum Lorentz factor $10^{3}$, B-field $10^{5}\rm{G}$, expansion speed of plasmoid $\beta_{\rm exp} = 0.1$, non-thermal slope $s=2$ corresponding to $\kappa=3$. The left panel shows particle distributions in momentum space, re-normalised to illustrate the difference in shapes; the right panel shows Lorentz factor units taking the same normalisation, to highlight the difference in number density of particles in the non-thermal tails of each distribution. Solid lines neglect cooling, dashed lines show the solution of equation (\ref{eq:FP_steadystate}) for each injected distribution.}
    \label{fig:particles}
\end{figure*}

\section{Kariba library: particle distributions}
\label{sec:particles}

All particle distributions in \texttt{Kariba} are calculated in momentum space and in the co-moving frame of the emitting region, in units of dimensionless momentum $\varrho = p/m_{\rm e}c$, in order to capture both the relativistic and trans-relativistic regimes. For each value of $\varrho$, the corresponding Lorentz factor is $\gamma(\varrho) = \sqrt{\varrho^{2}+1}$; therefore in the relativistic limit, $\varrho \approx \gamma$. Each type of particle distribution is supported by a separate C++ class, inherited from a base \texttt{Particles} class.

Some definitions and methods are common for all particle distribution classes. First, for a particle number density $n$, in units of particles per unit volume, the normalisation $N_{\rm 0}$ (in units of ${\rm cm}^{-3}$) of the particle distribution (regardless of its shape) is always defined as:
\begin{equation}
    N_{\rm 0} = \frac{n}{\int N(\varrho)d\varrho},
    \label{eq:norm}
\end{equation}
where $N(\varrho)$ is the (dimensionless) functional form of the particle distribution, independent of normalization (for example, in the case of a power-law distribution $N(\varrho)=\varrho^{-s}$). Second, after calculating the particle distribution in units of dimensionless momentum, the code automatically computes and stores the particle distribution in Lorentz factor units (which can then be used as an input in the classes treating radiative processes) by taking $N(\varrho) d\varrho = N(\gamma) d\gamma$, so that:
\begin{equation}
    N(\gamma) = N(\varrho)\frac{d\varrho}{d\gamma}.
\end{equation}
All (non-thermal) distribution classes include a method to calculate the effects of adiabatic and radiative cooling, once the particles have reached a steady state. The radiative loss term is defined as in \cite{Ghisellini98} :
\begin{equation}
        \dot{\varrho}_{\rm rad} = -\frac{4 \sigma_{\rm t}c U_{\rm rad}}{3m_{\rm e} c^{2}}\varrho\gamma,
\label{eq:rad_dot}
\end{equation}
where $\sigma_{\rm t}$ is the Thomson cross section and $U_{\rm rad}$ is the appropriate  energy density; $U_{\rm rad}= U_{\rm b}$ for the case of synchrotron cooling. This form for radiative losses assumes that radiative cooling always occurs far from the Klein-Nishina regime, in which the cross section is reduced (e.g. \citealt{RybickiLightman, deKool89}). This is generally true for synchrotron losses, but may not be correct for inverse Compton scattering. However, jetted objects in which inverse Compton scattering losses dominated significantly over synchrotron losses tend to be limited to the most powerful blazars (discussed more in depth in sec.\ref{sec:blazars}). The leptons responsible for the emission in these sources tend to have fairly low Lorentz factors (as discussed later, but see also e.g. \citealt{Fossati98,Ghisellini17}), in which case the scattering still occurs in the Thomson regime. Therefore, the error introduced by neglecting the Klein-Nishina regime in inverse Compton losses is generally very small. Furthermore, our choice to use equation (\ref{eq:rad_dot}) means that our code currently ignores the effect of synchrotron self-absorption on the loss term \citep[e.g.][]{Ghisellini98,Katarzynsky06,Zdziarski14}.
Similarly, the adiabatic loss term is defined as:
\begin{equation}
    \dot{\varrho}_{\rm ad} = -\frac{\beta_{\rm exp}c}{r}\varrho,
\label{eq:ad_dot}
\end{equation}
where $\beta_{\rm exp}$ is the expansion speed of the emitting region and $r$ its radius. As in \cite{Boettcher13}, $\beta_{\rm exp}$ can be thought of as a parameter which absorbs the uncertainty in the electron diffusion coefficient, the importance of adiabatic losses within the emitting region, and the nature of the expansion (equation (\ref{eq:ad_dot}) is correct for 3D expansion, but requires an additional factor of 2/3 in the case of 2D expansion which users can incorporate into $\beta_{\rm exp}$, should they choose to). 

The effects of cooling on the steady-state particle distribution in momentum space, assuming constant injection and neglecting the spatial advection of particles, can be computed (for a given injection term) by solving the equation:
\begin{equation}
    N(\varrho) = \frac{\int Q(\varrho^{\prime}) d\varrho^{\prime}}{\dot{\varrho}_{\rm ad} + \dot{\varrho}_{\rm rad}};
\label{eq:FP_steadystate}
\end{equation}
note that for large particle momentum, $\varrho \approx \gamma$ and equation (\ref{eq:FP_steadystate}) reduces to the standard relativistic form. The definition in equation (\ref{eq:ad_dot}) slightly overestimates the adiabatic loss term for $\varrho \leq 1$; as a result, equation (\ref{eq:FP_steadystate}) slightly underestimates the number of particles at the low-energy end of the distribution, as shown by the dashed lines in fig.\ref{fig:particles}. However, the contribution of these particles to the total observed radiation is generally negligible, and as a result this small inaccuracy has no impact on the SEDs computed by our code.

The current implementation of the particle distribution classes has two main limitations. First, the treatment detailed below suffers severe numerical issues for very low average particle momenta ($\approx 2-3$ keV and below), causing the normalisations in the code to diverge. The second limitation is that we do not explicitly include pair creation/annihilation, and its effects on the particle distribution. This will be supported in future releases; until then, it is up to the user to ensure a given set of input parameters is physically self-consistent (for example, a sufficiently low optical depth to allow $\gamma$-rays to escape the source un-absorbed). The jet model presented in this work automatically prints a warning on the terminal when runaway pair production may occur for the chosen set of input parameters.

\subsection{Supported distributions}
\label{sec:distributions}

The library supports five different types of particle distributions (with class names \texttt{Thermal, Powerlaw, Bknpower, Mixed} and \texttt{Kappa}), allowing it to be applied to a variety of regimes in which purely thermal, purely non-thermal, or both types of particles are present in the emitting region. The \texttt{Thermal} and \texttt{Mixed} classes described here are evolutions of the code first introduced in \cite{Markoff01} and further developed in \cite{Connors17} and \cite{Lucchini21}, and are reported here for completeness. The \texttt{Thermal} distribution does not support the solution highlighted in equation (\ref{eq:FP_steadystate}) to compute the combined effects of radiative and adiabatic cooling. The reason for this choice is that for the thermal distribution only, we already assume that the particles have reached their steady state. For any other particle distribution, on the other hand, users can either specify the steady state distribution through equation (\ref{eq:powerlaw}), (\ref{eq:bknpower}), (\ref{eq:mixed}) or (\ref{eq:kappa}), or use the same equations to assume an injection term, and then include the effects of cooling through equation\ref{eq:FP_steadystate}. In this latter case, the final distribution $N(\varrho)$ is re-normalised appropriately in order to conserve the total particle number density.

Purely thermal particles are described by the Maxwell-J{\"u}ttner distribution in momentum space:
\begin{equation}
    N(\varrho) = N_{\rm 0} \varrho^2e^{-\frac{\gamma(\varrho)}{\Theta}},
\end{equation}
where $\Theta = kT_{\rm e}/m_{\rm e}c^{2}$ is the temperature of the leptons in units of $m_{\rm e}c^{2}$, $\gamma(\varrho)=\sqrt{\varrho^{2}+1}$ is the Lorentz factor corresponding to the dimensionless momentum $\varrho$, and $k$ is the Boltzmann constant. In this case, equation (\ref{eq:norm}) reduces to $N_{\rm 0} = n/m_{\rm e}c^{3}\Theta K_{\rm 2}(1/\Theta)$, where $K_{\rm 2}(1/\Theta)$ is the modified Bessel function of the second kind. 

Purely non-thermal particles can be described either by a simple power-law with an exponential cutoff:
\begin{equation}
    N(\varrho) = N_{\rm 0} \varrho^{-s}e^{-\varrho/\varrho_{\rm max}}, \quad \varrho \geq \varrho_{\rm min},
    \label{eq:powerlaw}
\end{equation}
where $s$ is the power-law slope, $\varrho_{\rm min}$ and $\varrho_{\rm max}$ are the minimum and maximum momenta, or by a smoothly broken power-law with an exponential cutoff:
\begin{equation}
    N(\varrho) = N_{\rm 0} \frac{(\varrho/\varrho_{\rm brk})^{-s_1}}{1+(\varrho/\varrho_{\rm brk})^{-(s_1+s_2)}} e^{-\varrho/\varrho_{\rm max}}, \quad \varrho \geq \varrho_{\rm min},
    \label{eq:bknpower}
\end{equation}
where $\varrho_{\rm brk}$ is the break momentum, and $s_1$ and $s_2$ are the slopes of the particle distribution before and after the break.  A hybrid thermal+non-thermal particle distribution can be roughly approximated by taking $s_1 = -2$ (for large $\varrho$, $\gamma\approx\varrho$ and the Maxwell-J{\"u}ttner distribution scales as $\gamma^{2}$ before the thermal cutoff) and setting $\varrho_{\rm brk}$ at the average momentum of the corresponding Maxwellian distribution.

Finally, two types of mixed distributions are included. The first is a simple mixed distribution, in which the total number density is divided between a Maxwell-J{\"u}ttner pool and a non-thermal tail:
\begin{equation}
N(\varrho) = \left\{\begin{aligned} 
(1-f_{\rm nth})N_{\rm 0,th}\varrho^2e^{-\frac{\gamma(\varrho)}{\Theta}} 
& \quad \varrho<\varrho_{\rm min} \\
(1-f_{\rm nth})N_{\rm 0,th}\varrho^2e^{-\frac{\gamma(\varrho)}{\Theta}} + & \\ f_{\rm nth}N_{\rm 0,nth}\varrho^{-s}e^{-\frac{\varrho}{\varrho_{\rm max}}} & \quad  \varrho\geq\varrho_{\rm min},  \end{aligned}\right.
\label{eq:mixed}
\end{equation}\\
where the distribution is normalized so that $N_{\rm 0} = N_{\rm 0,th} + N_{\rm 0,nth}$ follows the definition in equation (\ref{eq:norm}). The minimum momentum of the non-thermal tail $\varrho_{\rm min}$ is taken to be the average momentum of the thermal pool. In this way, the power-law component contains a fraction $f_{\rm nth}$ of the total particle number density $n$, and the rest are in the thermal pool, similarly to the behaviour observed in particle-in-cell simulations \cite[e.g.][]{Sironi11,Sironi13,Sironi14,Sironi15,Crumley19}. The second hybrid distribution supported is the relativistic $\kappa$ distribution, which is commonly used to smoothly join a thermal and non-thermal distribution \citep[e.g.][]{Livadiotis13}, particularly when post-processing GRMHD simulations \citep[e.g.][]{Davelaar18}:
\begin{equation}
     N(\varrho) = N_{\rm 0} \gamma(\varrho)\sqrt{\gamma(\varrho)^{2}-1}\left(1+\frac{\gamma(\varrho) -1}{\kappa \Theta}\right)^{-\kappa-1},
     \label{eq:kappa}
\end{equation}
where $\Theta$ is the dimensionless temperature of the thermal particles, and the $\kappa$ index is related to the typical power-law slope $s$ by $\kappa = s+1$.

We note that while all of these prescriptions can mimic a hybrid thermal/non-thermal population, each has its own set of limitations which users need to be mindful of. These limitations are highlighted in fig.~\ref{fig:particles}. First, the mixed distribution (equation (\ref{eq:mixed})) only remains smooth between the thermal and non-thermal parts if the former dominates the particle number density, shown by the blue line. This regime roughly captures a plasma in which particle acceleration is not very efficient. If instead $f_{\rm nth} \geq 0.3$ (which can be thought of as indicating more efficient particle acceleration), shown by the orange line, a discontinuity between the two branches appears near the peak of the Maxwellian. In this case, the broken power-law or $\kappa$ distributions are more appropriate. However, these two descriptions differ noticeably from each other. The broken power-law prescription, shown by the red line, does not capture the shape of the Maxwellian peak very well, but beyond it the normalization matches that of the orange mixed distribution. The $\kappa$-distribution, shown by the green line, has a broader shape near the peak, similar to the Maxwellian. However, the greater width of the distribution also results in more particles being channelled in the non-thermal tail. This behaviour becomes more important for increasing $\Theta$, and its result is to predict the largest luminosity (due to the increased average Lorentz factor of the particles) out of all of these prescriptions, for a given set of input parameters. Finally, we note that the effects of cooling, computed such that the cooling break $\rho_{\rm b}$ is near the peak of the distributions (shown by the dashed lines), do not mitigate these conclusions significantly. 

\section{Kariba library: radiation}
\label{sec:radiation}

Similarly to the particle distributions, the radiation methods for different spectral component are inherited from a base \texttt{Radiation} class. The derived classes currently implemented are \texttt{BBody, Compton, Cyclosyn}, and \texttt{ShSDisk}. The \texttt{Cyclosyn} and \texttt{Compton} classes support both spherical and cylindrical emitting regions. Regardless of the radiative mechanism (thermal or non-thermal), the main output of every class is two arrays of photon energies (in units of $\rm{erg}$) and two of specific luminosities (in units of $\rm{erg\,s^{-1}\,Hz}$), tracking both observer and lab frame quantities. All of these classes have been presented in \cite{Lucchini21} and are described here for completeness; however, the \texttt{Compton} class has received several significant improvements, detailed below.

\subsection{Thermal components}
\label{sec:rad_thermal}

Currently, two thermal components are included in \texttt{Kariba}: a generic black body spectrum (\texttt{BBody}) and a Shakura-Sunyaev disc (\texttt{ShSDisk}), which may either be truncated or extend all the way to the innermost circular stable orbit.

\texttt{BBody} requires a temperature $T_{\rm bb}$ and luminosity $L_{\rm bb}$ to be specified. The specific luminosity then is calculated as:
\begin{equation}
    L_{\rm bb}(\nu) = \frac{L_{\rm bb}}{\sigma_{\rm sb}T_{\rm bb}^{4}c^{2}} \frac{2h\nu^{3}}{e^{h\nu/kT_{\rm bb}}-1},
\end{equation}
where $h$, $\sigma_{\rm sb}$ are the Planck and Stefann-Boltzmann constants, respectively.

\texttt{ShSDisk} requires as input the inner truncation radius $r_{\rm in}$, the outer radius $r_{\rm out}$ (both measured in units of the gravitational radius of the black hole $r_{\rm  g}$), either a luminosity $L_{\rm d}$ or a temperature $T_{\rm in}$ at the innermost radius, and the viewing angle $\theta$. $T_{\rm in}$ and $r_{\rm in}$ connected through:
\begin{equation}
    T_{\rm in} = \left[\frac{L_{\rm d}}{2\pi\sigma_{\rm sb}(r_{\rm in}r_{\rm  g})^{2}}\right]^{1/4},
\end{equation}
which assumes that only one side of the disk is seen by the observer. The specific luminosity is:
\begin{equation}
    L_{\rm d}(\nu) = \cos(\theta)\int_{r_{\rm in}r_{\rm  g}}^{r_{\rm out}r_{\rm  g}} 4\pi r \frac{h\nu^{3}}{c^{2}(e^{h\nu/kT(r)}-1)} d r,
\end{equation}
and the temperature profile follows the standard $T(r)\propto (r/r_{\rm in}r_{\rm  g})^{-3/4}$ scaling, assuming non-zero torque at the disk boundary. Finally, we take the disc scale height to be:
\begin{equation}
H/R = \max (0.1,  L_{\rm d}/L_{\rm Edd}),
\label{eq:diskH}
\end{equation}
where $L_{\rm Edd} = 1.26\cdot10^{38}(M/M_{\sun})\rm{erg\,s^{-1}}$ is the Eddington luminosity of the black hole (with mass $M$ measured in Solar units). This factor was originally introduced to have a more realistic geometry of the disk and a more accurate estimation of the seed photon energy density; in general, its impact on the SED is minimal. We note here that this is an extremely simplistic model for the disk emission, not including disk irradiation \citep[e.g.][]{Gierlinski08} or colour corrections \citep[e.g.][]{Kubota98}. As such, inferring physical conclusions on the nature of the disk in an accreting system should not be the goal of fitting this model component.

\subsection{Cyclo-synchrotron}

The treatment of cyclo-synchrotron emission is similar to that of \cite{Blumenthal70}, with a minor modification to account for cyclotron emission in the near-relativistic regime. 

For particle Lorentz factor $\gamma > 2$ (synchrotron) we use the standard emissivity:
\begin{equation}
    j_{\rm s}^{\prime}(\nu^{\prime},\gamma) =  \frac{\sqrt{3}e^{3}B\sin{i}}{m_{\rm e}c^{2}} \frac{\nu^{\prime}}{\nu_{\rm s}^{\prime}}\int_{\nu^{\prime}/\nu_{\rm s}^{\prime}}^{\infty} K_{5/3}(y)dy,  
\label{eq:syn}
\end{equation}
where $\nu^{\prime}$ is the emitted frequency in the co-moving frame of the emitting region, $\gamma$ the Lorentz factor of the emitting electrons, $B$ is the magnetic field in the emitting region, $i$ the electron pitch angle, $e$ the electron charge, and $\nu_{\rm s} = 3eB\gamma^{2}/4\pi m_{\rm e}c$ is the scale synchrotron frequency. The code currently assumes an isotropic distribution of pitch angles, and averages over them.

For particle Lorentz factor $\gamma \leq 2$ instead we use the cyclotron emissivity of \cite{Ghisellini98}:
\begin{equation}
    j_{\rm c}^{\prime}(\nu^{\prime},\gamma) = \frac{4\varrho^2}{3}
\frac{\sigma_{\rm T}c U_{\rm b}}{\nu_{\rm l}}\frac{2}{1+3\varrho^2}
\exp\left[\frac{2(1-\nu^{\prime}/\nu_{\rm l}^{\prime})}{1+3\varrho^2}\right],  
\label{eq:cyclo}
\end{equation}
where $\nu^{\prime}$ is the emitted frequency in the co-moving frame of the emitting region, $\varrho$ is the dimensionless particle momentum, $\sigma_{\rm t}$ the Thomson cross section, $U_{\rm b} = B^{2}/8\pi$ the magnetic energy density in the emitting region, and $\nu_{\rm l} = eB/2\pi m_{\rm e} c$ is the Larmor frequency. Regardless of the form of $ j_{\rm c}^{\prime}(\nu^{\prime},\gamma)$, the total emissivity is the integral over the particle distribution, averaged over all pitch angles:
\begin{equation}
    j^{\prime}(\nu^{\prime}) = \int_{\gamma_{\rm min}}^{\gamma_{\rm max}} N(\gamma) j_{\rm c,s}^{\prime}(\nu^{\prime},\gamma) d\gamma,
\end{equation}
Similarly, following \cite{Blumenthal70} we take the absorption coefficient for a given emissivity to be:
\begin{equation}
 \alpha^{\prime}(\nu^{\prime}) = \int_{\gamma_{\rm min}}^{\gamma_{\rm max}} \frac{N(\gamma)}{\gamma^{2}} \frac{d}{d\gamma}\left( \gamma^{2} j_{\rm c,s}^{\prime}(\nu^{\prime},\gamma)\right) d\gamma.  
\label{eq:alfa1}   
\end{equation}
Finally, the co-moving specific luminosity is:
\begin{equation}
L_{\rm s}^{\prime}(\nu^{\prime}) = \frac{16 \pi V \nu^{\prime 2}}{c^{2}}\frac{j^{\prime}(\nu^{\prime})}{\alpha^{\prime}(\nu^{\prime})}\left(1-e^{-\tau^{\prime}(\nu^{\prime})} \right),    
\label{eq:syn-comoving}
\end{equation}
where $V$ is a re-normalising factor to account for different emitting region geometries; $V = R^{2}$ for a sphere of radius $R$ and $V=HR$ for a cylinder of radius $R$ and height $H$, and 
\begin{equation}
 \tau_{\rm s}^{\prime}(\nu^{\prime}) = \frac{ \alpha^{\prime}(\nu^{\prime}) R }{\delta f(\theta)} 
\end{equation}
is the cyclo-synchrotron optical depth, which approximates skin depth and viewing angle effects; we take $f(\theta) = 2\sin(\theta)/\pi$ for cylindrical emitting regions, and $f=3/2\pi$ for spherical ones, in order to correctly account for the observed emitting volume both in the optically thin and thick regimes. 

\subsection{Inverse Compton}
\label{sec:Kompton}

Performing full radiative transfer calculations describing the inverse Compton spectrum for a wide regime of optical depths ($\tau=10^{-5}-3$) and electron temperatures ($T_e = 20 - 2500$ keV) is a cumbersome problem that has been extensively studied in the literature \citep[e.g.][]{Sunyaev85,Hua95,Poutanen96, Zdziarski96,Zdziarski14}. Within this code, we do not aim at performing such calculations in detail, but rather to  approximate the shape of the inverse Compton spectrum while containing the run time to allow for an efficient multiwavelength fitting procedure. In this section we describe how we obtain such approximated spectrum. 

Similarly to the case of cyclo-synchrotron, we use the inverse Compton kernel of \cite{Blumenthal70} and adopt their notations here, which accounts for the full Klein-Nishina cross section for Lorentz factor $\gamma\gg1$. We stress the fact that this is not strictly correct for the entirety of the ($\tau,T_e$) ranges we want to cover, but we will use this to produce an initial spectrum and then apply a correction factor, when needed, depending on which combination of optical depths and electron temperature we are sampling. For each lepton interacting with a photon field with number density $N_{\rm ph,0}(\epsilon_{\rm 0})$ (in units of number of photons, per unit volume and initial photon energy), the scattered photon spectrum in the co-moving frame of the emitting region is:

\begin{figure*}
    \centering
    \includegraphics[width=\textwidth]{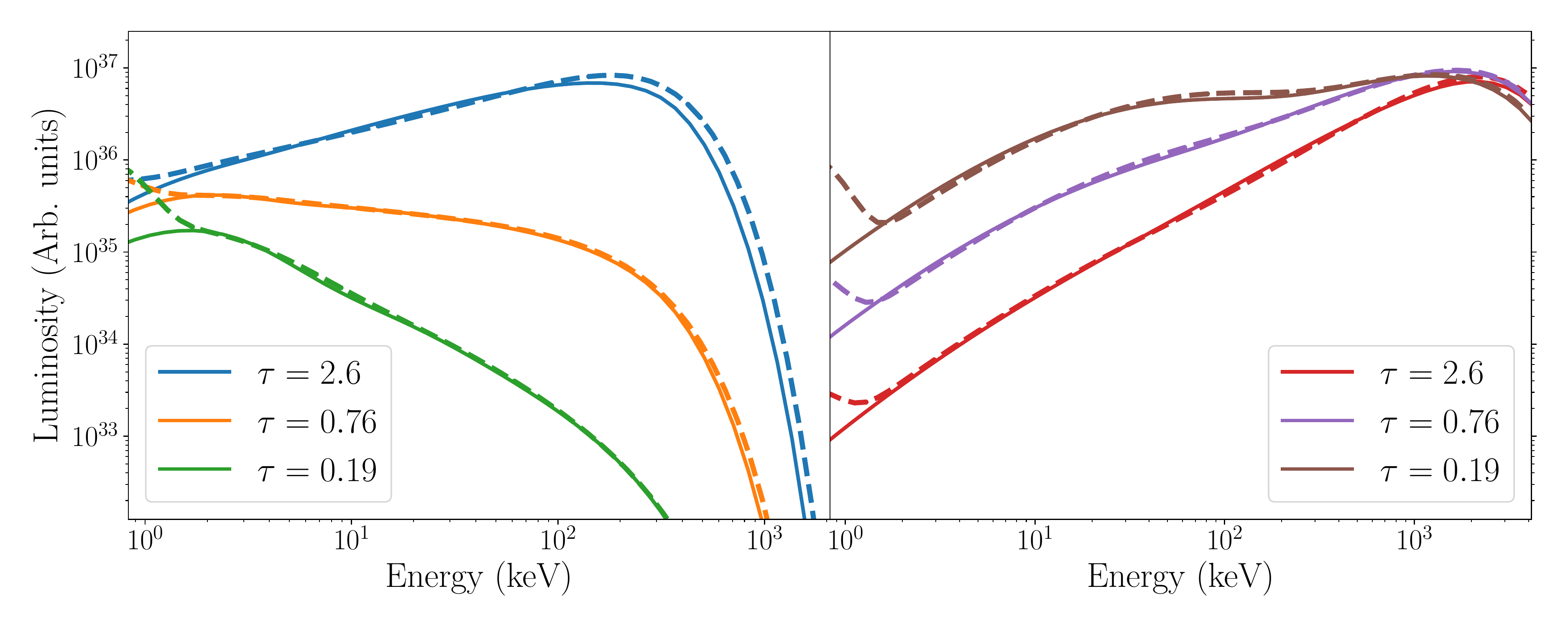}
    \caption{Examples of output Comptonisation spectra from \texttt{Kariba} (continous lines), compared with \textup{CompPS} (dashed lines), for a range of typical optical depths. The spectra have been re-normalised to the typical luminosity of a hard state XRB. Left panel: low temperature coronae ($T_{\rm e}=90\,\rm{keV}$); right panel: high temperature coronae ($T_{\rm e}=900\,\rm{keV}$). The two codes are in excellent agreement; the correction factors $f(T_{\rm e},\tau)$ in this case (for increasing optical depth) are $f=0.35,0.35,0.38$ for $T_{\rm e}=90\,\rm{keV}$ and $f=0.57,0.61,0.73$ for $T_{\rm e}=900\,\rm{keV}$.}
    \label{fig:corona}
\end{figure*}

\begin{equation}
    \frac{dN_{\rm ph}}{dt d\epsilon_{\rm 1}} = \frac{2\pi r_{\rm e}^{2} m_{\rm e}c^{3} N_{\rm ph,0}(\epsilon_{\rm 0}) d\epsilon_{\rm 0}}{\gamma \epsilon_{\rm 0}} F_{\rm IC}(q, \Gamma_{\rm e}), 
\label{eq:IC_kernel}
\end{equation}
where $\epsilon_{\rm 0} = h\nu$ and $\epsilon_{\rm 1}=h\nu'$ are the initial and final photon energies, and $r_{\rm e}$ is the classical radius of the electron, $\gamma$ is the electron's Lorentz factor. The factor $F_{\rm IC}(q, \Gamma_{\rm e})$ is defined as:
\begin{equation}
    F_{\rm IC}(q, \Gamma_{\rm e}) = 2q\ln{q}+(1+2q)(1-q)+\frac{(\Gamma_{\rm e}q)^{2}(1-q)}{2(1+\Gamma_{\rm e}q)},    
\label{eq:KN}
\end{equation}
and accounts for whether the scattering occurs in the Thomson or Klein-Nishina regimes, through the quantity $\Gamma_{\rm e} = 4\epsilon_{\rm 0}\gamma/m_{\rm e}c^{2}$ ($\Gamma_{\rm e} \ll 1$ and $\Gamma_{\rm e} \gg 1$ respectively). $q = E_{\rm 1} / (\Gamma_{\rm e} (1 - E_{\rm 1}))$ accounts for the photon energy gain from $\epsilon_{\rm 0}$ to $\epsilon_{\rm 1}$, and $E_{\rm 1} = \epsilon_{\rm 1}/\gamma m_{\rm e}c^{2}$ is the final photon energy in units of the initial electron energy. The spectrum for an individual scattering order is found by integrating over the particle and seed photon distributions:

\begin{equation}
\frac{dN_{\rm ph}}{dt d\epsilon_{\rm 1}} = \int \int N(\gamma) \frac{dN_{\rm ph}}{dt d\epsilon_{\rm 1}} d\gamma d\epsilon_{\rm 0},  
\label{eq:IC}
\end{equation}
which has units of total number of scatterings, per unit of outgoing photon energy, volume and time. This quantity is then multiplied by $h \epsilon_{\rm 1}$, and by the volume of the emitting region $\mathcal V$, in order to obtain a specific luminosity ($\rm erg\,s^{-1}\,Hz^{-1}$). In the version of the code presented here, if the Thomson optical depth $\tau=n_{\rm e}\sigma_{\rm t}R \geq 1$ then only emission from the volume of the outer shell of the emitting region is assumed to escape, down to the skin depth for which $\tau = 1$; previous versions were limited to the optically thin regime ($\tau \leq 1$).

To calculate each successive scattering order, the output of the last scattering computed through equation (\ref{eq:IC}) is then passed as the input spectrum back in the same equation, if necessary re-normalising the volume such that only photons in the region up to $\tau = 1$ are considered. The total output spectrum is computed each time as the sum over all scattering orders:
\begin{align}
\frac{dN_{\rm ph,  tot}}{dtd\epsilon_{\rm1}} = \sum_{i} \int \int N(\gamma) \frac{f(T_{\rm e},\tau)dN_{{\rm ph},i}}{dtd\epsilon_{\rm 1}} d\gamma d\epsilon_{\rm 1};
\label{eq:IC_scatters}
\end{align}
where the factor $f(T_{\rm e},\tau)$ is a re-normalisation factor that we introduce in order to roughly mimic the effects of a full radiative transfer calculation including escape probability, as well as pair production/annihilation. We estimated this correction factor by generating a table of spectra using the \texttt{CompPS}\footnote{We chose \texttt{CompPS} because it allows for a wider interval in both optical depth and temperature than other Comptonisation codes available in \textsc{Xspec}, and because it includes geometries that are nearly identical to the cases we consider in our model.} model, ranging between $\tau = 0.05-3$ and $T_{\rm e}=20-2500$ keV and taking the viewing angle to be 45$^{\circ}$, for both spherical and cylindrical geometries (in this case, we fixed the aspect ratio $H/R$ to 2). The range of optical depths and temperatures was chosen because these are the hard-coded limits for \texttt{CompPS} in \textsc{Xspec} for both spherical and cylindrical geometries; therefore, we expect the model output to be reliable. We note that both inclination and aspect ratio have a minor impact on the output of \texttt{CompPS}, particularly for the cylindrical geometry. (e.g. fig. 2 in \citealt{Poutanen96}). We did not account for either, and take $\theta = 45^{\circ}$ for the source inclination and $h/r=1$ for the aspect ratio for our benchmarks. The introduction of the correction factor based on the \texttt{CompPS} model allows us to extend the range of optical depths and electron temperatures that we can handle with our code while keeping the computational time fairly low. 

The value of the correction factor $f(T_{\rm e},\tau)$ is estimated by matching  the spectral index obtained by our code with that of \texttt{CompPS} for a given set of $\tau$ and $T_e$. We find that without this correction, equation (\ref{eq:IC_scatters}) tends to predict spectra that are slightly harder than \texttt{CompPS}, and therefore $f(T_{\rm e},\tau) \approx 0.1-0.9$ for most values of $T_{\rm e}$ and $\tau$. We then interpolate the table of correction factors to estimate the value of $f(T_{\rm e},\tau)$ for a given temperature and optical depth. For $\tau<0.05$, i.e. the original range of optical depths explored by older versions of \texttt{agnjet}, we always consider only one scattering order and therefore just the pure \cite{Blumenthal70} IC kernel, thus avoiding the need to include the correction factor. A comparison between Comptonisation spectra calculated with \texttt{Kariba} and \texttt{CompPS}, assuming a spherical corona and thermal seed photons, is shown in fig.\ref{fig:corona}. The codes are in excellent agreement for a wide range of temperatures and optical depths.

The total number of scattering orders can be set by the user; for typical applications (e.g. black hole coronae), using $\approx 15-20$ fully captures the energy gain of the photons; using fewer under-estimates the location of the exponential cutoff, and using more has no adverse effect on the output spectrum, as the code runs simply runs into a regime where no more energy is transferred from the electrons to the photons. In this case, the Compton scattering kernel essentially returns 0 every time, and the code continues integrating over it, so a fair amount of computational time is wasted without affecting the final spectrum.

\begin{figure*}
    \centering
    \includegraphics[width=0.49\textwidth]{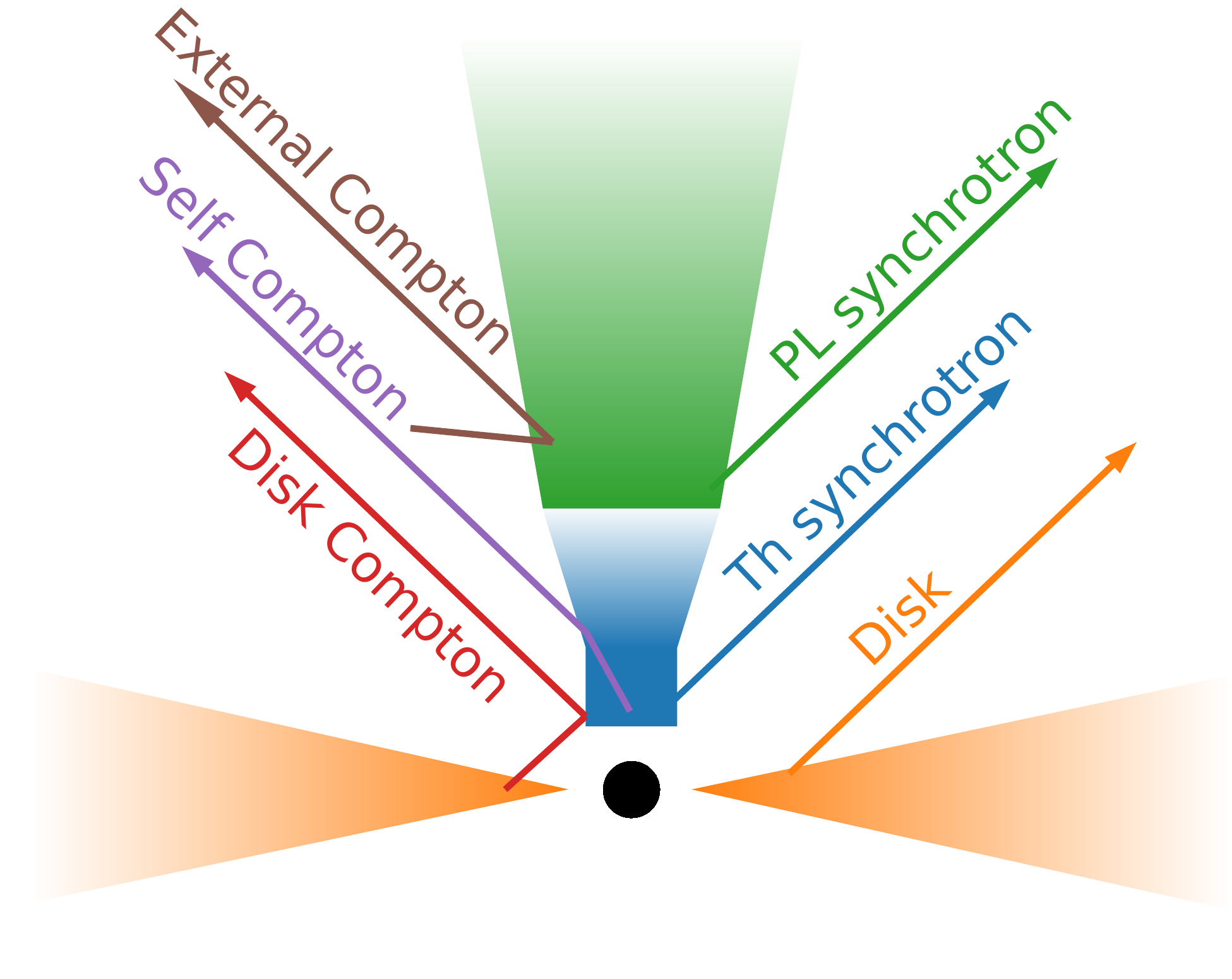}
    \includegraphics[width=0.49\textwidth]{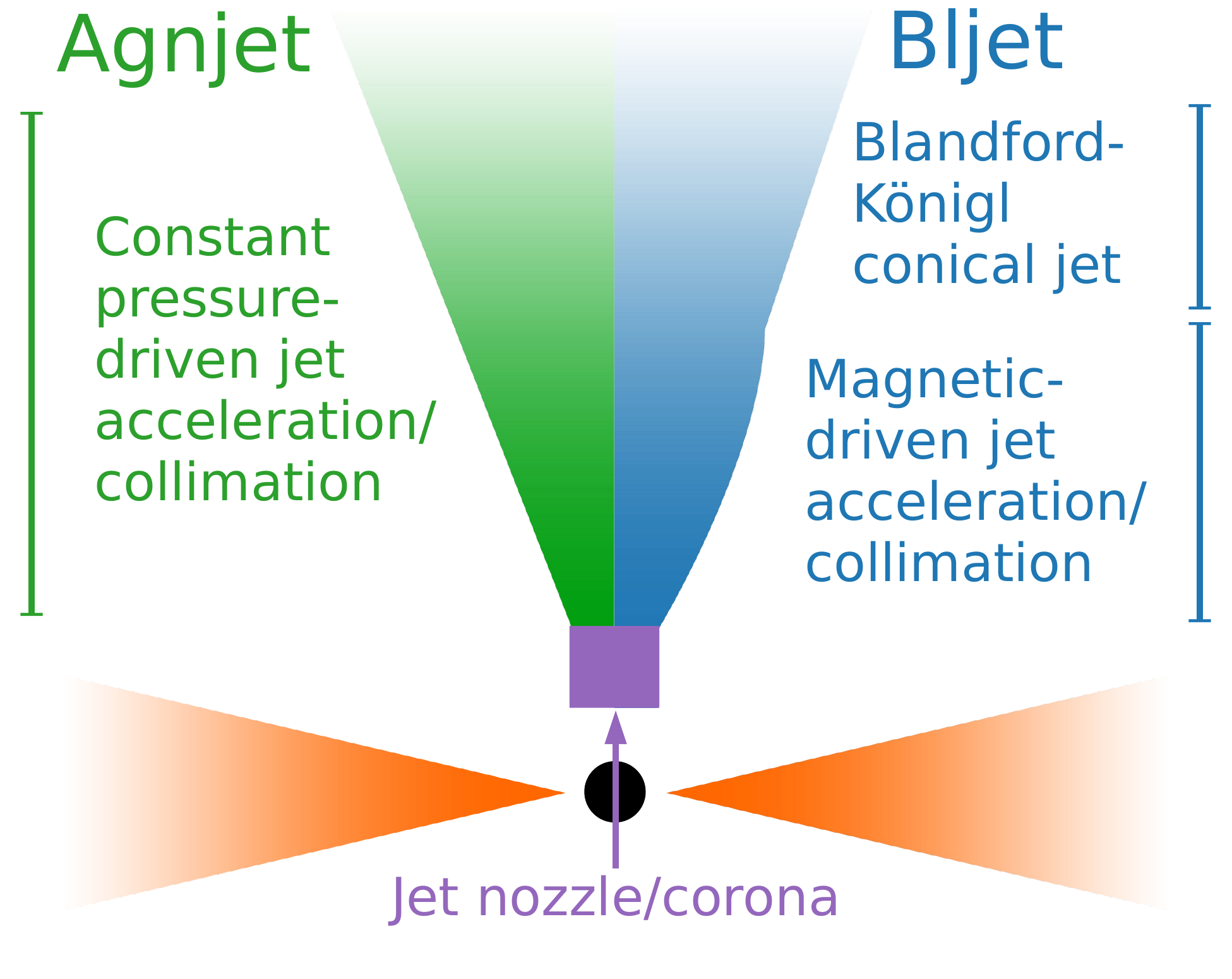}
    \caption{Model schematic of \texttt{BHJet}. The arrows in the left panel highlights the radiative components included in the model; darker and lighter opacity corresponds to regions which tend to be more or less bright, respectively. The disk produces standard black-body emission; the blue jet regions produce thermal synchrotron and thermal Comptonisation, and the green jet regions produces non-thermal synchrotron and inverse Compton emission. The right panel shows a sketch of the model flavours beyond the nozzle region (purple): green and blue indicate the \texttt{agnjet} and \texttt{bljet} model flavours, respectively.}
    \label{fig:schematic}
\end{figure*}

\subsection{Photon fields}
\label{sec:Photon_fields}
There are three methods to calculate the seed photons for inverse Compton scattering. These are not mutually exclusive, and multiple photon fields can be added to the seed photon distribution before computing the spectrum. 

The first method is for synchrotron photons produced in the same region, with co-moving luminosity $L^{\prime}(\nu)$ at a frequency $\nu$. In this case, we estimate the target photon distribution (in units of number of photons, per unit volume and per unit of photon energy) of equation (\ref{eq:IC_kernel}) as:
\begin{equation}
   N_{\rm ph,0}(\epsilon_{\rm 0}) = \frac{L_{\rm s}^{\prime}(\nu^{\prime})}{ch^{2}\nu^{\prime} R^{2}}, 
\end{equation}
where $R$ is the radius of the emitting region. 

The second method is for a black body with energy density $U^{\prime}_{\rm rad}$ and temperature $T^{\prime}_{\rm bb}$, calculated in the co-moving frame of the emitting region. For this case, the target photon field is calculated as:
\begin{equation}
    N_{\rm ph,0}(\epsilon_{\rm 0}) = \frac{2U^{\prime}_{\rm rad}\nu^{2}_{\rm 0}}{hc^{2}\sigma_{\rm sb}T^{4\prime}_{\rm bb}[\exp{(\epsilon_{\rm 0}/kT^{\prime}_{\rm bb})}-1]},
\end{equation}
where $\nu_{\rm 0} = \epsilon_{\rm 0}/h$ is the initial photon frequency. This method can be used to approximate a variety of photon fields, such as the dust torus or broad line region in blazars, or the companion star in X-ray binaries, provided the user computes the appropriate co-moving temperatures and energy densities.   

Finally, the most complex method currently included is that of photons from an optically thick, geometrically thin disc described in sec.\ref{sec:rad_thermal}. Because we neglect fully treating radiative transfer, in this case the photon distribution is computed on the symmetry axis of the system, at a height $z$. In this sense, taking $z>0$ can be thought of as a lamp-post corona, while $z=0$ roughly mimics a slab or hot flow-type corona. Regardless of the value of $z$, the photon distribution is computed by integrating the temperature profile along the disc radius:
\begin{equation}
N_{\rm ph,0}(\epsilon_{\rm 0}) = \int_{\alpha_{\rm min}}^{\alpha_{\rm max}} \frac{4\pi \nu^{2}(\delta)}{hc^{3}\left(e^{h\nu(\delta) / kT(\alpha, \delta)} -1\right)} d\alpha;
\end{equation}
the extremes of the integral are $\alpha_{\rm min} = \arctan(r_{\rm in}r_{\rm  g}/z)$, $\alpha_{\rm max} = \arctan(r_{\rm out}r_{\rm  g}/(z-HR_{\rm out}/2))$ if $z<Hr_{\rm out}r_{\rm  g}/2)$ and $\alpha_{\rm max} = \pi/2 + \arctan(r_{\rm out}r_{\rm  g}/(z-Hr_{\rm out}r_{\rm  g}/2))$ otherwise, in order to account for the change in viewing angle of all the disc regions. $\nu(\delta)$ and $T(\alpha, \delta)$ are the photon frequencies and disc temperatures for each viewing angle, accounting for Doppler beaming if the emitting region is moving with respect to the disc.

\subsection{Doppler boosting}
Both the cyclo-synchrotron and inverse Compton classes track both co-moving and observer frame luminosities with the standard Doppler transformations:

\begin{equation}
L_{\rm c,s}(\nu) = \delta^{\alpha} L_{\rm c,s}^{\prime}(\nu^{\prime});   \quad \nu = \delta\nu^{\prime}.
\label{eq:boost}
\end{equation}
The factor $\alpha$ depends on the assumed emission geometry. If the emitting region is spherical, the code takes $\alpha=3$, appropriate for a plasmoid moving with respect to the observer. If the emitting region is taken to be cylindrical, the code instead assumes that it is part of a compact jet, in which case $\alpha=2$ \citep{Lind85}.

Additionally, both classes can account simultaneously for the emission of both the main emitting region, observed at a viewing angle $\theta$, and its counterpart, observed at a viewing angle $\pi - \theta$, by computing the appropriate boosting factors and summing both components as appropriate. This allows a user to easily account for the presence of a counter-jet. By default, the  \texttt{Kariba} constructors assume a static source with $\delta = 1$ and no counter-jet present.

\section{BHJet model flavours}
\label{sec:BHJet}
\texttt{BHJet} is a family of steady state, time independent, scale-invariant, multiwavelength jet models designed to fit the SED of jetted accreting black holes across a wide interval (but generally sub-Eddington) in jet power and black hole mass. They all share a similar treatment for the jet launching region and particle distribution, and differ mainly in the assumptions made regarding the jet dynamics. A simple sketch of the model is shown in fig.~\ref{fig:schematic}.

\subsection{Basic assumptions}
\label{sec:BHJet_basics}

Following \cite{Falcke95}, all model flavours assume that \{a fraction $q_{\rm j}$ of the accretion rate $\dot{M}_{\rm acc}$ powers two polar jets, so that the mass flow rate through both is $\dot{M_{\rm j}} = q_{\rm j} \dot{M}_{\rm acc}$. $q_{\rm j} \dot{M}_{\rm acc}$ can be related to the co-moving lepton energy density in the jet-launching region, which we call the jet nozzle, through:
\begin{equation}
U_{\rm e,0} = \bf{\frac{q_{\rm j} \dot{M}_{\rm acc}}{2\pi r_{\rm 0}^{2} \gamma_{\rm 0}\beta_{\rm 0}c^{3} f_{\rm eq}(\beta_{\rm p},\eta,\langle\gamma\rangle)},}
    \label{eq:jetrat}
\end{equation}
where the factor 2 accounts for the launching of two jets. $R_{\rm 0}$ is the radius of the jet nozzle, which is a cylinder with aspect ratio $h_{\rm r}=z_0/r_{\rm 0} $\footnote{$h_{\rm r}$ is set to 2 by default. Users can modify it by accessing the source code, however this results in the inverse Compton spectra being slightly inaccurate due to the correction discussed in sec.~\ref{sec:Kompton}. Therefore, $h_{\rm r}$ is not a fitted parameter, unlike in previous versions of the model.}. The initial speed of the jet is assumed to be $\beta_{\rm 0} =  \sqrt{\Gamma(\Gamma-1)/(\Gamma+1)}\approx 0.43$, which is the sound speed for a relativistic gas with adiabatic index $\Gamma = 4/3$ \citep{Crumley17}. The corresponding initial Lorentz factor is $\gamma_{\rm 0} = 1.1$. This parameter has a very minor effect on the SED, mostly affecting the boosting/de-boosting of the disc photons when they are inverse-Compton scattered, and therefore it is never left free during spectral fits. From now on, we will use both the terms corona and jet nozzle interchangeably regardless of whether the nozzle is compact, similarly to a lamp-post, or extended (our model can replicate either geometry), except in sec.\ref{sec:GX}. We note that if the magnetisation (defined later in this section) in the nozzle is on the order of unity, then this region can be thought of as the interface between inflowing and outflowing material, and essentially captures the emission from both while still allowing us to couple its properties with the compact jet. The equipartition factor $f_{\rm eq}(\beta_{\rm p},\eta,\langle\gamma\rangle)$ depends on the model flavour, and in its more general form it is defined as:
\begin{align}
    f_{\rm eq}(\beta_{\rm p},\eta,\langle\gamma\rangle) &  = \frac{U_{\rm e,0} + U_{\rm b,0} + U_{\rm p,0}}{n_{\rm e}}
    \nonumber
    \\ & =  \langle \gamma \rangle m_{\rm e} c^{2}\left(1+\frac{1}{\beta_{\rm p}} + \frac{m_{p}}{\eta \langle \gamma \rangle m_{\rm e}} \right),
\end{align}
where $U_{\rm e,0}$ is the energy density of the injected leptons with number density $n_{\rm e,0}$ and $\langle \gamma \rangle$ their average Lorentz factor, $U_{\rm b,0} = B_{\rm 0}^{2}/8\pi$ is the energy density of the magnetic field at the jet base, and $U_{\rm p,0} = n_{\rm p,0} m_{\rm p} c^{2}$ is the energy density of the injected protons, which we always assume to be cold. The equipartition parameter $\beta_{\rm p}$ is defined as $\beta_{\rm p} = U_{\rm e,0}/U_{\rm b,0}$ (in analogy with the standard plasma-$\beta$ parameter in plasma physics), $\eta = n_{\rm e}/n_{\rm p}$ quantifies the jet matter content. We define an injected power as $N_{\rm j} = q_{\rm j} \dot{M}_{\rm acc} c^{2}$ for convenience, because a) it effectively absorbs the uncertainty on the unknowns $q_{\rm j}$ and $\dot{M}_{\rm acc}$ in a single model parameter and b) it can be readily expressed in units of the Eddington luminosity, which provides users an immediate estimate for the power required by the model. However, $N_{\rm j}$ should \textsl{not} be thought of as a direct measure of the total (kinetic+magnetic+internal)  power of the full outflow, but rather as a model normalization on the order of the total jet power. This is for two reasons: first, the total initial jet power differs from $N_{\rm j}$ by a small multiplicative factor of $\gamma_0\approx1$ \citep[e.g.][]{Crumley17}. Second, similarly to a standard \cite{Blandford79} model, we do not account for the power required to re-accelerate the radiating leptons in the jet; depending on the model flavour and parameter used, this power may or may not be negligible with respect to $N_{\rm j}$. We discuss the latter issue further in sections \ref{sec:agnjet} and \ref{sec:bljet}. We also note that unless users choose parameters that are inconsistent with the assumptions of each model flavor, $N_{\rm j}$ is within a factor of (at most) a few of the total power. Equation (\ref{eq:jetrat}) can be solved to find the number density of the leptons at the jet base, as a function of the model input parameters:
\begin{equation}
    n_{\rm e,0} = \frac{N_{\rm j}}{2\pi r_{\rm 0}^{2}\gamma_{\rm 0}\beta_{\rm 0}c f_{\rm eq}(\beta_{\rm p},\eta,\langle\gamma\rangle)}.
    \label{eq:ne}
\end{equation}
Beyond the nozzle, the jet begins expanding and accelerating. Regardless of the details of the collimation and acceleration process, all flavours of the model assume that the number of particle is conserved, so that:
\begin{equation}
    n(z) = n_{\rm e,p,0} \left(\frac{\gamma_0 \beta_0}{\gamma(z) \beta(z)}\right)\left(\frac{r_{\rm 0}}{r(z)}\right)^{2},
\label{eq:n(z)}
\end{equation}
where $n_{\rm e,p,0}$ is the initial number density of either leptons or protons, $\beta(z)$ and $\gamma(z)$ the jet speed along the $z$ axis, $r(z)$ the jet radius at a distance $z$. The velocity and collimation profiles are set by the chosen model flavour. 

\begin{figure}
    \centering
    \includegraphics[width=0.49\textwidth]{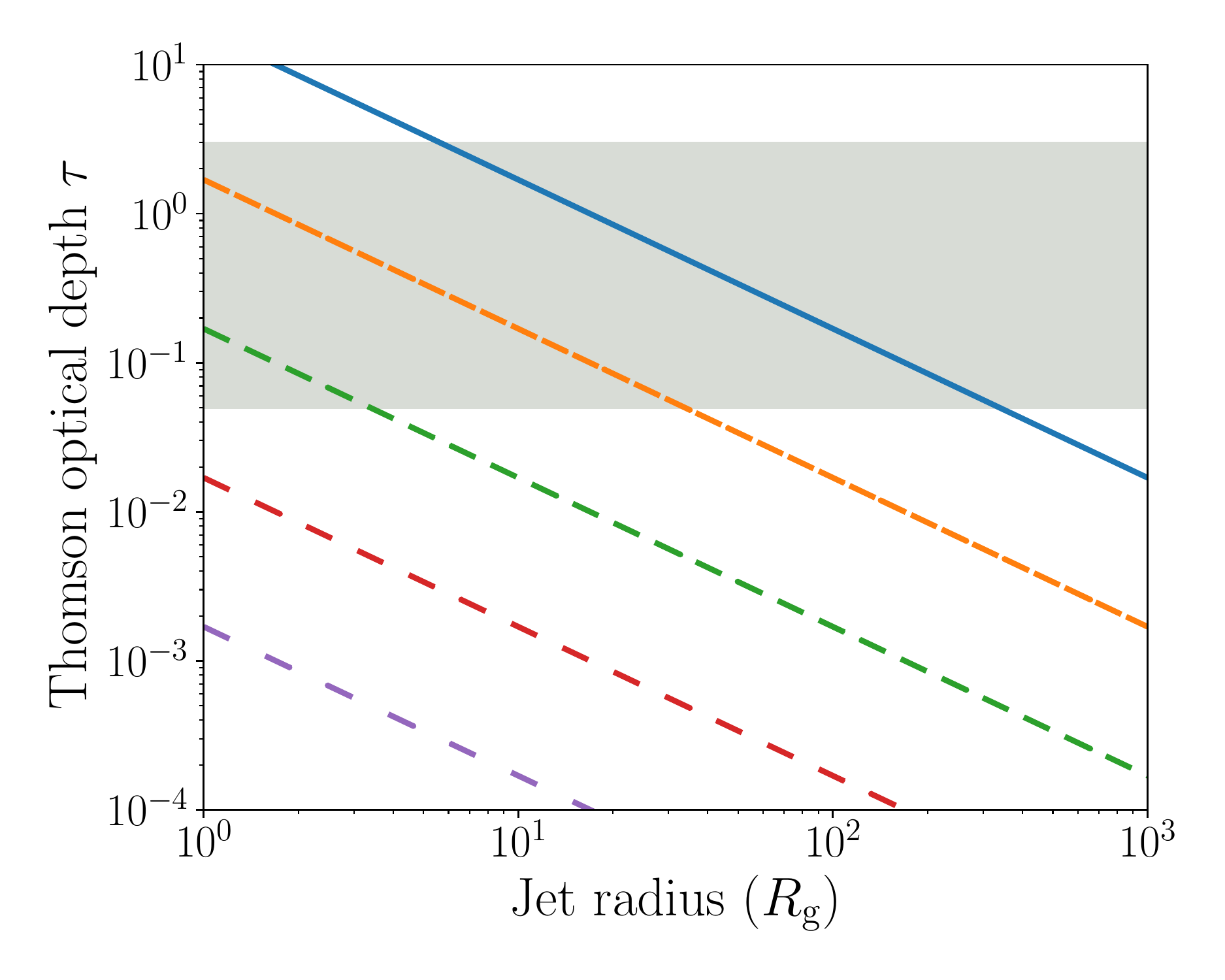}
    \caption{Jet optical depth $\tau$ as a function of radius, for varying jet powers, assuming $\eta = 10$, $\langle\gamma\rangle = 3$, and $\beta_{\rm p} = 0.01$, and negligible jet acceleration. More sparsely dashed lines correspond to decreasing jet powers of $L_{\rm Edd}$ (blue line), $10^{-1}\,L_{\rm Edd}$ (orange line), $10^{-2}\,L_{\rm Edd}$ (green line), $10^{-3}\,L_{\rm Edd}$ (red line), $10^{-4}\,L_{\rm Edd}$ (purple line), respectively. The grey shaded area corresponds to the typical range of coronal optical depths.}
    \label{fig:tau}
\end{figure}

The corresponding Thomson optical depth of the jet is:
\begin{equation}
    \tau(z) = n_{\rm e}(z) r(z) \sigma_{\rm t},
    \label{eq:tau_base}
\end{equation}
where $\sigma_{\rm t}$ is the Thomson cross section and $r(z)$ is set by the model flavour. One can get a sense of how the optical depth would vary as a function of jet power and radius, by assuming for simplicity that no acceleration occurs. In this case, by combining equation (\ref{eq:jetrat}) and \ref{eq:n(z)} we see that $\tau(z) \propto r^{-1}(z)$, meaning that the optical depth drops very quickly as the jet expands. The scaling of $\tau$ versus radius, for a wide range of jet powers and typical parameters, is shown in fig.~\ref{fig:tau}. In this simple scenario $\tau(z)$ does not depend on black hole mass, because from equation (\ref{eq:ne}) one can see that $n_{\rm e}\propto N_{\rm j}/r_{\rm 0}^{2} \propto M_{\rm bh}/M_{\rm bh}^{2} \propto M_{\rm bh}^{-1}$. As a result, $\tau(z) \propto n_{\rm e}(z) r(z) \propto M_{\rm bh}^{-1} M_{\rm bh} \propto M_{\rm bh}^{0}$. We note (again, assuming negligible jet acceleration for simplicity) that the radius plotted here can be understood either as the initial jet radius $r_{\rm 0}$, or as the radius of the jet at a distance $z$ from the black hole. 

This plot highlights several features of the model (assuming fixed $\beta_{\rm p}$ and $\eta$). First, in order to achieve the optical depth of a canonical black hole corona ($\tau \approx 0.1-1$) with a mild pair content (10 pairs per proton, in fig.\ref{fig:tau}), the jet power should be in the range $10^{-2}-10^{-1}\,L_{\rm Edd}$, typical of a fairly luminous XRB hard state or FR\,\RN{2}-type AGN. Second, it is interesting to note how jet power, optical depth and jet radius vary as a function of each other. For a fixed jet power, $\tau$ drops as the radius of the emitting region increases. For a fixed radius, $\tau$ increases as the jet power increases. Finally, for a fixed $\tau$, as the radius of the emitting region increases, so does the total power required. This conclusion is independent of the model flavour (and indeed, it applies to any jet model beyond \texttt{BHJet}), as it is purely a consequence of the generalised jet power defined in equation (\ref{eq:jetrat}).

Because the optical depth is expected to drop rather quickly as the jet expands, equation (\ref{eq:tau_base}) implies that thermal Comptonisation over multiple scatterings becomes less important along the jet axis. Therefore, the location where the bulk of the coronal inverse Compton X-ray emission occurs must be very close ($\approx$ tens of $R_{\rm g}$ at most) to the black hole, where the jet is first launched. Note that this is not the case for single-scattering, non-thermal Comptonisation as, for example, in blazars. We discuss this aspect of the model further in sec.\ref{sec:blazars}. 

\emph{These considerations result in one common prediction for the \texttt{BHJet} family of models:} while at moderate and high jet powers, Comptonisation in the corona can produce significant X-ray emission, in low power sources synchrotron emission from non-thermal electrons accelerated downstream ($\approx 10^{2-3}\,R_{\rm g}$) should dominate any potential high-energy emission. 

\begin{figure*}
    \centering
    \includegraphics[width=\textwidth]{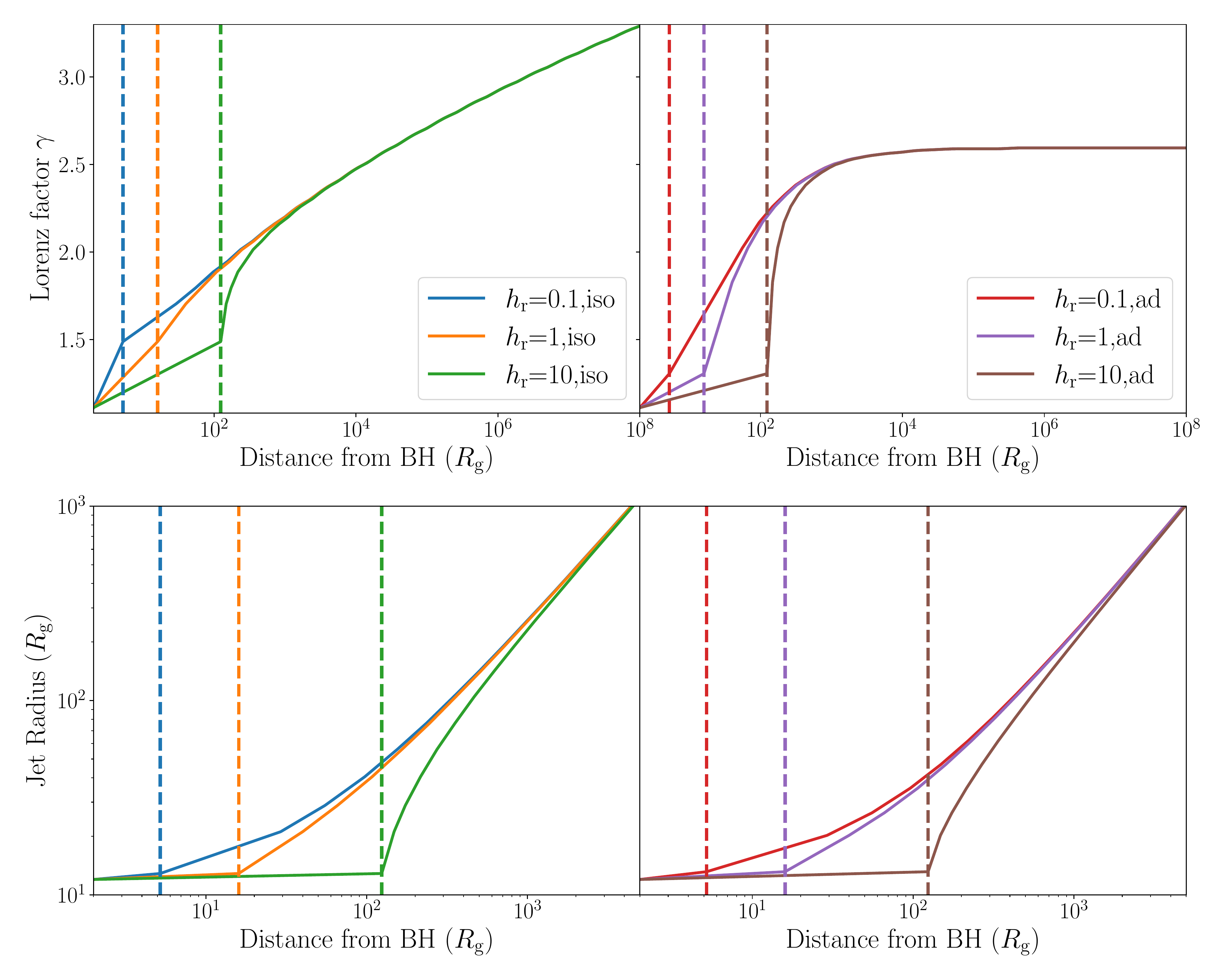}
    \caption{Jet speed and collimation profiles (top and bottom row) for the isothermal and adiabatic cases (left and right columns) of the \texttt{agnjet} model flavour, as a function of distance from the black hole $z$, with varying nozzle aspect ratios $h$ and assuming the initial jet radius is $r_{\rm 0}=12\,R_{\rm g}$. The vertical dashed lines correspond to the location where the nozzle ends and bulk acceleration begins. Regardless of the value of $h$, all models quickly turn into a conical outflow with Lorentz factor $\gamma\approx2-3$.}
    \label{fig:agnjet}
\end{figure*}

\subsection{Agnjet: pressure-driven jets}
\label{sec:agnjet}
The \texttt{agnjet} model flavour describes a mildly relativistic, pressure-driven jet; the dynamical properties were first presented in \cite{Falcke95,Falcke95b,Falcke99} and further refined in \cite{Crumley17}. In this regime, magnetic fields do not affect the dynamics of the outflow, the jet is efficiently accelerated (meaning its Lorentz factor becomes comparable to the terminal Lorentz factor over relatively short distances), and the power carried by the jet is of the order of the initial rest-mass energy. 

The model supports both adiabatic\footnote{The adiabatic profile is currently fully self-consistent only if the radiating particles are purely thermal. Furthermore, high ($T_{\rm e} \geq 1000\,\rm{keV}$) initial electron temperatures are required to avoid numerical issues.} and quasi-isothermal jets. The first case means that as the jet expands and propagates downstream of the launching point, the particles in it cool adiabatically and are not re-accelerated or re-heated. In this regime the jet internal energy can be written as:

\begin{equation}
U_{\rm j}(z) = n_{\rm p,0} \left(\frac{\gamma_{\rm 0}\beta_{\rm 0}}{\gamma_{\rm j}(z)\beta_{\rm j}(z)}\right)^{\Gamma} \left(\frac{r_{\rm 0}}{r(z)  }\right)^{2\Gamma};
\label{eq:Uj_ad}
\end{equation}
note that this equation is identical to the definition of the jet internal energy of \citep{Crumley17} by setting the $\zeta$ parameter from that work to be unity, which is always assumed in \texttt{agnjet}. $\Gamma = 4/3$ is the adiabatic index of the jet (we always assume that it can be treated as a relativistic fluid), $\beta_{\rm j}(z)$ is the jet velocity along the $z$ axis and $\gamma_{\rm j}(z)$ the corresponding Lorentz factor. In this case, $T_{\rm e} \propto n_{\rm e}^{\Gamma} \propto (\gamma_{\rm j}(z)\beta_{\rm j}(z))^{1-\Gamma}z^{2-2\Gamma}$, implying that the particles cool rapidly as they stream down the jet. In the second case, the work done by the particles as the jet expands is offset by an unspecified acceleration mechanism, such as internal shocks, re-accelerating the particles as they stream along the $z$ axis. In this case, we have:

\begin{figure*}
    \centering
    \includegraphics[width=0.49\textwidth]{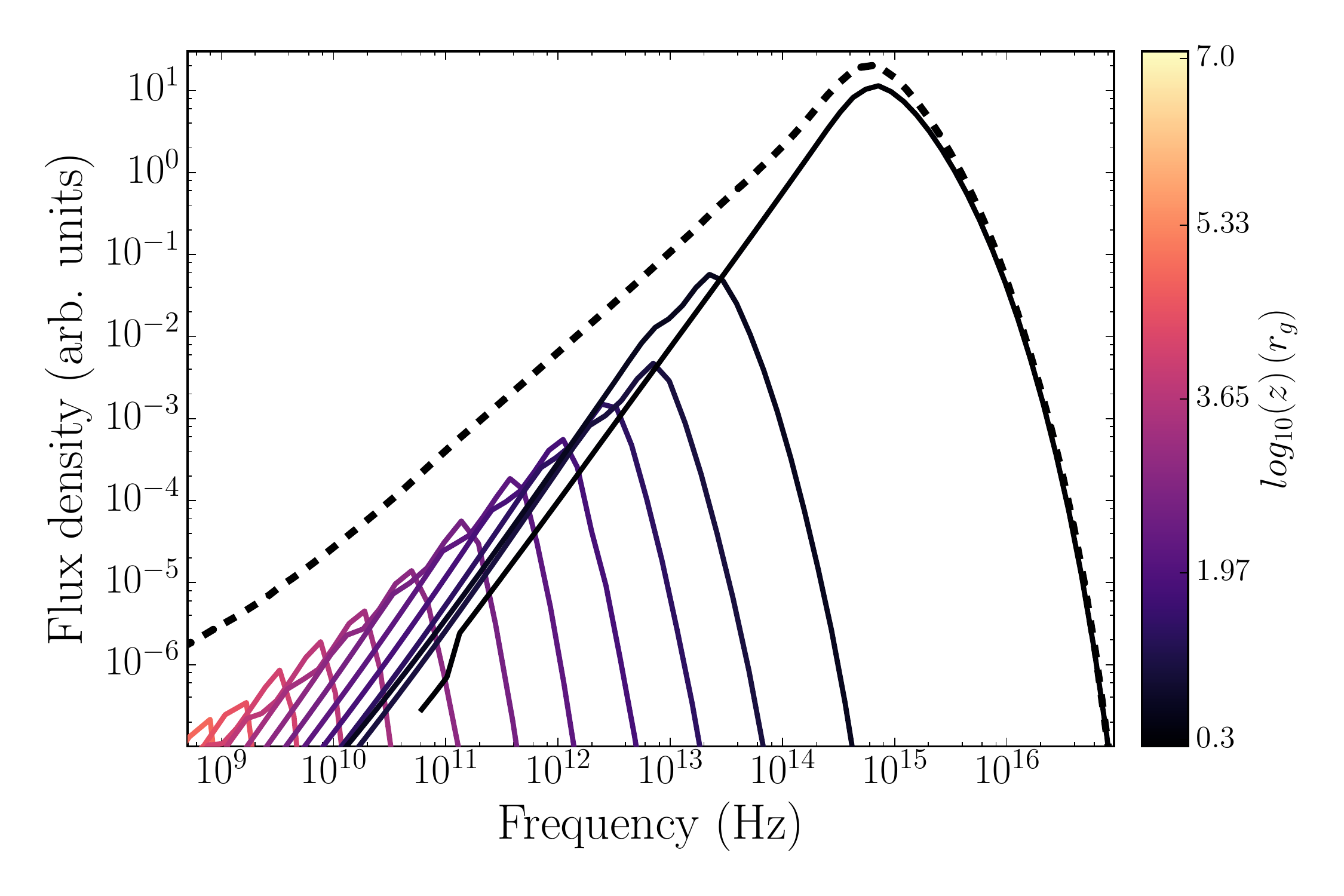}
    \includegraphics[width=0.49\textwidth]{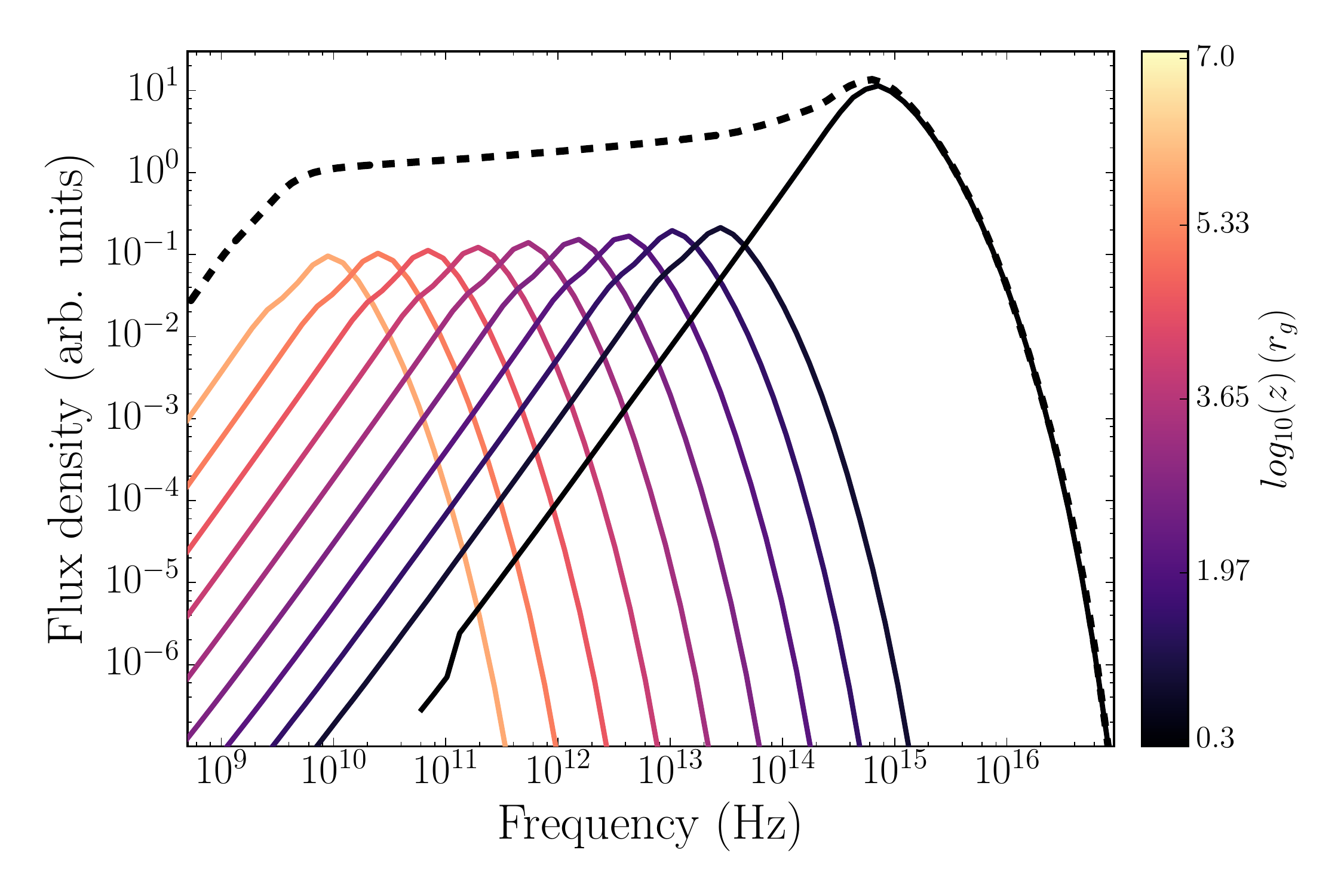}
    \caption{A comparison of the cyclo-synchrotron spectra computed in \texttt{agnjet}, in the adiabatic (left panel) and isothermal (right panel) cases. The color scale corresponds to different zones along the jet, only some of which are plotted here for clarity. The dashed line is the sum of the emission from each region of the jet. For both SEDs the viewing angle is taken to be 45$^{\circ}$, the initial electron temperature is $T_{\rm e} = 2500\,\rm{keV}$, and only thermal particles are present throughout the jet. For the adiabatic profile, the optically thick spectral index  (defined as $F(\nu) = \nu^{\alpha}$) between 10 and 100 GHz is extremely inverted ($\alpha = 1.2$), while it is essentially flat ($\alpha=0.1$) for the isothermal case. The optically thin part is mostly unchanged.}
    \label{fig:agjnet_sed}
\end{figure*}

\begin{equation}
  U_{\rm j}(z) = n_{\rm p,0} \left(\frac{\gamma_{\rm 0}\beta_{\rm 0}}{\gamma_{\rm j}(z)\beta_{\rm j}(z)}\right)^{\Gamma} \left(\frac{r_{\rm 0}}{r(z)  }\right)^{2};
\label{eq:Uj_iso}  
\end{equation}
here, $T_{\rm e }\propto (\gamma_{\rm j}(z)\beta_{\rm j}(z))^{1-\Gamma}$; as long as the jet only reaches mildly relativistic Lorentz factors, the particles do not cool significantly. Indeed, \cite{Crumley17} showed that this regime is almost identical to the fully isothermal case, in which no adiabatic cooling is present and $T_{\rm e} \propto \rm{const}$; therefore, we use the two interchangeably. 

Regardless of the importance of adiabatic losses, the model assumes that starting from the top of the nozzle (at a height $z_{\rm 0}$) the jet expands laterally at the sound speed. The radius of the jet is:
\begin{equation}
    r(z) = r_{\rm 0} + (z-z_{\rm 0})\frac{\gamma_{\rm 0}\beta_{\rm 0}}{\gamma_{\rm j}(z)\beta_{\rm j}(z)}.
    \label{eq:agnjet_radius}
\end{equation}
The jet velocity profile is found by substituting equation (\ref{eq:n(z)}), either equation (\ref{eq:Uj_ad}) or equation (\ref{eq:Uj_iso}), and equation (\ref{eq:agnjet_radius}) in the 1-D Euler equation \citep{Pomraning73,Crumley17}:

\begin{align}
    \gamma_{\rm j}(z)\beta_{\rm j}(z)n(z)\frac{d}{d z}\left[\gamma_{\rm j}(z)\beta_{\rm j}(z)\left(m_{\rm p}c^{2}+\frac{\Gamma U_{\rm j}(z)}{n(z)}\right) \right] = &  
    \nonumber
    \\-(\Gamma -1)\frac{d U_{\rm j}(z)}{d z}, & 
    \label{eq:Euler}
\end{align}
which imposes conservation of momentum along the $z$ axis. We note that this treatment for the jet dynamics does not account for the energy required to re-accelerate particles in the isothermal case. Therefore, the injected jet power defined in equation (\ref{eq:jetrat}) is not the total power carried by the outflow, which is larger by a factor $\approx 2-3$ if only leptons are accelerated, and much larger if protons contribute to the emission \citep{Crumley17,Kantzas20}. \textsl{As a result of neglecting the power required to re-accelerate the particles, the \texttt{agnjet} model flavor does not conserve energy, for the same reason that the standard \cite{Blandford79} model does not}. The only model parameter that has an impact on the solutions of equation (\ref{eq:Euler}) and (\ref{eq:agnjet_radius}) is the aspect ratio $h_{\rm r}=z_{\rm 0}/r_{\rm 0}$ of the nozzle. Different solutions for the isothermal and adiabatic (left and right columns, respectively) cases are shown in fig.~\ref{fig:agnjet}. The regions where the velocity and collimation profiles are most affected are between $\approx 10^{1}-10^{2}\,R_{\rm g}$; however, these regions tend to have a negligible effect on the SED. Note that varying $h$ also results in less self-consistent Comptonisation spectra, as discussed in sec.~\ref{sec:Kompton}, although this effect is negligible for values of $h$ near unity.

The original \texttt{agnjet} model \citep{Falcke00,Markoff01,Markoff01b} went on to assume that the jet carried one proton per electron, thus $n_{\rm e,0} = n_{\rm p,0} = n_{\rm 0}$ throughout the jet. With this choice, the factor $f_{\rm eq}(\beta_{\rm p},\eta,\langle\gamma\rangle)$ in equation (\ref{eq:ne}), which sets both the initial lepton and proton number density, is: 
\begin{equation}
    f_{\rm eq}(\beta_{\rm p},\eta=1,\langle\gamma\rangle) = \langle\gamma\rangle m_{\rm e}c^{2}  \left(1+\frac{1}{\beta_{\rm p}}+\frac{m_{\rm p}}{m_{\rm e}} \right).
\end{equation}

Versions of the model from 2004 and onward, including the refinements presented in \cite{Crumley17}, instead set $U_{\rm p,0} = U_{\rm e,0}+U_{\rm b,0}$, therefore assuming implicitly that the jet carries a few pairs per proton. In this case,  $f_{\rm eq}(\beta_{\rm p},\eta,\langle\gamma\rangle)$ is:

\begin{equation}
    f_{\rm eq}(\beta_{\rm p},\eta,\langle\gamma\rangle) = 2\langle\gamma\rangle m_{\rm e}c^{2}\left(1+\frac{1}{\beta_{\rm p}}\right),
\end{equation}
the initial proton number density is:
\begin{equation}
    n_{\rm p,0} = \left(1+\frac{1}{\beta_{\rm p}}\right)\frac{\langle\gamma\rangle m_{\rm e}}{m_{\rm p}}n_{\rm e,0}
\end{equation}
and the ratio of lepton to proton number density $\eta$ is:

\begin{equation}
    \eta =  \frac{n_{\rm e,0}}{n_{\rm p,0}} = \frac{m_{\rm p}}{\langle\gamma\rangle m_{\rm e}} \frac{1}{1+\frac{1}{\beta_{\rm p}}}
\end{equation}
In the code discussed in this paper, the user has the ability to choose either option. Regardless of the matter content, the initial magnetic field is:
\begin{equation}
    B_{\rm 0} = \sqrt{\frac{8\pi n_{\rm e,0}\langle\gamma\rangle m_{\rm e}c^{2}}{\beta_{\rm p}}}
\end{equation}
and the magnetic field along the jet axis is:

\begin{equation}
    B(z) = B_{\rm 0}\frac{r_{\rm 0}}{r(z)}\left(\frac{\gamma_{\rm 0}\beta_{\rm 0}}{\gamma_{\rm j}(z)\beta_{\rm j}(z)}\right)^{\Gamma/2}.
    \label{eq:agnjet_bfield}
\end{equation}

\begin{figure*}
    \centering
    \includegraphics[width=\textwidth]{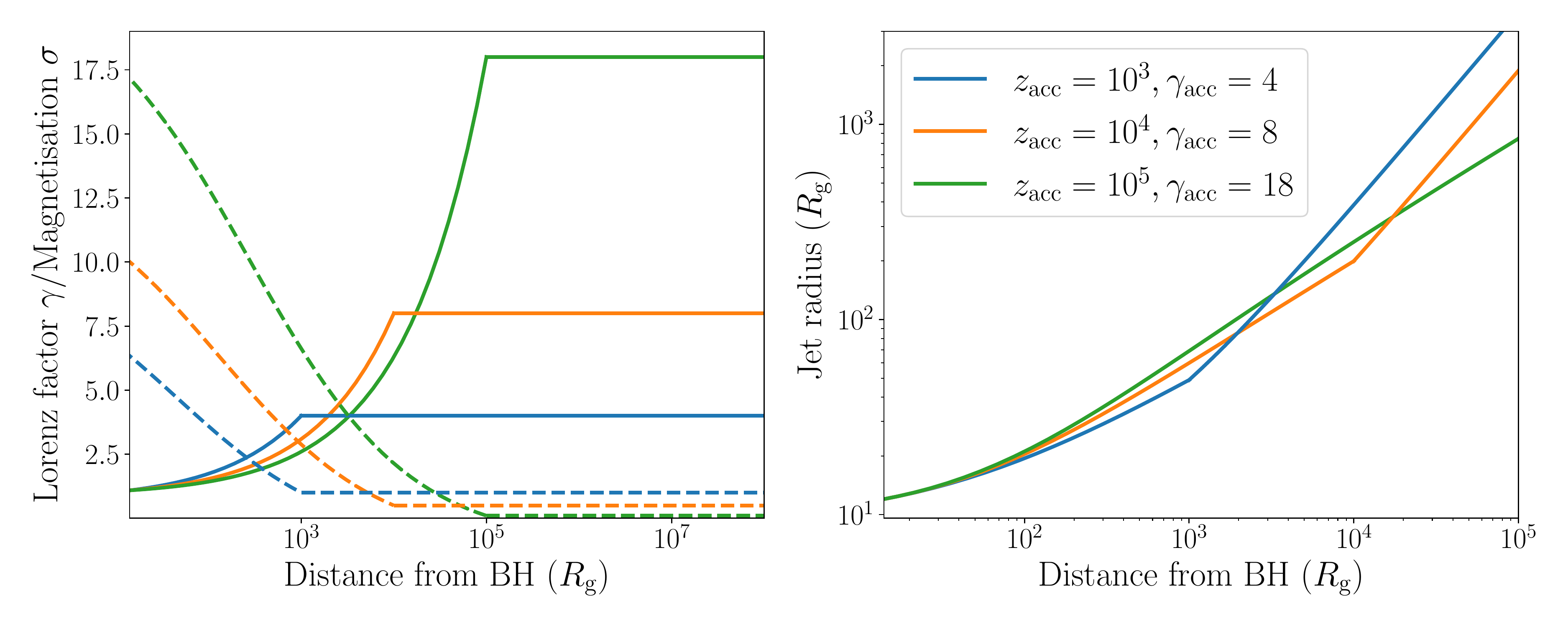}
    \caption{Left panel: jet speed and magnetisation for different realisations of the \texttt{BHJet} model flavour, as a function of distance from the black hole $z$. The blue, green and orange lines correspond to $z_{\rm acc} = 10^{3}$, $10^{4}$, $10^{5}$ $R_{\rm g}$, $\gamma_{\rm acc} = 4$, $8$, $18$, $\sigma_{\rm diss}=1$, $0.5$, $0.1$, respectively. Continuous lines correspond to the jet Lorentz factor, and dashed lines to the jet magnetisation. In all cases, the initial magnetisation $\sigma_{\rm 0}$ is of the order of the final Lorentz factor $\gamma_{\rm acc}$, with $\sigma(z)$ dropping smoothly as $\gamma(z)$ increases up to $z_{\rm acc}$. Right panel: jet collimation profile as a function of distance from the black hole, for the same parameters shown in the left panel.}
    \label{fig:bljet}
\end{figure*}

The choice between the adiabatic and isothermal profiles has a very large impact on the SED, as shown in fig.~\ref{fig:agjnet_sed}. Due to the extreme cooling in the adiabatic case, the optically thick spectrum of the jet is very inverted ($F(\nu)\propto \nu^{\alpha}$, and $\alpha \approx 1$), leading to very faint radio emission for a given IR/optical/UV luminosity.  Therefore, this velocity profile is not appropriate for modelling a typical compact jet, being more similar to the ``dark jet'' scenario \citep{Drappeau17}. Instead, the isothermal jet profile leads to a standard flat ($F(\nu)\propto \nu^{\alpha}$, and $\alpha \approx 0.1$) optically thick spectrum. This difference is mainly caused by the drop in temperature along the jet for the adiabatic jet profile: because the synchrotron luminosity is proportional to the average squared Lorentz factor $\langle\gamma^{2}\rangle$, if the temperature of the emitting particles along the jet drops over distance, the luminosity also decreases and the spectrum becomes inverted. Therefore, some mechanism (such as internal shocks) needs to maintain the average energy in the electrons constant along the jet. These assumptions reproduce the well-known result of \cite{Blandford79}. Because the energy required to re-energise the particles is unaccounted for in the injected power $N_{\rm j}$, this quantity should be thought of as a re-normalisation factor for the model, rather than a physical estimate of the jet power. The latter is expected to be higher than $N_{\rm j}$ by a factor of a few, up to roughly an order of magnitude, \citep{Markoff05,Crumley17}, depending on the model parameters chosen.

\subsection{Bljet: magnetic-driven jets}
\label{sec:bljet}

The \texttt{bljet} model flavour allows for a somewhat more self-consistent treatment of the magnetic fields carried in the jet, and of their role in setting the outflow dynamics. Compared to \texttt{agnjet}, this updated treatment allows for the jet to reach an arbitrarily high Lorentz factor. Additionally the total energy budget in the outflow is naturally closer to the injected power $N_{\rm j}$, as the former is always assumed to be dominated either by the magnetic field or the bulk kinetic energy carried by the protons \citep{Lucchini19a}.

\texttt{Bljet} assumes that the jet nozzle is highly magnetised, and that jet acceleration is powered by the conversion of magnetic field into bulk kinetic energy, in broad agreement with the predictions of ideal MHD as well as global GRMHD simulations \citep[e.g.][]{Beskin06,Komissarov07,Tchekhovskoy09}. The jet acceleration profile is assumed to be parabolic, in agreement with VLBI observations of several AGN \cite[e.g.][]{Hada13,Mertens16,Boccardi16,Nakamura18}, up to a maximum Lorentz factor $\gamma_{\rm acc}$, which is reached at a distance $z_{\rm acc}$ from the black hole:

\begin{equation}
    \gamma(z) = \gamma_0 + \left(\gamma_{\rm acc}-\gamma_0\right)\frac{z^{1/2} - z_{0}^{1/2}}{z_{\rm acc}^{1/2} - z_{0}^{1/2}}.
    \label{eq:bljet_gamma}
\end{equation}
The jet opening angle $\theta(z)$ is taken to be inversely proportional to this Lorentz factor (in agreement with observations of AGN jets, e.g., \citealt{Pushkarev09,Mertens16,Pushkarev17}), and the jet radius is computed for each value of $\theta(z)$:
\begin{align}
&\theta(z) = \frac{\rho}{\gamma(z)}\\
&r(z) = r_0+(z-z_0)\tan(\theta(z));
\label{eq:bljet_collimation}
\end{align}
a typical value of $\rho$ inferred from VLBI surveys is $0.15$. By varying $\rho$, $\gamma_{\rm acc}$ and $z_{\rm acc}$, it is possible to model VLBI imaging data together with multi-wavelength SEDs \citep[e.g][]{Lucchini19b}, and the jet can reach arbitrarily large Lorentz factors.

The strength of the magnetic field as the jet is accelerating is set by imposing energy conservation in the jet bulk acceleration process. This is done by solving the Bernoulli equation \citep[e.g.][]{Konigl80}:
\begin{equation}
    \gamma(z)\frac{\omega(z)}{n(z)} = \rm{const.},
    \label{eq:bernoulli}
\end{equation}
where $\omega(z) = U_{\rm p}(z) + U_{\rm e}(z) + P_{\rm e}(z) + U_{\rm b}(z) + P_{\rm b}(z)$ is the total enthalpy carried by the jet, assuming that the protons remain cold and have negligible pressure. We define as the magnetisation of the jet the ratio of magnetic to particle enthalpy:

\begin{figure*}
    \centering
    \includegraphics[width=\textwidth]{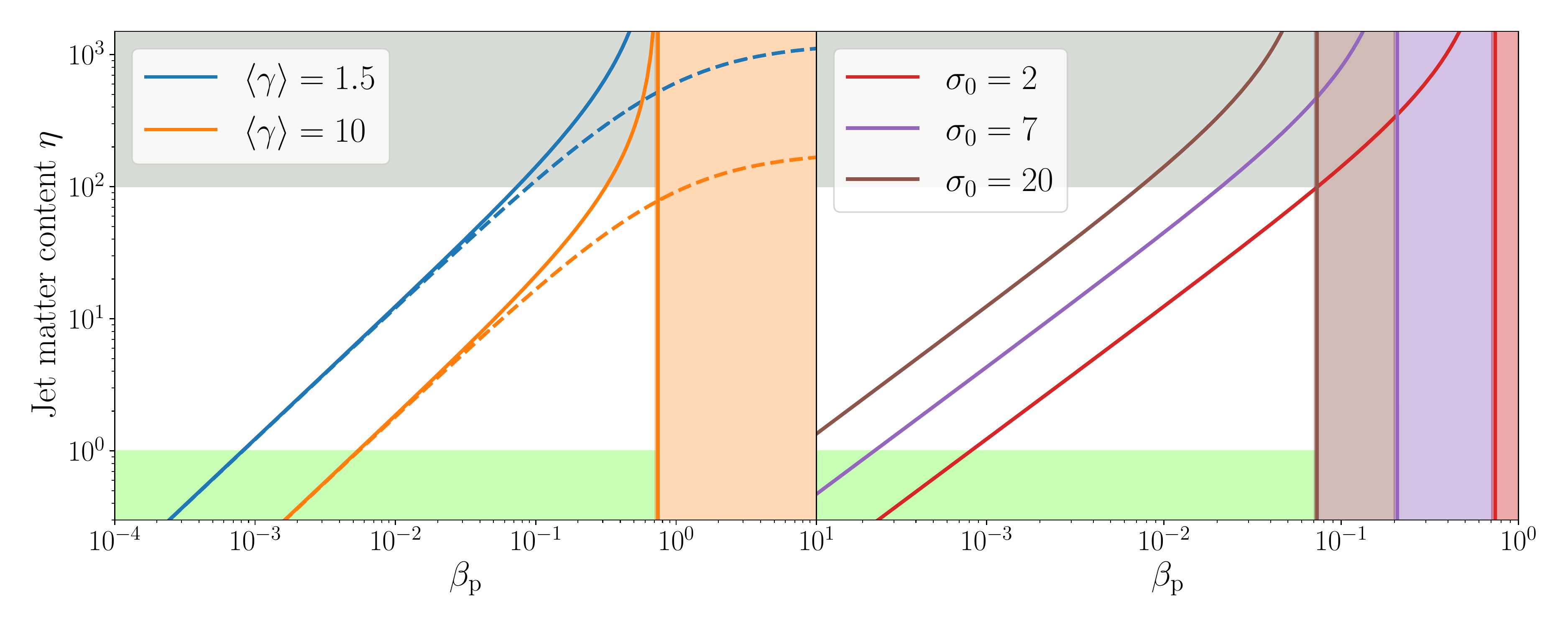}
    \caption{Allowed parameter space for $\beta_{\rm p}$ and jet matter content $\eta$ for \texttt{BHJet} model flavours. Shaded regions correspond to forbidden areas of the parameter space. Left: comparison of \texttt{bljet} (continuous lines) vs \texttt{agnjet} (dashed lines), in the low initial magnetisation regime ($\sigma_{\rm 0} = 2$ for \texttt{bljet} and $1.8$ for \texttt{agnjet}), for different initial electron average Lorentz factors. Right: pair content with varying initial magnetization $\sigma_{\rm 0}$ in \texttt{bljet}, with identical $\langle\gamma\rangle = 1.5$. The gray shaded region indicates pair-dominated jets, which are currently forbidden by the \texttt{BHJet} model. The green shaded regions correspond to magnetically-dominated, charged jets. This un-physical regime can be avoided by taking low pair average Lorentz factors, for both model flavours (left panel), or by taking large values of $\sigma_{0}$ with \texttt{bljet} (right panel). The orange, red, purple and brown shaded regions indicate that the jet base is not magnetically-dominated, and therefore it is always forbidden to \texttt{bljet}, but it is accessible to \texttt{agnjet}.}
    \label{fig:BHJet_content}
\end{figure*}

\begin{equation}
    \sigma(z) = \frac{U_{\rm b}(z)+P_{\rm b}(z)}{U_{\rm p}(z)+U_{\rm e}(z)+P_{\rm e}(z)}.
    \label{eq:sigma}
\end{equation}
In \texttt{bljet} it is always assumed that the contribution of the leptons to the total energy budget is always negligible, so that the second and third terms in the denominator of equation (\ref{eq:sigma}) are much smaller than the first. In this case, our definition of $\sigma$ reduces to the standard definition, $\sigma = B^{2}/4\pi nm_{\rm p}c^{2}$. When this happens, equation (\ref{eq:bernoulli}) evaluated at $z_{\rm 0}$ and $z_{\rm acc}$ simplifies to:

\begin{equation}
    \sigma_0 = \left(1+\sigma_{\rm acc} \right) \frac{\gamma_{\rm acc}}{\gamma_0}-1.
    \label{eq:sigma_sol1}
\end{equation}
With the exception of thermal Comptonisation at the jet base, the bulk of the emission from the model commonly originates in the vicinity of the most beamed region, near $z_{\rm acc}$. Therefore, in order to allow the model some freedom in predicting the emission from this region, we take the magnetisation at $z_{\rm acc}$, $\sigma_{\rm acc}$, as a free parameter. Equation (\ref{eq:sigma_sol1}) then can be used to determine what the initial magnetic content of the jet needs to be, in order to reach a Lorentz factor $\gamma_{\rm acc}$, at a distance $z_{\rm acc}$ from the black hole, with a leftover magnetisation $\sigma_{\rm acc}$. Knowing the initial magnetisation $\sigma_{\rm 0}$, the magnetisation along the jet is:
\begin{equation}
    \sigma(z) = \frac{\gamma_0}{\gamma(z)}\left(1+\sigma_0 \right)-1
    \label{eq:sigma_sol2}
\end{equation}
and the definition of $\sigma(z)$ can be inverted to find the corresponding magnetic field:

\begin{equation}
B(z) = \left[4\pi \sigma(z) \left(n_{\rm p}(z)m_{\rm p} c^{2}+\Gamma\langle\gamma_e\rangle n_{\rm e}(z)m_{\rm e} c^{2}\right)\right]^{1/2},  
\label{eq:bljet_bfield}
\end{equation}
where we note that for this derivation to be valid, it is necessary that $n_{\rm p}m_{\rm p} c^{2}\gg \Gamma n_{\rm e}\langle\gamma\rangle m_{\rm e} c^{2}$. Beyond $z_{\rm acc}$, the jet assumes the standard \cite{Blandford79} profile, with constant Lorentz factor $\gamma_{\rm acc}$, constant opening angle $\theta_{\rm acc}=\rho/\gamma_{\rm acc}$, a magnetic profile consistent with a toroidal magnetic field $B(z)\propto z^{-1}$ and continuous particle re-acceleration throughout the jet. \textsl{As long as users make sure that $n_{\rm p}m_{\rm p} c^{2}\gg \Gamma n_{\rm e}\langle\gamma\rangle m_{\rm e} c^{2}$ holds, energy is conserved throughout the jet, because both the kinetic or magnetic energy carried by the jet are far larger than the internal energy of the particles. As a result, the radiating leptons can be re-accelerated without affecting the bulk properties of the outflow. This behaviour is the key difference between \texttt{agnjet} and \texttt{bljet}}. Three possible acceleration and collimation profiles are shown in fig.\ref{fig:bljet}. In the bulk acceleration region the jet becomes roughly parabolic once $z\gg z_{\rm 0}$, with smaller acceleration distance $z_{\rm acc}$ resulting in slightly more collimated outflows in the inner region. Similarly, larger terminal Lorentz factors $\gamma_{\rm acc}$ result in more collimated jets once they reach the outer, conical regions. Figures \ref{fig:agnjet} and \ref{fig:bljet} also show that \texttt{bljet} predicts smaller opening angles throughout the jet, compared to \texttt{agnjet}, regardless of input parameters. For example, in the former, the jet radius reaches $r(z)\approx 100 R_{\rm g}$ at a distance $z \approx 3000\,R_{\rm g}$. In the latter, this already occurs at $z \approx 500 R_{\rm g}$. 

Similarly to \texttt{agnjet}, \texttt{bljet} supports both jets that carry exactly one electron per proton, or jets with a mild pair load, set by the plasma-$\beta_{\rm p}$ parameter (although we stress that this is purely for convenience, and does not imply any strict causality between the two quantities). In either case, the factor $f_{\rm eq}(\beta_{\rm p},\eta,\langle\gamma\rangle)$ which sets the initial lepton number density is:

\begin{equation}
    f_{\rm eq}(\beta_{\rm p},\eta,\langle\gamma\rangle) = m_{\rm p}c^{2}+\langle\gamma\rangle \eta m_{\rm e}c^{2}\left(1+\frac{1}{\beta_{\rm p}}\right)
    \label{eq:feq_bljet}
\end{equation}
and the ratio of lepton to proton number density $\eta$ is:
\begin{equation}
    \eta = \frac{n_{\rm e,0}}{n_{\rm p,0}} = \frac{m_{\rm p}}{\langle \gamma \rangle m_{\rm e}} \frac{\beta_{\rm p}\sigma_{\rm 0}}{(2-\Gamma_{\rm ad} \beta_{\rm p}\sigma_{\rm 0})}.
    \label{eq:pair_bljet}
\end{equation}
The user can set this ratio to unity before running the code, in which case the appropriate $\beta_{\rm p}$ is computed from equation\ref{eq:pair_bljet} and substituted in equation\ref{eq:feq_bljet} to compute the initial number density. Regardless of the matter content assumed in the jet, it is recommended that $\beta_{\rm p}$ be kept frozen during spectral fits in order to avoid model degeneracies.

\subsection{Model flavour jet composition comparison}

Fig.\ref{fig:BHJet_content} shows a comparison of the matter content in \texttt{agnjet} and \texttt{bljet} as a function of $\beta_{\rm p}$, $\langle\gamma\rangle$ (left panel) and $\sigma_{\rm 0}$ (right panel), which through the empirical treatment detailed in the previous section combine to set the relative contribution of the protons, leptons and magnetic field carried by the jet. The goal of this section is to highlight that with both model flavours, users should exercise care in setting both of these parameters.

The first important consideration, highlighted in the left panel, is that the pair content in \texttt{bljet} with low initial $\sigma_{\rm 0}\approx 2$ is almost the same as that in \texttt{agnjet}, as long as $\beta_{\rm p}$ is chosen appropriately, although the two can deviate slightly for high $\langle\gamma\rangle$ and $\beta_{\rm p} \leq 1$.

Values of $\beta_{\rm p}$ near unity (implying that the outflow is near equipartition throughout) are allowed in \texttt{agnjet}, but not in \texttt{bljet}. This is caused by two effects. First, if the pair content is sufficiently high, these particles could carry a significant portion of the jet kinetic energy, invalidating the derivations in sec.\ref{sec:bljet}. These regions of parameter space, indicated by the gray shaded area, are currently forbidden to \texttt{bljet} (but could be explored with a further improvement of the jet dynamics). Second, if $\beta_{\rm p}$ is large enough, equation \ref{eq:pair_bljet} returns a negative value, resulting in  un-physical, charged jets. These regions are the orange, brown, purple and red areas; the exact location depends on the value of $\sigma_{\rm 0}$. This high $\beta_{\rm p}$ case invalidates the basic assumption of \textup{bljet} of a magnetically-dominated jet base, and therefore cannot be explored with any possible extension of this model flavour. It can however be probed by using \texttt{agnjet}. 

Finally, for sufficiently low values of $\beta_{\rm 0}$, both model flavours require $\eta<1$; this also implies un-physical, charged jets. Regions near this limit of extremely low $\beta_{\rm p}$ can be explored by \texttt{bljet}, provided that $\sigma_{\rm 0}$ is large enough (as shown in the right panel), but not by \texttt{agnjet}. 

Fig.\ref{fig:BHJet_content} highlights how the two model flavours are complementary: \texttt{bljet} allows the study of proton-heavy, magnetically dominated jets, while \texttt{agnjet} can treat pair-loaded, mildly magnetised, pressure-dominated outflows.

Finally, we note that within \texttt{BHJet} large pair contents should be handled with care. Because \texttt{CompPS} accounts for pair balance when computing the total spectrum \citep{Poutanen96}, the corrections to the Comptonisation kernel discussed in sec.\ref{sec:Kompton} do account for any potential pair production when computing the spectra. However in the present version of the code we do not track this potential additional population of leptons that may be produced in the jet. Doing so would require an extensive rework of the model, as we always assume that the number of particle is conserved (equation (\ref{eq:n(z)})). Instead, we assume that any positron/electron pair carried by the jet is injected in the nozzle region, and the rest of the outflow is sufficiently optically thin that no further pair process takes place. Additionally, the pair content of the jet is highly degenerate with the injected power. This can be shown by writing the jet power as a function of pair content for the bljet flavour, as discussed in \cite{Lucchini21}:
\begin{align}
N_{\rm j}(\eta) = & 2\gamma_0 \beta_0 c \pi r_0^{2} U_{\rm e,0}\times  \nonumber \\
    & \left(1+\frac{\sigma_{\rm 0} m_{\rm p} + \Gamma_{\rm ad} \sigma_{\rm 0} \eta \langle \gamma \rangle m_{\rm e}}{2 \eta \langle \gamma \rangle m_{\rm e}} + \frac{m_{\rm p}}{\eta \langle \gamma \rangle m_{\rm e}} \right).
\label{eq:Nj_vs_eta}
\end{align}
This equation shows that $N_{\rm j}$ is a monotonically decreasing function of $\eta$, so it is easy to tune $\eta$ so that a given jet power $N_{\rm j}$ can match observations. Note that this is less of an issue for \texttt{agnjet}, as $\beta_{\rm p}$ also directly sets the strength of the magnetic field in the nozzle (equation (\ref{eq:agnjet_bfield})), but some amount of degeneracy between the two still remains \citep[e.g.][]{Connors17}. Therefore, if possible, users should try to avoid fitting $N_{\rm j}$ and $\beta_{\rm p}$ at the same time when using \texttt{bljet}, and the values of $N_{\rm j}$ inferred from fits should always be interpreted very carefully, as they only provide lower limits to the jet power for a given value of $\beta_{\rm p}$.

\subsection{Additional features}
\label{sec:BHJet_additional}

All model flavours allow for the inclusion of a Shakura-Sunyaev disc, a black body, or both, as described in sec.~\ref{sec:Photon_fields}, in addition to the continuum emission from the whole jet. 

Regardless of model flavour, the particle distribution is always assumed to be thermal at the base of the jet, up to a distance $z_{\rm diss}$\footnote{When using \texttt{bljet}, $z_{\rm diss}$ is typically taken to be equal to $z_{\rm acc}$ (so that the location of non-thermal particle injection is highly beamed), but this need not be the case.} away from the black hole, which is a free parameter.  Here, the jet experiences an unspecified dissipation region (such as a shock or turbulent/shear regions) where we inject non-thermal particles in the form of a mixed distribution (using the \texttt{Kariba} classes \texttt{Mixed}, \texttt{Bknpower} or \texttt{Powerlaw}), channelling a fraction $f_{\rm nth}$ in a non-thermal tail with slope $s$, so that the number density of non-thermal particles is generally $n_{\rm nth,0}(z_{\rm acc}) = f_{\rm nth} n(z_{\rm acc})$. In order to avoid the issues highlighted in sec.\ref{sec:distributions}, we use the \texttt{Bknpower} class if $f_{\rm nth} >0.5$, setting the break momentum to be the average of a relativistic Maxwellian of the same temperature and the low energy slope to be $2$, as in a thermal distribution. Finally, if $f_{\rm nth} =1$, we inject a pure power-law distribution using the class \texttt{Powerlaw}, taking the minimum momentum to also be identical to the average Maxwellian distribution of the same temperature. Therefore, for large values of $f_{\rm nth}$, the fraction of non-thermal particles is fixed rather than a free parameter. Beyond $z_{\rm diss}$, a free parameter $f_{\rm pl}$ is used to smoothly decrease both the temperature of the thermal particles, and the fraction of non-thermal particles, so that:
\begin{equation}
n_{\rm nth}(z) = n_{\rm nth,0}\left(\frac{\log_{\rm 10}(z_{\rm diss})}{\log_{\rm 10}(z)}\right)^{f_{\rm pl}}
\label{eq:pldist}
\end{equation}
\begin{equation}
T_{\rm e}(z) = T_{\rm e}(z_{\rm diss})\left(\frac{\log_{\rm 10}(z_{\rm diss})}{\log_{\rm 10}(z)}\right)^{f_{\rm pl}},
\label{eq:tedist}
\end{equation}
where $T_{\rm e}(z_{\rm diss})$ and $n_{\rm nth,0}$ are the lepton temperature and the number density in non-thermal particle at $z_{\rm diss}$, respectively. The effect of $f_{\rm pl}$ is to  suppress the synchrotron emissivity along the jet axis, allowing the model to match an arbitrarily inverted spectrum by deviating slightly from the assumption of an isothermal jet. We note that the inversion of the radio spectrum is driven mainly by the $T_{\rm e}(z)$ decrease, if the temperature of the electrons is very high and relativistic, or the decrease of $n_{\rm nth}(z)$, if the temperature is low. We stress that the functional form to produce the spectral shape is purely phenomenological; equation (\ref{eq:pldist}) and (\ref{eq:tedist}) are purely convenient parameterisations to represent physics (like dissipation along the jet axis) that are not fully captured by the model.

Three additional free parameters, $f_{\rm heat}$, $f_{\rm \beta}$, and $f_{\rm sc}$, control the minimum, break, and maximum energy in the non-thermal tail. At $z_{\rm diss}$, the temperature of the particle distribution can be increased by a factor $f_{\rm heat}$, which is sometimes required by the data (e.g. \citealt{Lucchini19a}, \citealt{Kantzas22}). This parameter allows users to de-couple the minimum Lorentz factor $\gamma_{\rm min}$ of the non-thermal distribution from the temperature of the electrons in the nozzle; if the latter is mildly relativistic ($T_{\rm e}\leq 511\,\rm{keV}$) and with $f_{\rm heat} = 1$, then $\gamma_{\rm min} \approx 1$. As a rule of thumb, users should avoid values of $f_{\rm heat}$ high enough to result in $U_{\rm e}(z_{\rm acc})\approx U_{\rm p}(z_{\rm acc})$, but otherwise this parameter can take any value required by the data. Similarly to the jet pair content, we advise users to keep this parameter frozen to 1 (i.e., no additional heating) unless required, due to potential model degeneracies.

Starting from $z_{\rm diss}$, the model computes the steady state particle distribution along the jet following equation (\ref{eq:FP_steadystate}). The parameter $f_{\rm \beta}$ corresponds to the factor $\beta_{\rm exp}$  in equation (\ref{eq:ad_dot}), and can be used to set the relative importance of radiative versus adiabatic losses, thus setting the cooling break Lorentz factor $\gamma_{\rm brk}$ of the non-thermal distribution in equation (\ref{eq:FP_steadystate}). This description is analogous to that of \cite{Boettcher13}.

Finally, the maximum lepton Lorentz factor $\gamma_{\rm max}$ can be set in two ways. In previous versions of the model $\gamma_{\rm max}$ was set by balancing the cooling and acceleration timescales of the radiating electrons. The acceleration timescale is defined as:
\begin{equation}
     t_{\rm acc}(\gamma) = \frac{4\gamma m_{\rm e}c}{3f_{\rm sc}eB(z)},
\end{equation}
where $f_{\rm sc}$ is a free parameter, $e$ is the electron charge, and $B(z)$ is the strength of the magnetic field along the jet given by equation (\ref{eq:agnjet_bfield}) or (\ref{eq:bljet_bfield}). The maximum Lorentz factor (and its corresponding momentum) is then computed by solving:
\begin{equation}
    t_{\rm acc}^{-1}(\gamma_{\rm max}) = t_{\rm ad}^{-1}(\gamma_{\rm max}) + t_{\rm rad}^{-1}(\gamma_{\rm max}),
\end{equation}
where the cooling timescales are defined through equation (\ref{eq:ad_dot}) and \ref{eq:rad_dot}, taking $t_{\rm ad, rad} = \varrho/\dot{\varrho}_{\rm ad, rad}$. The radiation energy densities included in the radiative loss term are described below. This balance gives a maximum Lorentz factor:
\begin{align}
\gamma_{\rm max, pl}(z) = & \frac{-3m_{\rm e}c^{2}\beta_{\rm eff}}{8\sigma_{\rm t}U_{\rm rad}(z) r(z)} 
\nonumber
\\ & +\frac{1}{2}\sqrt{\left(\frac{-3m_{\rm e}c^{2}\beta_{\rm eff}}{4\sigma_{\rm t}U_{\rm rad}(z) r(z)}\right)^{2}+\frac{3f_{\rm sc}eB(z)}{4\sigma_{\rm t}U_{\rm rad}(z)}}.
\label{eq:max_gamma}
\end{align} 

Alternatively, in the newest version users can choose to either provide the value of $\gamma_{\rm max}$ directly, such that it is maintained throughout the jet. We will discuss the necessity of this flexibility in sec.~\ref{sec:blazars}. Similar to earlier versions of \texttt{agnjet}, \texttt{BHJet} allows for several photon fields to be included in the cooling rates, and to be used as seed photons for inverse-Compton scattering. Synchrotron cooling is always included, in which case $U_{\rm rad} = B(z)^{2}/8\pi$. For disc photons, we estimate the disc energy density by assuming that all the disc luminosity originates at the innermost radius $R_{\rm in}$, so that
\begin{equation}
U_{\rm rad}(z) = \frac{\delta^{2}_{\rm disc}(z) L_{\rm disc}}{4\pi D^{2}(z,R_{\rm in})},   
\end{equation}
where $L_{\rm disc}$ is the emitted disc luminosity (assuming for simplicity that $L_{\rm disc}$ is produced near $R_{\rm in}$, and non-zero torque at the boundary), $\delta_{\rm disc}$ is the (de)boosting factor for the disc photons as seen in the co-moving frame of the jet, and $D(z,R_{\rm in})$ is the Euclidean distance between the jet segment and the innermost radius of the disc. In the case of blazars, which are most relevant to this section, the energy density of either the magnetic fields, or the external photon fields described below, is much higher than that of the disc. 

Finally, two types of external fields can be included. The first case is that of a black body of energy density (in the co-moving frame) $U_{\rm rad}'$ and temperature $T_{\rm bb}$; this can represent, for example, the starlight of the host galaxy of an AGN or stellar companion in an XRB. When computing the luminosity from inverse Compton scattering of external photon fields, we use the Doppler factor $\delta$ rather than the Lorentz factor $\Gamma$ to convert between rest frame and co-moving quantities, following \cite{Dermer95}. Alternatively, the latest version of the model presented in this work allows for the inclusion of both a broad line region and torus, following a prescription similar to \cite{Ghisellini09}. The broad line region and torus are assumed to be located respectively at a distance:

\begin{table*}
\centering
\begin{tabular}{@{}cp{13.5cm}}
\hline
Source Parameters: &  \\ \bigstrut
$M_{\rm bh}$ & Mass of the black hole, in units of $M_{\sun}$ \\ \bigstrut
$D_{\rm lum}$ & Luminosity distance to the source, in units of \rm{kpc} \bigstrut \\ 
$\theta$ & Source viewing angle \bigstrut \\ 
$z_{\rm red}$ & Source redshift \bigstrut \\ \hline
disc Parameters: &  \\ \bigstrut
$L_{\rm disc}$ & disc luminosity, in units of $L_{\rm Edd}$ \bigstrut \\
$r_{\rm in}$ & Innermost disc radius, in units of $R_{\rm g}$ \bigstrut \\ 
$r_{\rm out}$ & Outer disc radius, in units of $R_{\rm g}$ \bigstrut\\ \hline
Main jet parameters: &  \bigstrut\\
$N_{\rm j}$ & Injected jet power, in units of $L_{\rm Edd}$\bigstrut \\
$r_{\rm 0}$ & Jet base radius, in units of $R_{\rm g}$\bigstrut \\
$z_{\rm diss}$ & Location of initial non-thermal particle acceleration, in units of $R_{\rm g}$ \bigstrut \\ 
$z_{\rm acc}$ & End of bulk jet acceleration if $\rm{velsw}>1$, unused otherwise \bigstrut\\
$z_{\rm max}$ & Largest distance considered to be in the compact jet \bigstrut \\
$\sigma_{\rm acc}$ & Leftover magnetisation at the end of bulk jet acceleration if $\rm{velsw}>1$, unused otherwise \bigstrut \\ 
$\beta_{\rm p}$ & Plasma beta parameter $U_{\rm e}/U_{\rm b}$ in the nozzle, can be used to indirectly set the pair content. If using \texttt{bljet}, then setting $p_{\rm beta}=0$ sets the jet to carry one electron per proton.  \bigstrut\\
$\rm{velsw}$ & If \rm{velsw}=0 or 1 then \texttt{agnjet} is used, with the adiabatic or isothermal profile respectively. Otherwise, \texttt{bljet} is used and $\gamma_{\rm acc} = \rm{velsw}$ \bigstrut \\ \hline
Particle distribution parameters: &  \bigstrut\\ 
$T_{\rm e}$ & Lepton temperature at the jet base, in units of \rm{keV} \bigstrut \\
$s$ & Slope of the non-thermal lepton distribution  \bigstrut \\ 
$f_{\rm nth}$ & Fraction of lepton number density channelled into the non-thermal tail \bigstrut\\
$f_{\rm heat}$ & Shock heating parameter; increases $T_{\rm e}$ by a factor $f_{\rm heat}$ at $z=z_{\rm diss}$ \bigstrut \\
$f_{\rm \beta}$ & Adiabatic timescale parameter; sets the break Lorentz factor $\gamma_{\rm b}$ in the non-thermal tail \bigstrut \\ 
$f_{\rm sc}$ & Acceleration timescale parameter, if $f_{\rm sc}<1$; sets the maximum Lorentz factor $\gamma_{\rm max}$ of the non-thermal tail. If $f_{\rm sc }\geq 10$, then $\gamma_{\rm max} = f_{\rm sc}$ throughout the jet  \bigstrut\\
$f_{\rm pl}$ & Phenomenological parameter used to fit inverted radio spectra \bigstrut\\ \hline
Inverse Compton Parameters: &  \\ \bigstrut
\rm{compsw} & If \rm{compsw}=0 only disc and synchrotron photons are scattered; if \rm{compsw}=1 a black body is added, if \rm{compsw}=2 the BLR and DT are included \bigstrut \\
$\rm{compar_1}$ & Sets the temperature of the black body $T_{\rm bb}$, in units of Kelvin, if \rm{compsw}=1. Sets the fraction of disc photons reprocessed in the BLR if  \rm{compsw}=2. Unused otherwise \bigstrut \\ 
$\rm{compar_2}$ & Sets the luminosity of the black body $L_{\rm bb}$, in units of $\rm{erg\,s^{-1}}$, if \rm{compsw}=1. Sets the fraction of disc photons reprocessed in the DT if  \rm{compsw}=2. Unused otherwise \bigstrut \\
$\rm{compar_3}$ & Sets the energy density of the black body, in uni ts of $\rm{erg\,cm^{-3}}$, if \rm{compsw}=1. Unused otherwise \bigstrut \\ \hline
Code Parameters: & \bigstrut \\ 
$h=2$ & Jet nozzle aspect ratio; no longer a free parameter due to the updates in the Comptonisation code \bigstrut \\
$\rm{nz}=100$ & Number of zones in which the jet is divided, only impacts computational time \bigstrut \\
$\rm{IC_{check}}$ & Comptonisation cutoff check; neglects computing the IC spectrum of a zone if it is estimated to be negligible \bigstrut \\
$z_{\rm min}=2$ & Nozzle height above the black hole, in units of $R_{\rm g}$  \bigstrut \\
$\beta_{\rm 0}=0.4$ & Initial jet speed, in units of $c$ \bigstrut \\
$z_{\rm cut}=1000$ & Distance at which the code switches from considering zones with equal height and radius, to a logarithmically-binned grid over the jet axis \bigstrut \\ \hline
\end{tabular}
\caption{\texttt{BHJet} model parameter summary.}
\label{tab:pars}
\end{table*}

\begin{equation}
    z_{\rm BLR} = 10^{17}L_{\rm d,45}^{1/2}\,\rm{cm}
    \label{eq:z_BLR}
\end{equation}
\begin{equation}    
    z_{\rm DT} = 2.5\cdot10^{18}L_{\rm d,45}^{1/2}\,\rm{cm},
    \label{eq:z_DT}
\end{equation}
where $L_{\rm d,45}$ is the disc luminosity in units of $10^{45}\,\rm{erg\,s^{-1}}$. The two photon fields reprocess a fraction $f_{\rm BLR}$ and $f_{\rm DT}$ of disc photons, respectively. When this happens, the observed disc luminosity in the model is reduced by a factor $f_{\rm BLR} + f_{\rm DT}$ (the emitted luminosity used to compute Comptonisation spectra and cooling rates remains unchanged). Both photon fields are treated as simple black bodies; the broad line region having a (co-moving) temperature equal to the (boosted) Lymann-$\alpha$ frequency, $T_{\rm BLR}^{\prime} = 13.6\cdot\delta(z) \,\rm{eV}$; the DT having $T_{\rm DT}^{\prime}=370\cdot\delta(z) \,\rm{K}$. When $z\leq z_{\rm BLR}$ or $z\leq z_{\rm DT}$ each photon field is boosted in the jet frame, so for the energy density of each we take:

\begin{equation}
    U_{\rm BLR}^{\prime}(z) = \frac{17\delta(z)^{2}}{12}\frac{f_{\rm BLR}L_{\rm d}}{4\pi z_{\rm BLR}^{2}}\quad \quad z\leq z_{\rm BLR},
    \label{eq:UBLR_boost}
\end{equation}
\begin{equation}
    U_{\rm DT}^{\prime}(z) = \frac{17\delta(z)^{2}}{12}\frac{f_{\rm DT}L_{\rm d}}{4\pi z_{\rm DT}^{2}}\quad \quad z\leq z_{\rm DT},
    \label{eq:UDT_boost}
\end{equation}
where $\delta(z)$ is the Doppler factor in each jet segment and $L_{\rm d}$ is the disc luminosity. Note that because in both cases $z_{\rm BLR/DT}\propto L_{\rm disc}^{1/2}$, the energy densities of both are constant in the observer frame as long as $z<z_{\rm BLR/DT}$. At large distances (we take $z\geq 3z_{\rm BLR/DT}$) from the location of the BLR/DT, the photon field is strongly de-boosted. In this case, the radiation energy density is \citep{Ghisellini09}:
\begin{align}
    U_{\rm BLR}^{\prime}(z) = & \frac{f_{\rm BLR}L_{\rm d}}{4\pi z_{\rm BLR}^{2}}\frac{\delta^{2}}{3\beta(z)}  [2(1-\beta(z)\mu_{\rm 1,BLR})^{3}-
    \nonumber \\
    &(1-\beta(z)\mu_{\rm 2,BLR})^{3}-(1-\beta(z))^{3}], \quad z\gg z_{\rm BLR}
    \label{eq:UBLR_deboost}
\end{align}
\begin{align}
    U_{\rm DT}^{\prime}(z) = & \frac{f_{\rm DT}L_{\rm d}}{4\pi z_{\rm DT}^{2}}\frac{\delta^{2}}{3\beta(z)}  [2(1-\beta(z)\mu_{\rm 1,DT})^{3}-
    \nonumber \\
    &(1-\beta(z)\mu_{\rm 2,DT})^{3}-(1-\beta(z))^{3}], \quad z\gg z_{\rm DT},
    \label{eq:UDT_deboost}
\end{align}where 

\begin{figure*}
    \centering
    \includegraphics[width=\textwidth]{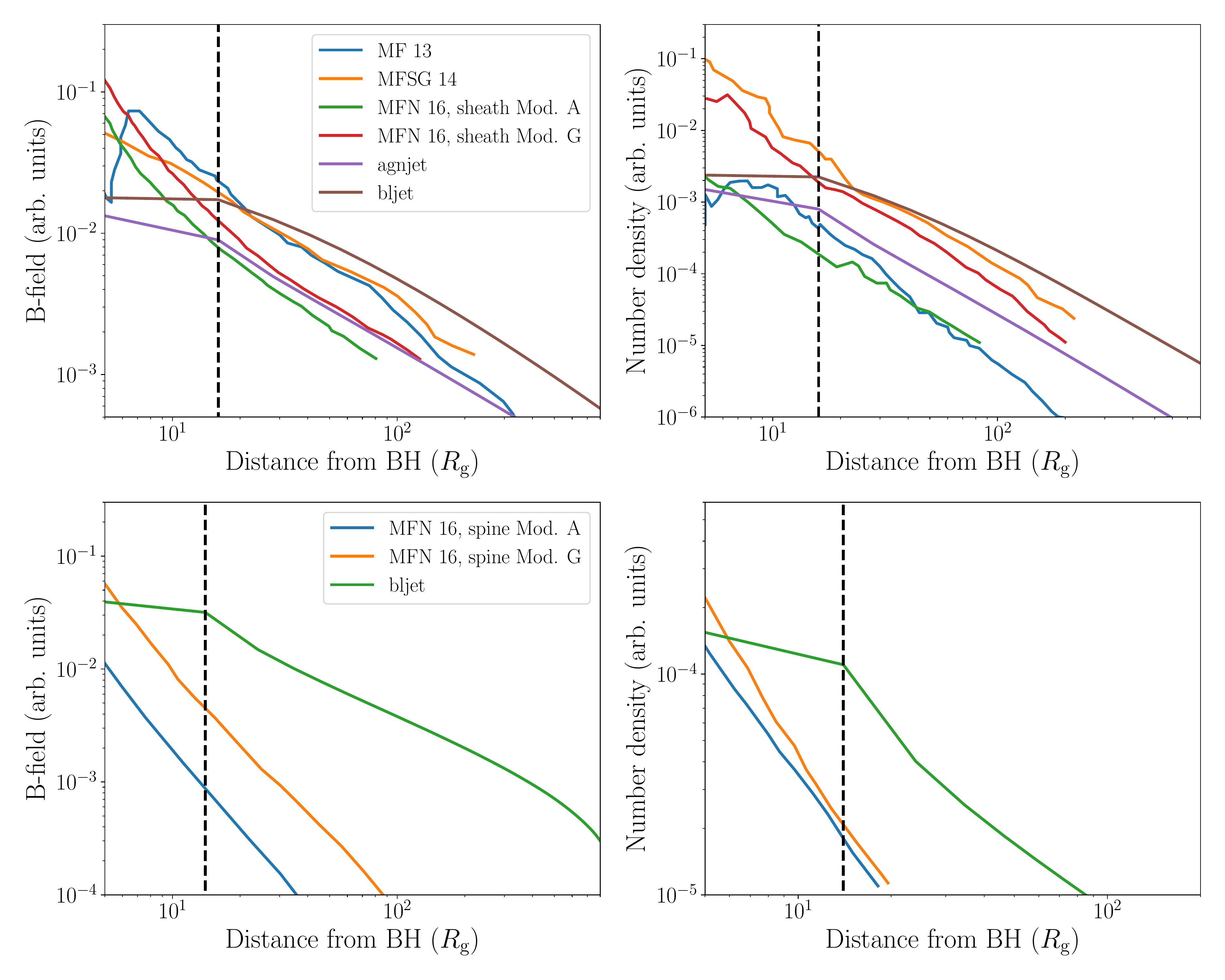}
    \caption{Magnetic field (left column) and number density (right column) scaling against distance from the black hole in both our model and GRMHD simulations. Top row: comparison between both model flavours in \texttt{BHJet} with GRMHD models of \protect\cite{Monika13} (blue line), \protect\cite{Monika14} (orange line), and \protect\cite{Monika16b} (green and red line, for low and high BH spins, respectively). Here for \texttt{bljet} we took $\gamma_{\rm acc}=3 $ and $z_{\rm acc} = 5500\,\rm{R_g}$. Both model flavours are in good agreement with the results of simulations beyond the end of the nozzle (shown by the dashed line). Bottom row: identical comparison, but focusing on the spine region from \protect\cite{Monika16b}, and using only \texttt{bljet} with $\gamma_{\rm acc} = 20 $ and $z_{\rm acc} = 1000\,\rm{R_g}$. Here, the model requires much higher number density and magnetic field throughout the outflow than what is observed in simulations. }
    \label{fig:mhd_comparison1}
\end{figure*}

\begin{equation}
    \mu_{1,\rm{BLR/DT}} = \frac{1}{\sqrt{1+\left(\dfrac{z_{\rm BLR/DT}}{z}\right)^{2}}}
\end{equation}
\begin{equation}
    \mu_{2,\rm{BLR/DT}} = \sqrt{1-\left(\dfrac{z_{\rm BLR/DT}}{z}\right)^{2}}
\end{equation}
are geometrical coefficients which account for the solid angle under which the BLR/DT are viewed, $\beta(z)$ is the speed of the jet at a distance $z$ from the black hole, and $z_{\rm BLR/DT}$ are defined in equation (\ref{eq:z_BLR}) and \ref{eq:z_DT}. For distances between $z_{\rm BLR/DT}$ and $3z_{\rm BLR/DT}$, the exact value of the co-moving energy density depends on the details of the geometry of the BLR or DT. For simplicity, we interpolate linearly between the energy density at the start of the BLR and at $z=3z_{\rm BLR/DT}$ where we first switch to equation (\ref{eq:UBLR_deboost}) and \ref{eq:UDT_deboost}:
\begin{align}
    U_{\rm BLR,DT}^{\prime}(z) = & U_{\rm BLR,DT}^{\prime}(z_{\rm BLR,DT}) +  \frac{z-z_{\rm BLR}}{2z_{\rm BLR}}\times
    \\
    & \left(U_{\rm BLR,DT}^{\prime}(3z_{\rm BLR,DT})-U_{\rm BLR,DT}^{\prime}(3z_{\rm BLR,DT})\right) \nonumber 
\end{align}

Finally, we require the source distance luminosity $D_{\rm lum}$ (and, for extra galactic sources, the redshift $z_{\rm red}$) to be known, so that the total observed bolometric flux $F$ for a given luminosity $L$ is given by the standard formula $F = L/4\pi D^{2}_{\rm lum}(1+z_{\rm red})^{2}$ and the frequencies can be shifted by a factor $1+z_{\rm red}$.

All model parameters are reported in table \ref{tab:pars}. Note that the code parameters require re-compilation and can not be changed at run time; they have a negligible effect on the SED and are included only for completeness.

\section{Comparison with ideal GRMHD simulations of jets}
\label{sec:GRMHD}

The strength of a multi-zone approach similar to \texttt{BHJet} can easily be illustrated by comparing the properties of our jets with the results of GRMHD simulations. In GRMHD, the jet can roughly be divided in two separate regions: a low density, highly magnetised central region, typically referred to as the jet spine, surrounded by a mass-loaded, less magnetised sheath of plasma \citep[e.g.][]{McKinney06,Penna13,Nakamura18}. In general, the outflow near the sheath reaches only mildly relativistic speeds, while the plasma inside the spine can accelerate to high Lorentz factors \citep{Chatterjee19}. Observations of jetted AGN are in broad agreement with this picture \citep[e.g,][]{Nagai14,Mertens16,Hada16,Giovannini18}. 

\begin{figure}
    \centering
    \includegraphics[width=0.49\textwidth]{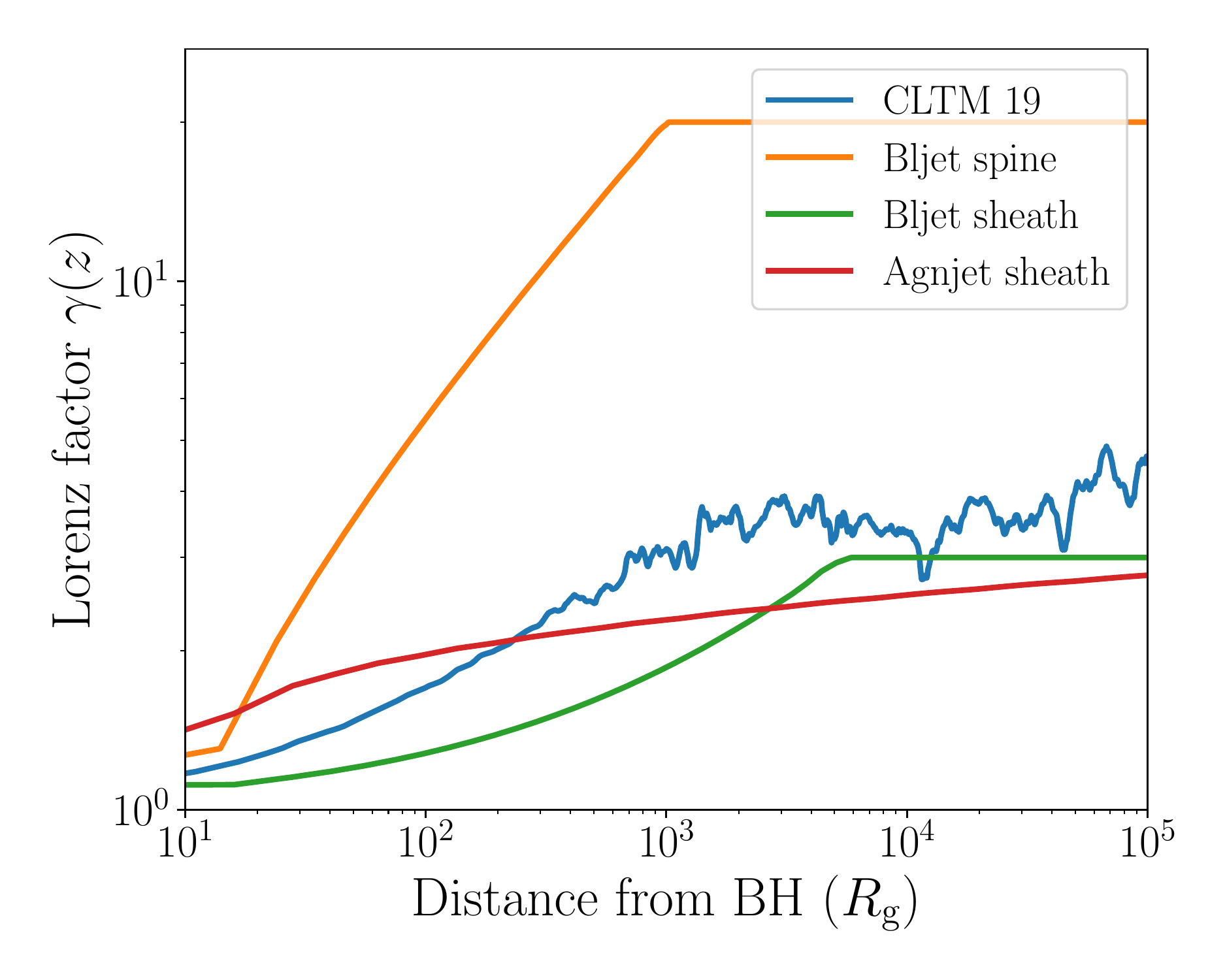}
    \caption{Comparison of jet speeds between the GRMHD simulations of \protect\cite{Chatterjee19} (blue line, computed by choosing a streamline in the spine), and \texttt{BHJet} for different model parameters and flavours. As in fig.\ref{fig:mhd_comparison1}, we took $\gamma_{\rm acc}=3 $ and $z_{\rm acc} = 5500\,\rm{R_g}$, and $\gamma_{\rm acc} = 20 $ and $z_{\rm acc} = 1000\,\rm{R_g}$, for the sheath-like and spine-like models with \texttt{bljet}. The discrepancy between the simulation result and the \texttt{bljet} spine model, as inferred by modelling blazar SEDs, is evident.}
    \label{fig:mhd_comparison2}
\end{figure}

The \texttt{BHJet} family of models can capture either a mildly relativistic, mass-loaded, Blandford-Payne driven sheath or a highly relativistic, highly magnetised, Blandford-Znajek driven spine, essentially assuming that only one at a time dominates the observed emission. In the former case, both model flavours are appropriate, provided that a user sets a low terminal velocity when using \texttt{bljet} ($\gamma_{\rm acc}\geq 2-3$). This typically happens when \texttt{BHJet} is used to model low-luminosity AGN or X-ray binaries \citep[e.g.][]{Markoff01,Markoff05,Maitra09,Connors17,Lucchini21}. In the latter case, only \texttt{bljet} is appropriate, and requires the user to set a high ($\gamma_{\rm acc}\geq 4-5$) terminal Lorentz factor for the jet. This regime has been invoked for modelling blazars and other jet-dominated AGN \citep{Lucchini19a,Lucchini19b}.

We are particularly interested in comparing quantities that have a large impact on the final SED, focusing on the region beyond the jet nozzle. Fig.~\ref{fig:mhd_comparison1} shows how the magnetic field strength and particle number density scales in \texttt{BHJet}, for different model flavours and initial parameters, compared to the work of \cite{Monika13}, \cite{Monika14} and \cite{Monika16b}. These authors used 2D and 3D GRMHD simulations to predict the radio flux and shape of the radio spectrum originating in the compact jet (focusing on Sgr\,A*), as well as to explore the effects of black hole spin on scale-invariant models for the radio emission of jets; therefore, these simulations are an ideal benchmark for our model. 

The top panel of fig.\ref{fig:mhd_comparison1} shows a comparison between the sheath region in simulations, and both model flavours, using model parameters typically required by blazars (for the spine) and X-ray binaries or LLAGN (for the sheath). These are discussed in more detail in sec.~\ref{sec:applications}. In this regime, both the magnetic field and number density in \texttt{BHJet} scale with distance very similarly to simulations, and therefore we expect that the predicted SEDs of both should be roughly similar. The same is not true when comparing \texttt{bljet} with a jet spine, as highlighted in the bottom panels. In this regime, both quantities drop very quickly with distance in simulations, and this behaviour is not replicated in our model. We note, however, that the dynamics of the spine region in GRMHD is strongly affected by the artificial particles introduced via a numerical floor density, and the processes that might provide more realistic mass-loading (such as pair production in the BH ergosphere, e.g. \citealt{Neronov07,Rieger08,Levinson11,Moscibrodzka11,Broderick15}) are not captured by these simulations. At the same time, when changing between studying a sheath and a spine in \texttt{BHJet}, we assume naively that the dynamics of both can be described by the same simple formalism detailed in sec.~\ref{sec:bljet}. Finally, we note that while comparing the scaling of the magnetic field and number density is relatively straightforward, the same is not true for the electron temperature. This is because the simulations we compare our model to only included the contribution of the electrons after post-processing the simulations with phenomenological prescriptions similar to our isothermal models \citep[e.g.,][]{Monika16}. A more appropriate comparison for the energy in the electrons would be with simulations that include two-temperature fluids; however, these are still being developed.

The discrepancy between the spine region in simulations, and the models inferred from modelling powerful AGN, can be further highlighted by comparing the jet velocity profile along its axis. In particular, in fig.~\ref{fig:mhd_comparison2} we compare different model flavours and setups to the results of \cite{Chatterjee19}, who performed the highest resolution study of large-scale jet acceleration GRMHD simulations to date. At first look, it appears that the MHD result is in fair agreement with our models for the sheath, with the jet spine accelerating much more efficiently. However, the streamline chosen by \cite{Chatterjee19} to calculate the jet speed is located in the spine region.  Despite this, the terminal speed of the jet is fairly moderate ($\gamma \approx 4-5$), and is only reached at fairly large distances ($z\approx 10^{4}\,\rm{R_g}$) from the black hole. This is in sharp disagreement with models of the emitting region of blazars, including \texttt{bljet}: for example, when in the case of the canonical BL Lac PKS\,2155$-$304, \cite{Lucchini19a} found that the jet Lorentz factor needs to reach $\approx 15$ at a distance of $\approx 600-1700\,\rm{R_g}$ from the black hole. This value (and in general, observations of blazars, e.g. \citealt{Marscher08,Pushkarev09,Abdo11,Pushkarev17,Costamante18}) implies far more efficient jet acceleration than what is observed in GRMHD simulations.

In conclusion, we find that our model behaves fairly similarly to a jet sheath (which likely dominates the radio emission in off-axis sources, e.g. 3C84: \citealt{Giovannini18}, M87: \citealt{Mertens16}, Cen A: \citealt{Janssen21}) in numerical simulations. However, modelling sources in which the spine might be dominating the emission (such as blazars), requires jets that are more relativistic, and accelerated more efficiently to their terminal Lorentz factor, than what simulations currently seem to produce. 

\section{Example Applications}
\label{sec:applications}

In this section we apply the model to a variety of sources. Rather than provide a detailed physical picture of each, this section has the goal of exploring the parameter space of the model and highlighting some of its features. All the fits presented in this section were performed using the spectral tool \textsl{ISIS} \citep{Houck00}, version 1.6.2-47. All spectral fits include the contribution of absorbing neutral material along the line of sight, using the \texttt{tbabs} model \citep{Wilms00}.  We adopt the abudances of \cite{Wilms00} and set the photo-ionisation cross-sections according to \cite{Verner96}.

\begin{figure*}
    \centering
    \includegraphics[width=0.49\textwidth]{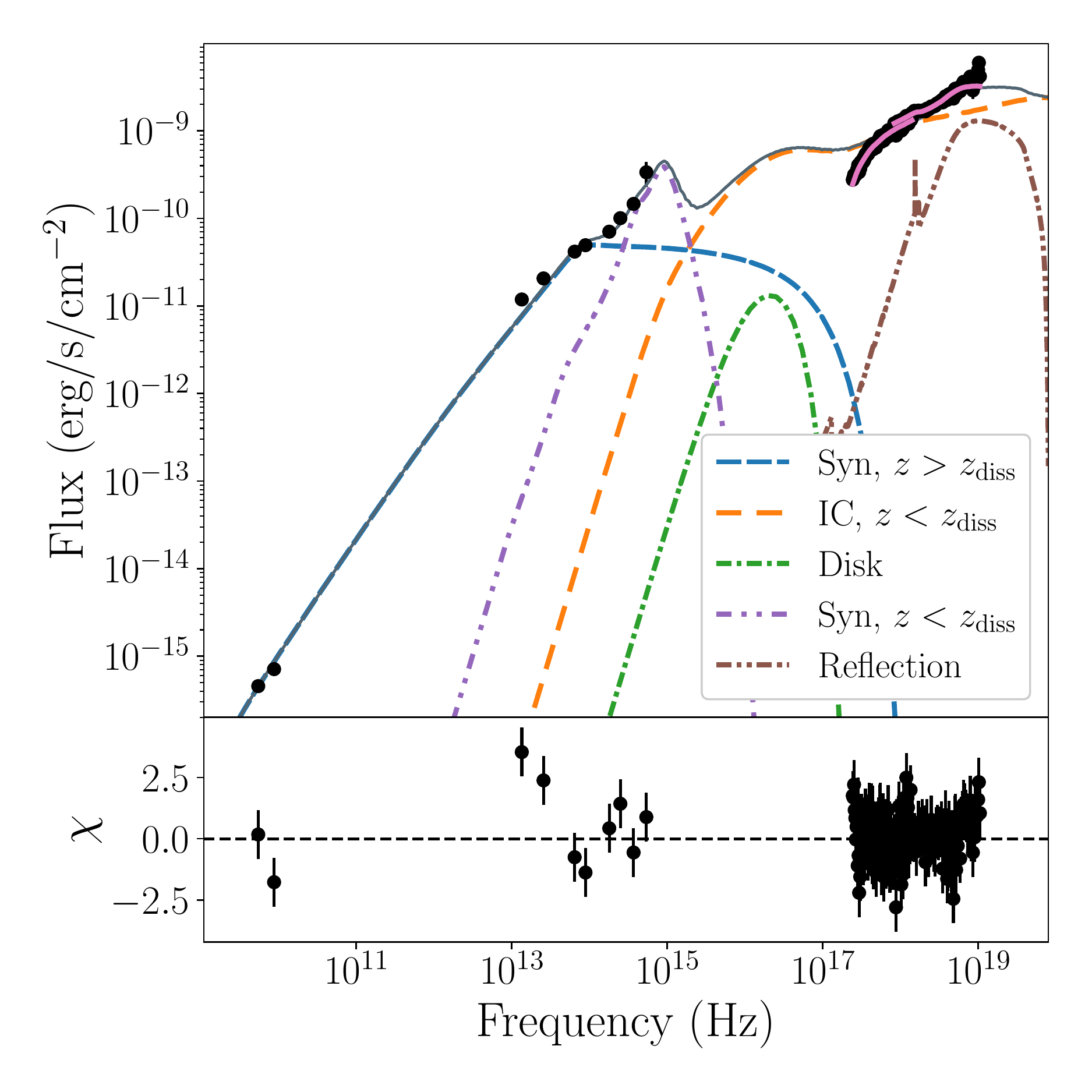}
    \includegraphics[width=0.49\textwidth]{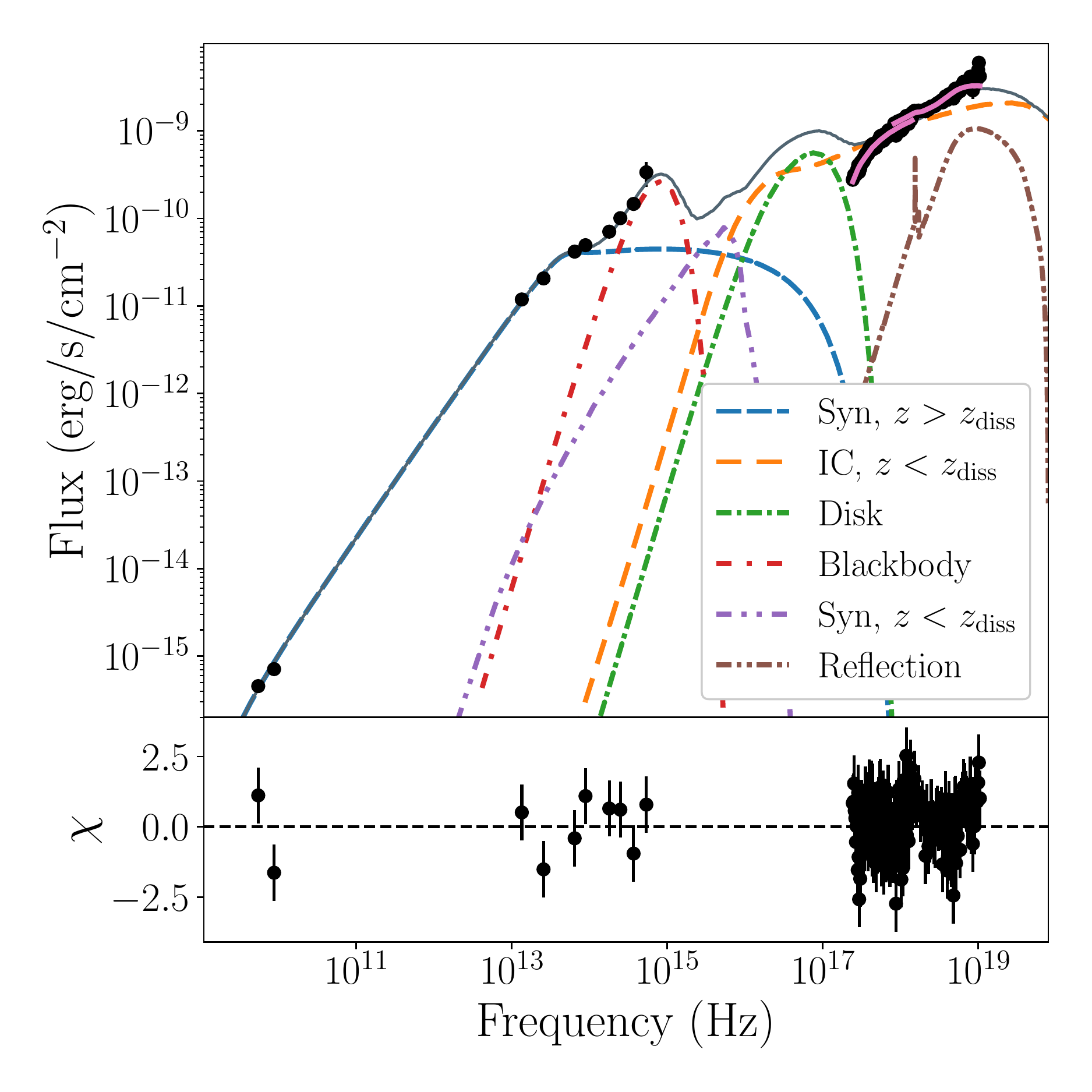}
    \caption{Jet-dominated fits of a bright hard state of BHXB GX~339-4, data from \protect\cite{Kantzas22}. Left panel: best fit with model A, using the \texttt{agnjet} model flavour. Right panel: fit best fit with model B, using the \texttt{bljet} model flavour. The $\chi^{2}/\rm{d.o.f}$ is $168.8/197$ and $191.1/195$, respectively. The blue line is non-thermal synchrotron emission from $z\geq z_{\rm diss}$; the orange line is Comptonisation of both disc and cyclo-synchrotron photons near the jet base, the green line is the disc emission, the red line in model B is a phenomenological black body to match the optical excess, the purple line is cyclo-synchrotron emission from near the jet base, and the brown line is the reflection spectrum.    }
    \label{fig:gx339_fits1}
\end{figure*}
\begin{table}
\begin{center}
\begin{tabular}{| l | c | c | c |}
\hline
Model &  A & B & C \\ 
\hline
$N_{\mathrm{j}}$ ($L_{\mathrm{Edd}}$) & $0.09$ & $0.13$ & $0.11$ \\
$r_{\rm 0}$ ($\mathrm{r_{g}}$) & $181$ & $5.5$ & $20$ \\
$z_{\mathrm{diss}}$ ($\mathrm{r_{g}}$)  & $3.2\cdot10^{3}$ & $1.4\cdot10^{4}$ & $1.5\cdot10^{5}$ \\
$T_{\mathrm{e}}$ ($\rm{keV}$) & $500$ & $109$ & $22$\\
$f_{\rm pl}$ & $8.8$ & $8.4$  & $19$\\
$s$ & $2^{*}$ & $2^{*}$ & $1.5$\\
$f_{\rm heat}$ & $2.6$ & $11.8$ & $70$ \\
$\beta_{\rm p}$ & $0.17$ & $0.02085^{*}$ & $0.02085^{*}$\\
$L_{\rm disc}$ & $7\cdot10^{-4}$ & $2.8\cdot10^{-2}$ & $3.6\cdot10^{-2}$ \\
$r_{\rm in}$ ($\mathrm{r_{g}}$) & $200$ & $86$ & $97$ \\
$L_{\rm bb}$ ($\rm{ergs}$)& // & $2.7\cdot10^{36}$ & $1.1\cdot10^{36}$\\
$T_{\rm bb}$ ($\rm{K}$) & // & $1.0\cdot10^{4}$ & $7\cdot10^{3}$\\
$\rm{rel\_refl}$ & $0.76$ & $0.58$ & $0.30$\\
$\rm{line\_norm}$ & $1.5\cdot10^{-3}$ & $1.6\cdot10^{-3}$ & $8.1\cdot10^{-3}$\\
$\rm{line\_E}$ ($\rm{keV}$) & $6.4^{*}$ & $6.4^{*}$ & $6.4^{*}$ \\
$\rm{line\_\sigma}$ ($\rm{keV}$) & $1\cdot10^{-3*}$ & $1\cdot10^{-3*}$ & $1.2$ \\
$\rm{nH}$ ($\rm{cm^{-2}}$) & $4\cdot10^{21*}$ & $4\cdot10^{21*}$ & $4\cdot10^{21*}$\\
\hline
$r_{\rm cor}$ ($\mathrm{r_{g}}$) & & & $145$\\
$T_{\mathrm{e, cor}}$ ($\rm{keV}$) & &  & $108$ \\
$\tau_{\mathrm{cor}}$  &  &  & $1.4$\\
\hline 
$\chi^{2}_{\rm red}$ & $0.96$ & $0.85$ &  $0.99$\\
\hline
\end{tabular}
\label{tab:gx339-fit}
\caption{Best-fitting parameters for GX339 -4. For model B we froze $\beta_{\rm p}$ to fix the pair content $\approx 10$ during the fit, and took $\gamma_{\rm acc}=3$, $\sigma_{\rm diss}=0.1$ and $z_{\rm acc}=z_{\rm diss}$, following \protect\cite{Lucchini21}. In both models we took $f_{\rm nth}=0.1$ and $r_{\rm out}=10^{5}\,R_{\rm g}$. Parameters marked with an asterisk were frozen.}
\end{center}
\end{table}

\begin{figure}
    \centering
    \includegraphics[width=0.49\textwidth]{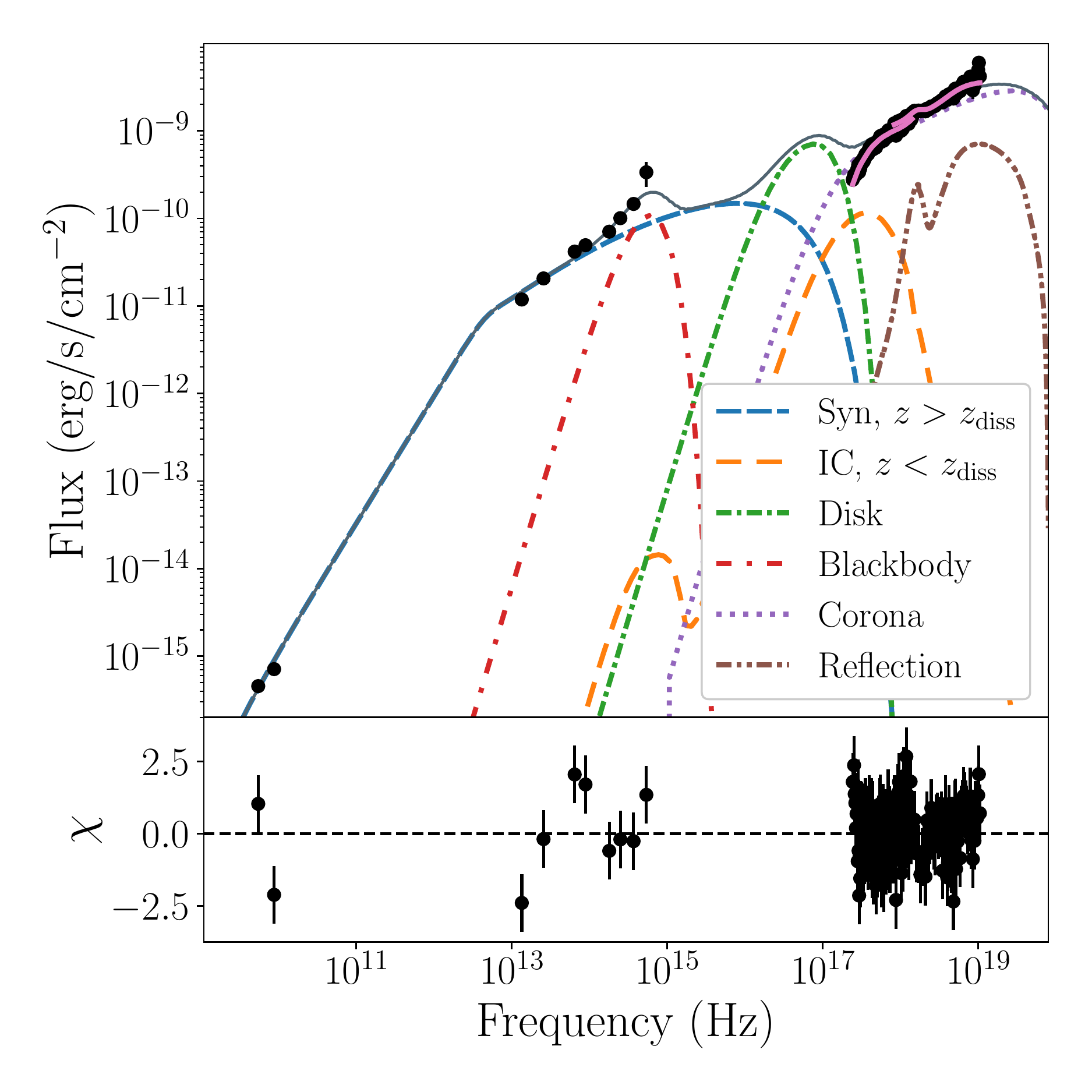}
    \caption{Jet+corona fit of GX 339-4. For Model C, the $\chi^{2}/\rm{d.o.f}$ is $187.8/89=0.99$. The colour convention is identical to fig.\ref{fig:gx339_fits1}, except for the purple line which here corresponds to the Comptonized coronal emission (the cyclo-synchrotron contribution is negligible for this model).     }
    \label{fig:gx339_fits2}
\end{figure}

\subsection{GX 339-4}
\label{sec:GX}

GX-339 is a low mass black hole X-ray binary, which exhibits a remarkably regular duty cycle: it goes into outburst every roughly 2-3 years, and as such it has been the target of multiple multi-wavelength campaigns over the years \citep[e.g.][]{Homan05b,Gandhi08,CadolleBel11,Corbel13}. Because of this regular behaviour, it is considered the ``benchmark source'' of this class of accreting compact objects. However, despite years of study, the physical properties of the source are not well constrained, and estimates of the black hole mass, distance and inclination all have large uncertainties. They range between $\sim 3-10\,M_{\sun}$ \citep{Hynes03,MunozDarias08,Heida17}, $\sim 6-15$ kiloparsecs \citep{Hynes04,Zdziarski04} and $15-50$ degrees \citep{Miller06,Reis08,Done10,Plant14,Plant15,Garcia15,Parker16}, respectively. The latest estimates generally favour larger distances, and a wide range of black hole masses \citep{Heida17,Zdziarski19}. In this section we take $M_{\rm bh} = 9.8M_{\sun}$, $d=8\,\rm{kpc}$ and $\theta=40^{\circ}$ and fix the column density  $\rm{n_H}$ to $4.3\cdot 10^{21}\,\rm{cm^{-2}}$, as in \cite{Connors19}. 

As is typical of BHBs, GX-339 shows bright radio emission in its hard and hard-intermediate states, attributed to the presence of a compact jet. Furthermore, both its variability and spectral characteristics suggest that the jet emission extends up to the IR and optical bands \citep[e.g.][]{Corbel02,Markoff05,Gandhi08,Vincentelli19,Tetarenko20}. In this section we focus on a single high-quality SED taken during the 2010 outburst. The data discussed here are the same that \cite{Kantzas22} modelled with a newer version of \texttt{BHJet} that is still undergoing development, which includes hadronic processes. Radio data were obtained by the Australian Telescope Compact Array (ATCA) on MJD 55263 \citep{Corbel13}. IR and optical data were obtained from the Wide-field Infrared Survey Explorer (WISE) on MJD 52666 and Small \& Moderate Aperture Research Telescope System (SMARTS) on MJD 55263, respectively \citep{Gandhi11}. The X-ray data include a soft X-ray spectrum from \Swift/XRT taken on MJD 55262, as well as a hard X-ray spectrum from \RXTE/PCA observation from MJD 55263 \citep{Corbel13}. We chose this SED because it is representative of a typical bright hard state of a low-mass BHB.

The goal of this section is to show that for relatively bright sources, the X-ray power-law component can be reproduced by inverse Comptonisation of both disc and cyclo-synchrotron radiation fields at the base of the jet, and to compare such a model with one in which the coronal emission originates from a separate region (like the hot accretion flow), as the latest IXPE observations of Cyg X-1 suggest \citep{Ixpe1}. We take $s=2$, $f_{\rm sc}=10^{-6}$ and $f_{\rm \beta}=0.1$ in our jet-dominated spectral fits; this causes the break and maximum Lorentz factor of the electrons to be fairly low ($\gamma_{\rm break} \approx 10^{1-2}$ and $\gamma_{\rm max} \approx 10^{3}$, respectively), which in turn suppresses the contribution of non-thermal synchrotron to the soft X-ray emission. This is the same regime that \cite{Connors19} explored with an older version of the model (the data considered here is SED \#12 in their work, and is handled identically except for the inclusion of the \Swift/XRT spectrum), and is consistent with both the fairly strong reflection features detected in the X-ray spectrum \citep{Markoff04}, and the behaviour of the X-ray continuum lags \cite{Connors19}. For the jet+hot flow fit, instead, we found we had to leave $s$ free to correctly reproduce the infrared/optical data. We present fits performed with both model flavours, although we stress that if the jet terminal Lorentz factor is small and the jet is mildly magnetically dominated  ($\sigma_{\rm 0}\approx\rm{a\,\,few}, \beta_{\rm p}\approx 10^{-1}-10^{-2}$), as is likely the case for X-ray binary jets, the two are fairly similar and the choice of model flavour has a minor impact on the parameters inferred from modelling. Regardless of model flavour, we include a phenomenological description for the reflection component using the \texttt{reflect} model \citep{reflect}, a gaussian line peaking at $6.4\,\rm{keV}$. We find that the model can fit the data very well in two different jet-dominated regimes, which we call model A and model B. The data is reproduced equally well if the X-rays are not dominated by the jet Comptonization component, which we call model C. Models A and B are shown in fig.~\ref{fig:gx339_fits1}, and Model C is shown in fig.~\ref{fig:gx339_fits2}. The best-fit parameters for all models are reported in tab.~\ref{tab:gx339-fit}.

In model A, shown in the left panel of fig.\ref{fig:gx339_fits1}, the electron temperature is very high ($T_{\rm e} = 500\,\rm{keV}$) and the jet base is very wide ($r_{\rm 0}=181\,\rm{R_g}$), resulting in very low optical depth $\tau \approx 0.1$. In this case, the thermal Comptonisation spectrum has some amount of curvature which may mimic a cutoff. Depending on the quality of available high-energy data, it may be possible to distinguish this curvature from a true exponential cutoff displayed from Models B and C, as is the case for the 2006 outburst of this source \citep{Motta09}; unfortunately, that is not the case for this particular SED as \RXTE/HEXTE had failed by this time. Additionally, we caution users that this regime may suffer from excessive compactness, leading to possible runway pair production \cite{Malzac09}; however, that is not the case for the fits presented here ($l\approx 10$ for $\theta\approx 1$, which is similar to many AGN coronae e.g. \citealt{Fabian15}). \cite{Connors19} found that in this case (model B1 in their paper), the model could not reproduce the data very well, which is not the case for our updated code. This discrepancy is mainly caused by the improvements included in the present version of the code, which were not present in their model. In particular, in the newest model the non-thermal particle distribution responsible for the radio through infrared emission has more freedom to match the data than in older works, thanks to introduction of the $f_{\rm heat}$ and $f_{\rm pl}$ parameters. This additional freedom in turn allows more freedom to adjust the temperature and optical depth of the jet base, resulting in a better fit of the data. In particular, in \cite{Connors19} the best fit of the same data with an older version of the model resulted in $\chi^{2}_{\rm red} = 3.2$, which improved to $\chi^{2}_{\rm red} = 2.5$ by adding an additional thermal Comptonisation component that dominates the X-ray continuum. Instead, with the updated model we find $\chi^{2}_{\rm red} \approx 0.96$ without invoking any further contributions to the SED (although we note that our SED has more data points, due to the inclusion of the \Swift/XRT spectrum). Additionally, in this high temperature, low optical depth regime, we find that the bulk of the optical emission can be attributed to cyclo-synchrotron emission in the jet base.
 
In model B, shown in the right panel of fig.\ref{fig:gx339_fits1}, the data can be well reproduced with a much more compact jet base ($r_{\rm 0}=5.5\,\rm{R_g}$) with lower temperature ($T_{\rm e} = 109\,\rm{keV}$), resulting in much higher optical depth ($\tau \approx 1.4$). The quality of the fit is slightly better compared to model A ($\chi^{2}_{\rm red} \approx 0.85$), although we note that in both cases we are over-fitting the data slightly. This regime is similar to that explored in \cite{Lucchini21}, and closely resembles a standard black hole corona; in particular, our model can be thought of as a physical realisation of an outflowing lamp-post corona \citep{Beloborodov99,Dauser13}. This finding is further strengthened by the changes in the Comptonisation code detailed in sec.~\ref{sec:Kompton}, which generally require higher optical depth (and therefore compact jet bases that more closely resemble a point source) to produce a given photon index, compared to older versions of \textup{BHJet}. Indeed, the size of the jet base inferred from our spectral modelling is a factor $\approx 2-3$ smaller than that found in \cite{Lucchini21}. Additionally, in this case the cyclo-synchrotron emission from the jet base is too faint to reproduce the optical data, which instead we reproduce phenomenologically with a black body. This optical excess could be caused, for instance, by disc irradiation, although we note that the optical luminosity is uncomfortably large for irradiation-dominated models \citep{Tetarenko20}.

Model C is shown in fig.\ref{fig:gx339_fits2}. In this case we added an additional coronal component characterised by an optical depth $\tau_{\rm cor}$, filled with thermal electrons of temperature $T_{\rm e, cor}$ and utilising the effective radius $R_{\rm cor}$ as a normalisation constant (which should not be confused with a physical coronal size due to the simplicity of the model), identically to \cite{Kantzas20}. We only consider disk seed photons and assume a spherical geometry to simplify this additional component. The Comptonised spectrum is computed identically to that of the jet using \texttt{Kariba}. We find that we can recover a good fit with coronal conditions similar to those of Model B, as long as we force the temperature of the electrons in the jet base to be low $T_{\rm e}\leq20\,\rm{keV}$, which suppresses their emission greatly. In this regime, the jet contribution to the Comptonized spectrum is in the form of a weak soft excess at $\approx$ a few keV. In terms of jet parameters, the main difference with respect to models A and B is the non-thermal particle slope $s$, which favours very hard values $s\approx1.5$, and the location of the dissipation region $z_{\rm diss}$, which increases by over one order of magnitude to $\approx 10^{5}\,\rm{R_g}$. Such a large distance of the acceleration region is in tension with X-ray to infrared time-lag measurements in GX 339-4 \citep{Gandhi11}, unlike models A and B. However, the model C constraints are driven by the suppression of the jet base cyclo-synchrotron emission in the optical range, which in turn demands a larger non-thermal synchrotron flux to match the data. In principle it should be possible to reconcile Model C with the infrared timing properties, by invoking a minor cyclo-synchrotron contribution from the hot flow; however, including a fully developed hot flow component in our jet model is beyond the scope of this initial public release.

Finally, we note that in all models we find that the disc has a significant truncation radius in this epoch ($R_{\rm in}\geq 80\,R_{\rm g}$). This is a common trend in fits with \texttt{BHJet} of X-ray binaries, and occurs when cyclo-synchrotron emission contributes a significant amount of soft photons for inverse Compton scattering. This causes the Comptonisation spectrum to extend down to UV frequencies, covering part of the soft excess typically attributed to the disc. We note however that the disc model we currently implemented is extremely simplistic, so these numbers should be interpreted with care. 

\subsection{M\,81$^{*}$: a case study of LLAGN}
\label{sec:M81}

M\,81 (or NGC\,3031) is a spiral galaxy and one of the nearest to us (3.5 Mpc, \citealt{Freedman94}), together with Centarus A, to host an active galactic nucleus at its centre. Its central mass has been estimated through spectroscopy with the Hubble Space Telescope (HST) to be $7\times10^{7}M_{\sun}$ \citep{Devereux03}. The nuclear region of M\,81, hereafter M\,81$^*$, harbours a faint AGN \citep[see e.g.][]{Ho99} with a luminosity on the order of $10^{37}$ erg s$^{-1}$in radio and  $\sim 10^{40}$ erg s$^{-1}$ in the optical and X-rays. Because of its weak emission together with low-ionisation nuclear emission-line region (LINER; \citealt{Heckman80}) and Seyfert 1 properties, M~81$^*$ is classified as a low-luminosity AGN (LLAGN) \citep[e.g.][and references therein]{Ho08}. Moreover, the radio-to-X-ray luminosity of M\,81$^*$ varies from values characteristic of radio-quiet Seyfert galaxies ($R_X \sim 1.8 \times 10^{-5}$) to level typical of radio-loud LLAGN ($R_X \sim 3.5 \times 10^{-4}$). This suggests that the source is an intermediate object or it can transit between the two classes \citep{Ros12}.

\begin{figure*}
    \centering
    \includegraphics[width=0.49\textwidth]{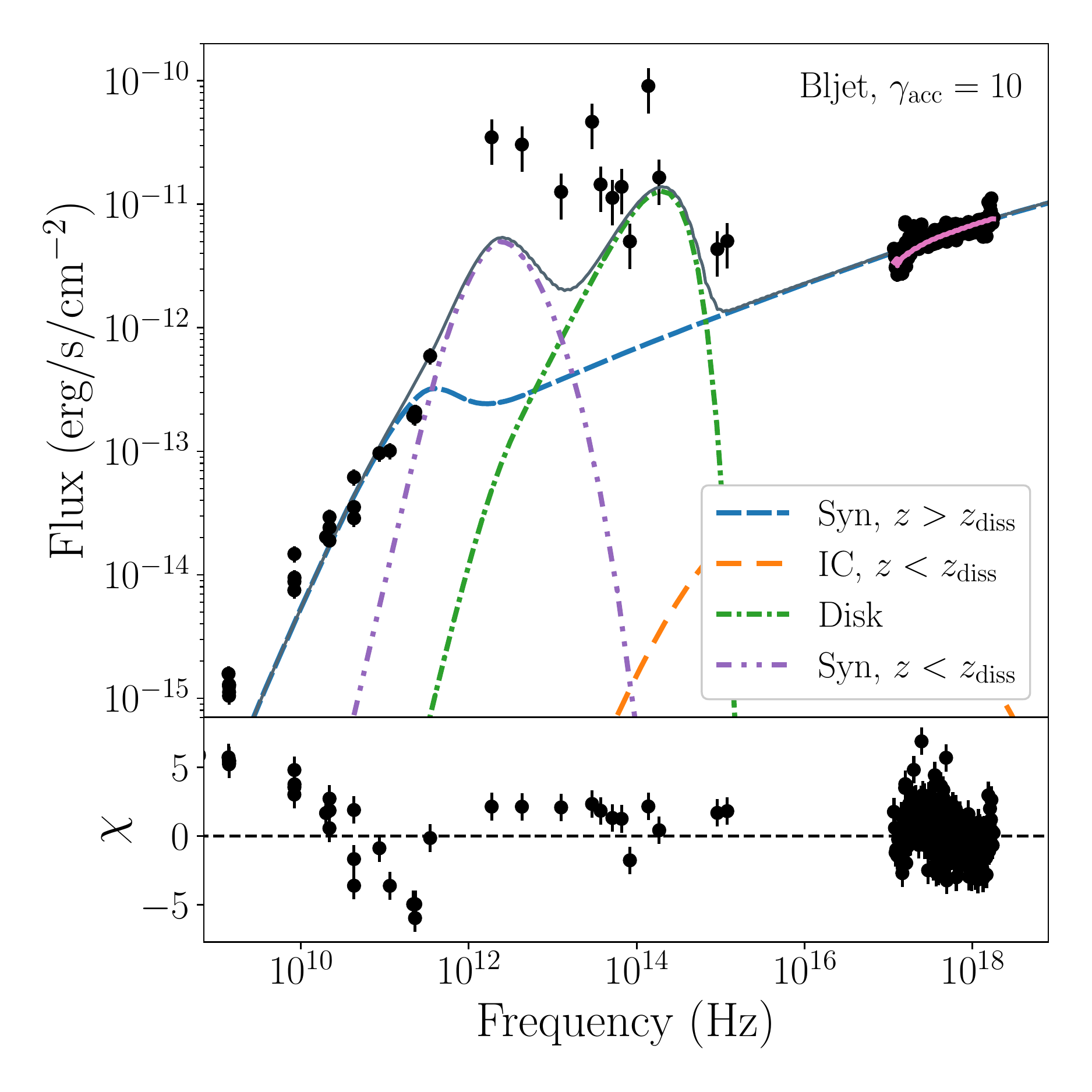}
    \includegraphics[width=0.49\textwidth]{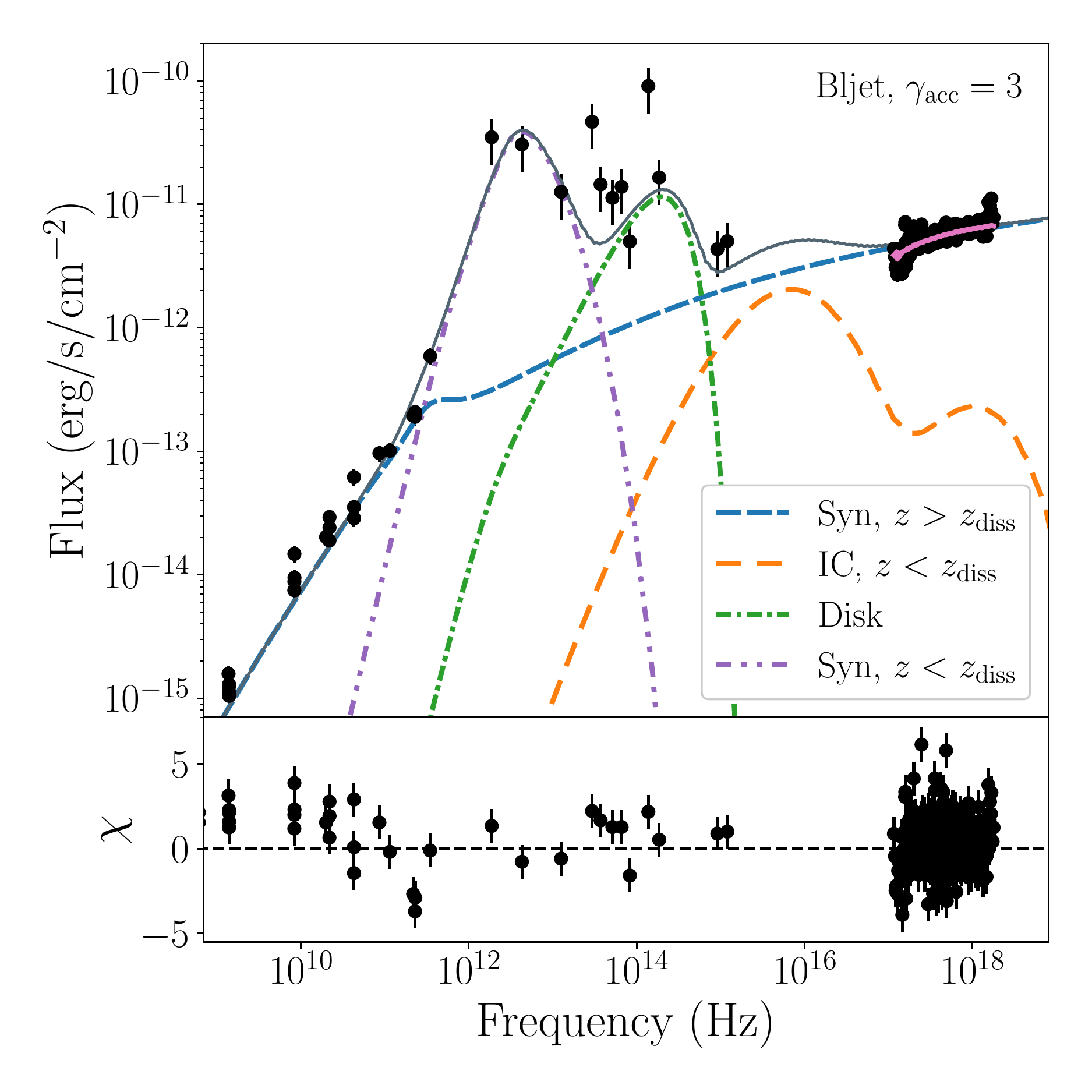}
    \includegraphics[width=0.49\textwidth]{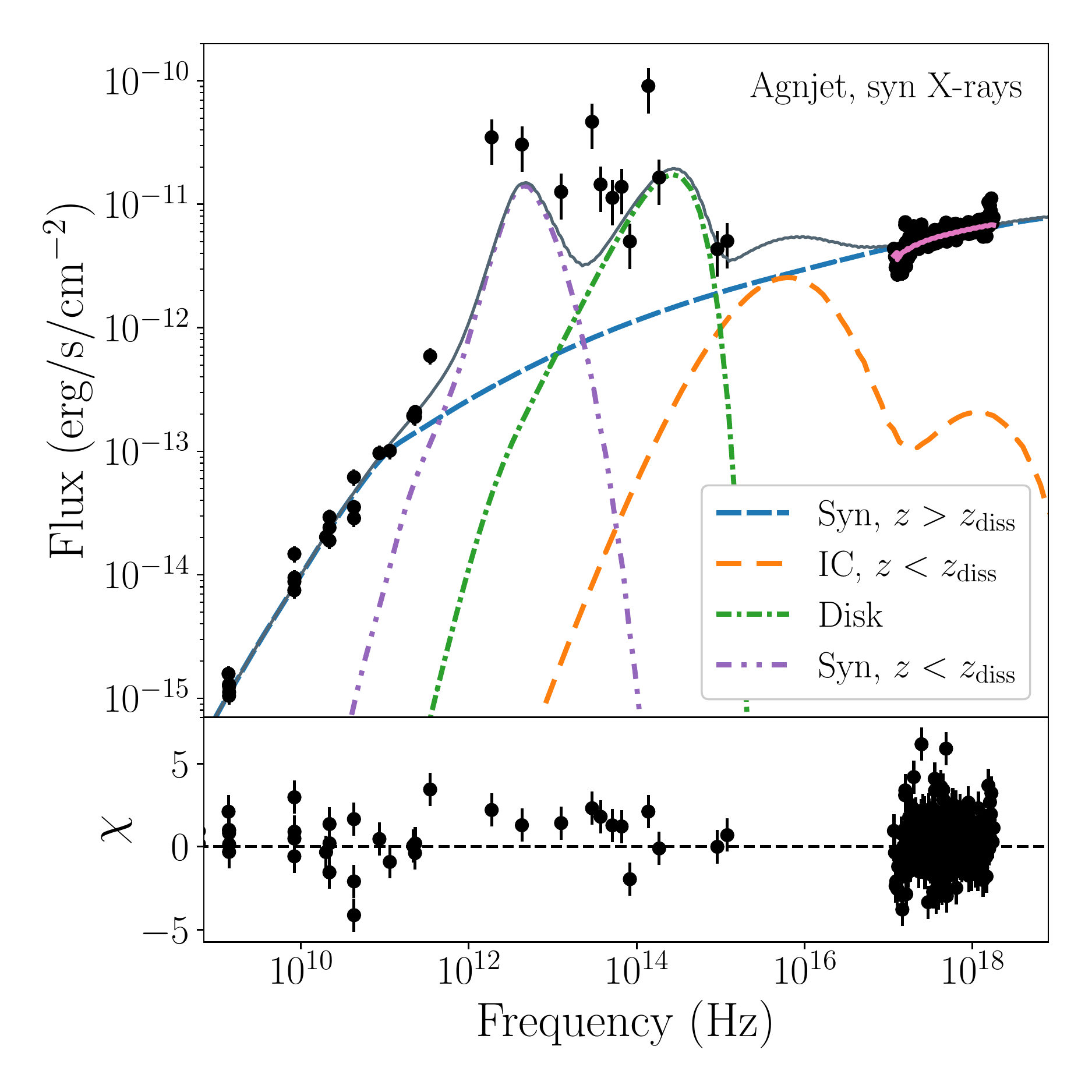}
    \includegraphics[width=0.49\textwidth]{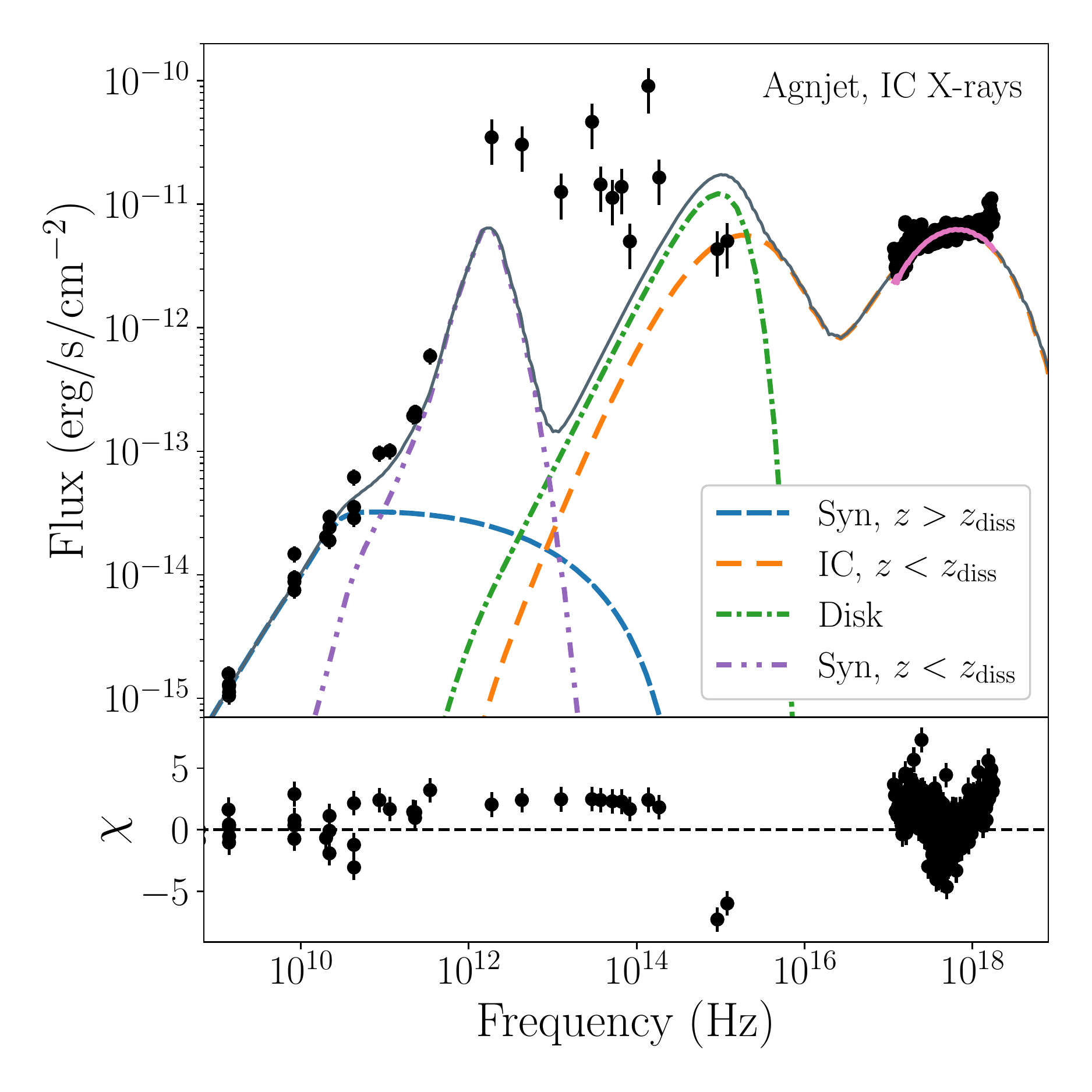}
    \caption{Model A (top left), B (top right), C (bottom left) and D (bottom right) for the SED of M\,81$^*$; for each, the $\chi^{2}/\rm{d.o.f.}$ are $1418.8/377$, $888.9/377$, $816.9/377$, $1846.0/377$, respectively. Only models B and C are in agreement with the full SED. Model A is inconsistent with the radio data as a result of the strong jet acceleration invoked, resulting in an inverted radio spectrum, and also underestimates the optical flux. Model D is inconsistent with both the optical fluxes and the shape of the X-ray spectrum.}
    \label{fig:M81}
\end{figure*}

The multiwavelength SED of the galaxy nucleus, which we refer to as M\,81$^*$, shows a characteristic flat/inverted spectrum in the radio band ($S_\nu \sim \nu^{\alpha}$, with $\alpha\sim 0.0-0.3$). A one-sided jet, likely responsible for the observed non-thermal emission, has been resolved in M\,81$^*$ \citep{Bietenholz00, Bietenholz04} which has been found to be precessing \citep{MartiVidal11}. A stationary radio core has been identified of the size of 138 AU, or $\sim$100 Schwarzschild radii \citep{Ros12}. There is no indication of the ``big blue bump'', typically associated with an optically thick/geometrically thin standard disc, in the optical band. The low column density ($n_H \sim 5\times 10^{20}$ cm$^2$) and nearly face on galactic disc reveal a steep IR/UV non-stellar spectrum ($\alpha \sim -2$) and X-ray radiation from the nuclear region \citep[][and reference therein]{Markoff08}. In the X-ray band, a power-law extends from 0.1--100 keV with an index $\Gamma\sim1.8-1.9$ \citep{Pellegrini00}. A collection of absorption and emission lines are consistent with some form of a radiatively inefficient accretion flow \citep{Young07}. 

Over the course of several multiwavelength campaigns \citep[e.g.][]{Markoff08,Miller10}, variability over timescales of months as well as intraday variability has been captured across the whole spectrum, likely connected with the ejection of optically thin synchrotron transient ejecta moving away from the source. 

In this section we use the multiwavelength data presented in \cite{Markoff08} where the M\,81$^*$ was observed with the Giant Meterwave Radio Telescope (GMRT), the Very Large Array/Very Large Baseline Array (VLA/VLBA), the Plateau de Bure Interferometer at IRAM, the Submillimeter Array (SMA), Lick Observatory and Chandra. A detailed description of the data can be found in \cite{Markoff08}, as well as fits to the SED using an older version of \texttt{agnjet}. We re-fit the data with the \texttt{bljet/agnjet} flavours of the model presented in this paper and compare to those results. We take $M_{\rm bh} = 7\cdot10^{7}\,M_{\sun}$ for the mass of the black hole, $d=3.48\rm{Mpc}$ for the source distance, and $\theta = 50^{\circ}$ for the source inclination. In the case of M81$^*$, there are no prominent reflection features, therefore we do not add that component to the model. We focus on two predictions of the model. First, in low-power sources we favour non-thermal synchrotron emission, rather than inverse Compton scattering, to dominate the X-ray emission. Second, we show that for off-axis sources, the shape of the radio spectrum can be used to indirectly constrain the jet acceleration profile.

\begin{table}
\begin{center}
\begin{tabular}{| l | c | c | c | c|}
\hline
Model & A & B & C & D\\
 \hline
$N_{\mathrm{j}}$ ($L_{\mathrm{Edd}}$) & $5.6\cdot10^{-5}$ & 
$1.9\cdot10^{-4}$ & $4.4\cdot10^{-5}$ & $9.2\cdot10^{-5}$\\
$r_{\rm 0}$ ($\mathrm{R_{g}}$) & $2.2$ & $3.4$  & $2.2$& $6.6$\\
$z_{\mathrm{diss}}$ ($\mathrm{R_{g}}$)  & $50$ & $53$  & $50$ & $200$\\
$z_{\mathrm{acc}}$ ($\mathrm{R_{g}}$)  & $1\cdot10^{5}$ & $120$  & // & //\\
$T_{\mathrm{e}}$ ($\rm{keV}$) & $2500$ & $2851$ & $2733$ & $2500$\\
$s$  & $1.5$ & $1.7$ & $1.7$ & $3$\\
$f_{\rm sc}$ & $2.5\cdot10^{-2}$ & $0.1$ & $2.5\cdot10^{-2}$ & $1.0\cdot10^{-9*}$ \\
$\beta_{\rm p}$ & // & // & $0.1$ & $5$\\
$L_{\rm disc}$ & $2.1\cdot10^{-5}$ & $1.9\cdot10{-5}$ & $1.9\cdot10^{-5}$ & $1.5\cdot10^{-5}$\\
$r_{\rm in}$ ($\mathrm{R_{g}}$) & $135$ & $125$ & $84$ & $5$\\
\hline
$\chi^2_{\rm red}$ & $3.7$ & $2.3$ & $2.2$ & $4.9$\\
\hline
\end{tabular}
\caption{Best-fitting parameters for M81$^*$. Models A and B assume one proton per electron in the jet, and therefore $\beta_{\rm p}$ is calculated from equation (\ref{eq:pair_bljet}). Parameters marked with an asterisk were frozen. All models have the same hydrogen column density ($N_{\rm H} = 2.38\cdot10^{20}$ cm$^{-3}$) estimated by modelling the X-ray spectra alone with an absorbed power-law model.}
\label{tab:m81}
\end{center}
\end{table}

Four different fits of the SED (named model A through D), highlighting different regions of parameter space, are shown in fig.~\ref{fig:M81}. The best-fitting parameters for each are reported in tab.~\ref{tab:m81}. Model A uses the \texttt{bljet} flavour, and represents a jet that accelerates to highly relativistic speeds ($\gamma_{\rm acc}=10$, reached at $z_{\rm acc}=10^{4}\,\rm{r_g}$), similarly to the blazar jets discussed in the following section. Model B also uses \texttt{bljet}, but in this case the terminal Lorentz factor reached is only mildly relativistic ($\gamma_{\rm max}=3$, at $z_{\rm acc}=120\,\rm{r_g}$). Models A and B assume that the jet carries one proton per electron, and therefore $\beta_{\rm p}$ is fixed. Models C and D use the \texttt{agnjet} flavour, and in these fits we let $\beta_{\rm p}$ free to vary. In model C we also keep $f_{\rm sc}$ free; together with a free $\beta_{\rm p}$, this should in principle allow the model to explore both synchrotron and inverse-Compton dominated regimes. However, we find that only non-thermal synchrotron models produce a satisfying fit to the data. This is highlighted in model D, in which we suppress the non-thermal X-ray emission by taking a low value of $f_{\rm sc}$. Out of these four models, only models B and C reproduce the data satisfactorily. The jet physics inferred from either fit are equivalent, despite the use of different model flavours. 

The best fit for each model set up is shown in fig.~\ref{fig:M81}. It is clear that model A is ruled out because of the shape of the radio spectrum; this is driven exclusively by the jet dynamics assumed in each model. If the jet is misaligned and accelerating to highly relativistic speeds, then both the Doppler factor and magnetic field strength decrease sharply along the jet axis; this results in a strongly inverted radio spectrum, in disagreement with the data. Note that if the viewing angle is smaller ($\theta \approx 15^{\circ}$, as is the case for M87), this conclusion does not hold. In this case, a highly relativistic magnetically-driven jet can successfully produce a flat radio spectrum \citep{Lucchini19b}, thanks to the larger Doppler factors involved. For the large viewing angle considered here, on the other hand, the model is in good agreement with the data only if a small terminal Lorentz factor is considered, as in model B. This finding suggests that for off-axis sources, the bulk of the radio emission originates in the outer jet sheath, rather than the inner jet spine, in agreement with both semi-analytical models of more powerful radio galaxies \citep{Ghisellini05} and post-processing of GRMHD simulations \citep[e.g.][]{Monika13,Monika14}. 

\begin{table}
\begin{center}
\begin{tabular}{| l | c |}
\hline
Parameter & Value\\
\hline
$M_{\rm bh}$ ($\rm{M_{\sun}}$) & $10^{9}$ \\
$\theta$ & $4^{\circ}$\\
$d\,(\rm Gpc)$ & $6.701$\\
$z$ & $1$\\
\hline
$L_{\rm disc}$ ($L_{\mathrm{Edd}}$) & $0.8$\\
$r_{\rm in}$ ($\mathrm{R_{g}}$) & $1$\\
$r_{\rm out}$ ($\mathrm{R_{g}}$) & $10^{5}$\\
$f_{\rm BLR}$ & $0.05$\\
$f_{\rm DT}$ & $0.5$\\
\hline
$N_{\mathrm{j}}$ ($L_{\mathrm{Edd}}$) & $0.4$ \\
$r_{\rm 0}$ ($\mathrm{R_{g}}$) & $10$ \\
$z_{\mathrm{diss}}$ ($\mathrm{R_{g}}$)  & $10^{3}$ \\
$z_{\mathrm{acc}}$ ($\mathrm{R_{g}}$)  & $10^{3}$ \\
$z_{\mathrm{max}}$ ($\mathrm{R_{g}}$)  & $10^{8}$\\
$\gamma_{\rm acc}$ & $20$\\
$\sigma_{\rm acc}$ & $0.01$\\
$s$  & $2.0$ \\
$f_{\rm \beta}$ & $0.1$\\
$f{\rm pl}$ & $0$\\
\hline 
$T_{\rm e}$ ($\rm{keV}$)& $511,1500$\\
$f_{\rm nth}$ & $0.1,0.5$\\
$f_{\rm sc}$ & $10^{-6*},5\times10^{3*}$ \\
$f_{\rm heat}$ & $1,20$\\
\hline
\end{tabular}
\caption{List of parameters used in sec.\ref{sec:blazars}. The top three panels indicate parameters that were kept constant for this section. Additionally, we assume the jet carries one proton per electron. \\
$^{*}$ for $f_{\rm sc}\leq 1$, we set the particle acceleration time scale and calculate the corresponding maximum particle Lorentz factor (equation (\ref{eq:max_gamma})); for $f_{\rm sc}>1$, we take a fixed maximum particle Lorentz factor along the jet, as discussed in the text.}
\label{tab:FSRQs}
\end{center}
\end{table}

We find that regardless of other parameters, the Thomson optical depth at the base of the jet is always very low for the limited jet powers ($N_{\rm j}\approx 10^{-4}-10^{-5}$) appropriate for LLAGN, despite the very small size of the jet nozzle ($r_{\rm 0} \approx 2-3\,\rm{R_g}$) found in all models. We find $\tau \approx 2\cdot 10^{-5}$, $10^{-4}$, $4\cdot 10^{-4}$, $3\cdot 10^{-3}$, for models A through D respectively. Such low optical depths suppress the thermal inverse-Compton emission, requiring non-thermal synchrotron to dominate the X-rays in order to fit the data. Regardless of model flavour, particle acceleration occurs very close to the black hole ($z_{\rm diss}\approx 100\,\rm{R_g}$, although this parameter is poorly constrained by the data). The non-thermal particle spectrum is very hard ($s\approx 1.7$), except for model D, in which we drove the fit towards an SSC-dominated scenario. We achieve this by effectively removing the non-thermal power-law from the X-ray band by starting the fit with an arbitrarily small $f_{\rm sc}$ (this has the secondary effect of making the exponent of the power-law a non-constrained parameter). This attempt to fine-tune the model in order to boost the optical depth and IC flux (mainly by increasing the jet pair content) results in a poor fit. This is because in this regime the model under-estimates the sub-mm/IR/optical cyclo-synchrotron thermal emission, while predicting an X-ray spectrum that is significantly more curved than the observed one.

\begin{figure*}
    \centering
    \includegraphics[width=0.49\textwidth]{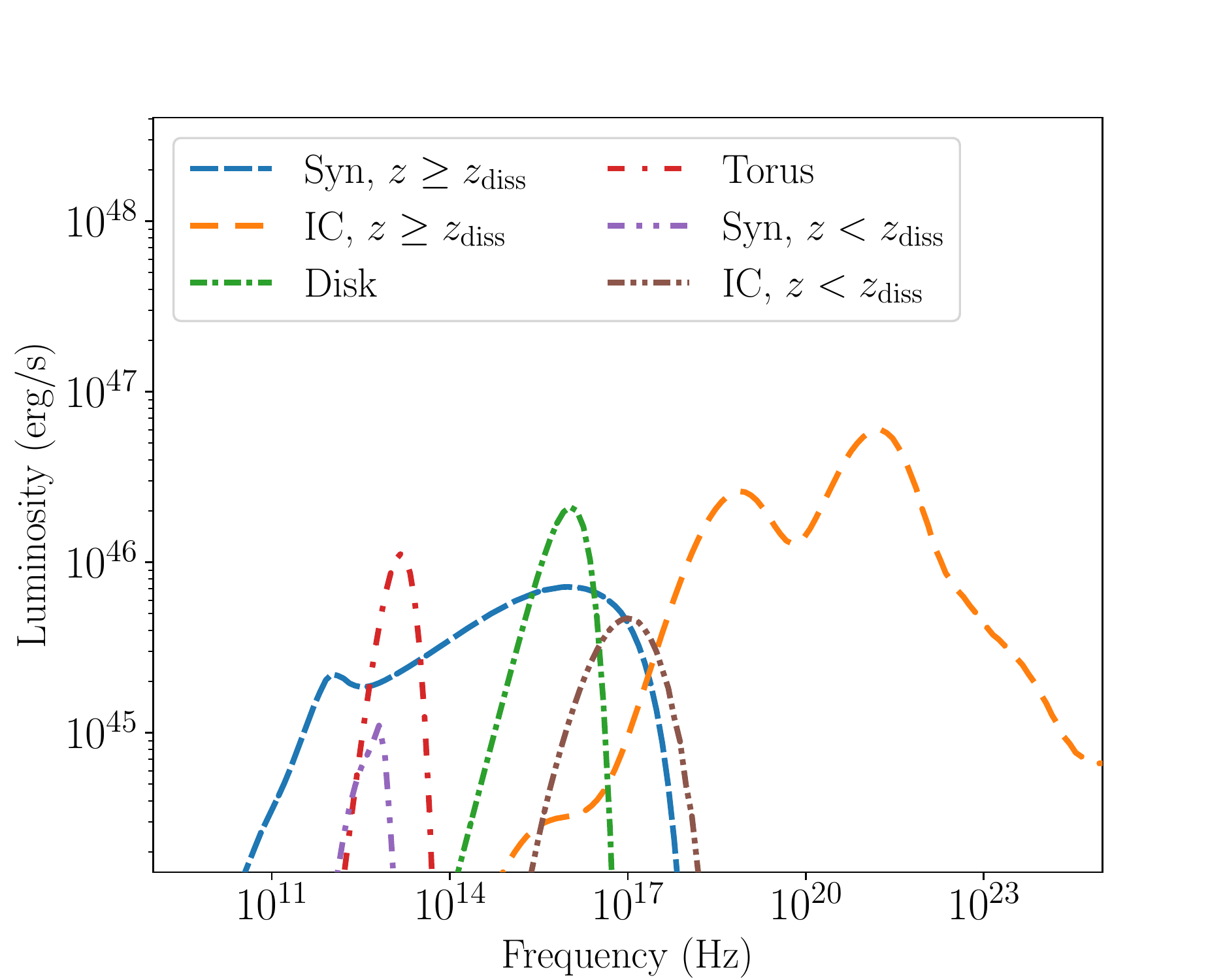}
    \includegraphics[width=0.49\textwidth]{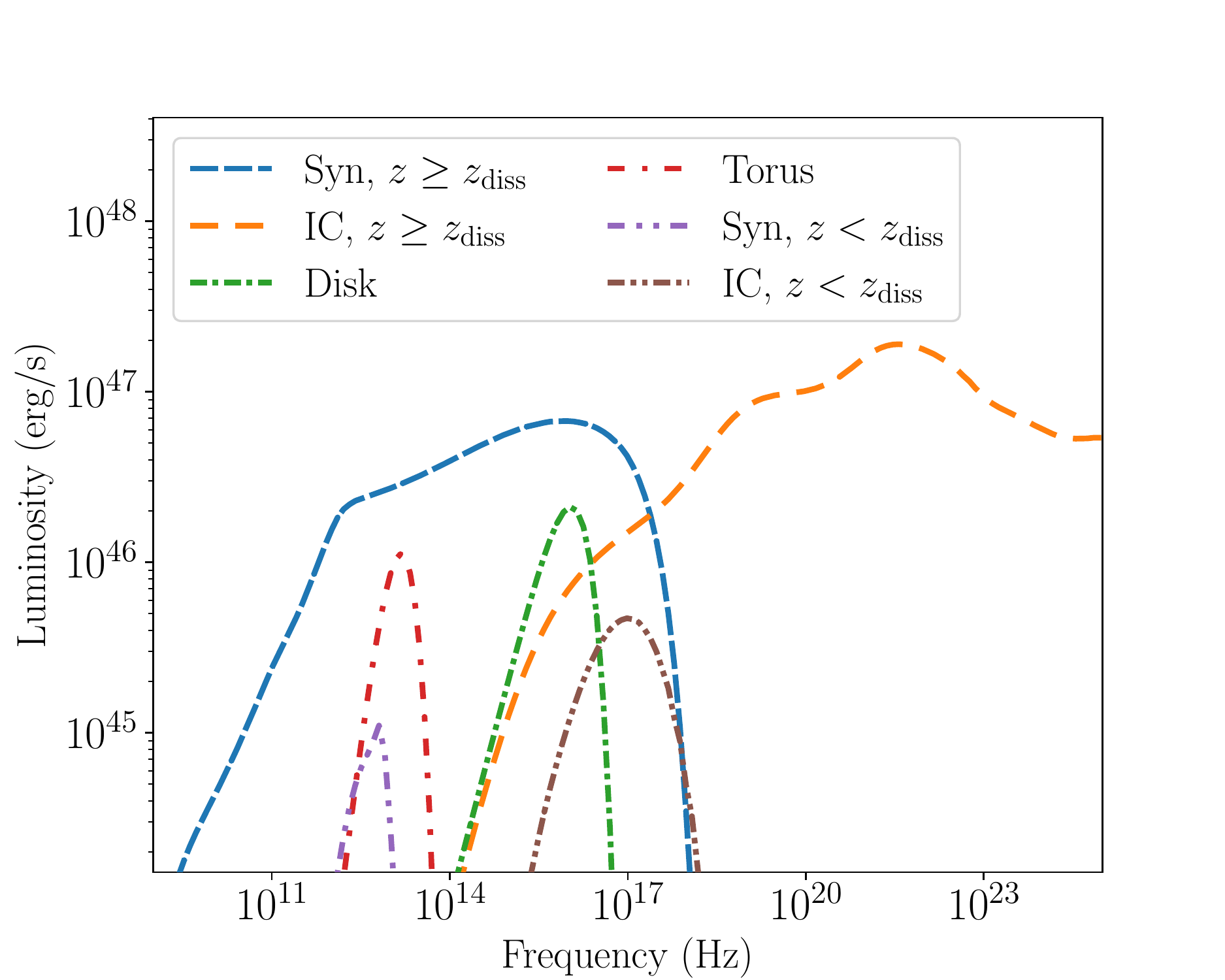}
    \caption{Left panel: SED computed with a low non-thermal particle fraction $f_{\rm nth}=0.1$. In this case, the two bumps in the non-thermal inverse Compton spectrum, shown in orange, are clearly visible. Right panel: identical model, but with $f_{\rm nth}=0.5$. In this case, the two features are much less prominent, and are produced by the sum of multiple components to the inverse Compton spectrum, rather than the shape of each individual component. Regardless of the value of $f_{\rm nth}$, the optically thin non-thermal synchrotron spectrum (dashed blue line)) peaks at UV/soft X-ray frequencies, which is much higher than what is observed in this class of sources.}
    \label{fig:FSRQ_plfrac}
\end{figure*}

Finally, all models share two characteristics regardless of model flavour. First, the temperature of the electrons at the jet base is consistently very high ($T_{\rm e}\approx 3000\,\rm{keV}$), in order to match the sub-mm bump present in the data. Second, a small excess is required in order to match the optical flux. This can be modelled with a very faint standard accretion disc ($L_{\rm disc}\approx 10^{-6}-10^{-5}\,\rm{L_{Edd}}$); however, due to the lack of optical data, the innermost radius is entirely unconstrained, with our models returning values ranging from a few to $118$ $\rm{R_g}$. However, at these low luminosities we would expect to find large truncation radii; as a result, models with $r_{\rm in}\approx 1$ are ruled out. We further note that in the case of model D, a small inner radius is necessary in order to supply the jet base with sufficient photons for inverse Compton scattering. While this is a somewhat artificial way of estimating the disc parameters, it further strengthens the case against SSC as the dominant X-ray radiative mechanism in this object. \cite{Markoff08} found an SSC-dominated fit for M81$^*$, by injecting a power-law of particles directly at the jet base (which is roughly consistent with the small values of $z_{\rm sh}$ found) and letting the nozzle aspect ratio $h$ free. They find $h\approx 10$ (which can no longer be probed with our updated radiative code, due to the changes discussed in sec.\ref{sec:Kompton}) and $r_{\rm 0} = 4$, implying a nozzle height of $40\,R_{\rm g}$. A large nozzle aspect ratio boosts the SSC flux by forcing the optical depth (and therefore the efficiency of Comptonisation) to remain constant, rather than decrease due to the jet's expansion. As a result, this fit is also consistent with a scenario in which SSC requires extreme parameters in order to produce a sufficient X-ray flux.

\subsection{Canonical high power blazars}
\label{sec:blazars}

In this section we discuss how our model can produce an SED compatible with that of a canonical high power flat spectrum radio quasar (FSRQ), defined as in \cite{Ghisellini09}: sources with large black hole masses ($M_{\rm bh}\approx 10^{9}\,\rm{M_{\sun}}$), high accretion rates ($\dot{M}\approx 0.1-1\,\dot{M}_{\rm Edd}$), and which launch powerful jets that are observed at a small viewing angles ($\theta \leq 10^{\circ}$). The optically thick, geometrically thin disks extends to the ISCO as in the standard Shakura-Sunyaev model, and supports a bright broad line region and/or dusty torus. The combination of large beaming factors and photon-rich environment surrounding the jet results in SEDs with a very large Compton dominance. In these sources, the crucial difference between single- and multi-zone jet models is that in the latter case, the cooling rates decrease dramatically along the jet. In this section we show that these large changes in cooling rates, combined with our standard assumptions (eq.\ref{eq:max_gamma}) on the acceleration rate, have a major impact on the resulting SED.

Most leptonic SED models for these objects find that the X-rays are dominated by synchrotron self-Compton emission, but it is unclear whether the UV photons of broad line region (inverse Compton/broad line region, or IC/BLR), or the infrared photons of the torus (inverse Compton/dusty torus, or IC/DT), make up the dominant source of seed photons for inverse Compton scattering in the $\gamma$-ray band \citep[e.g.][]{Dermer09,Ghisellini10,Boettcher13}. The more standard assumption, particularly when using a typical single-zone model, is to consider only the broad line region \citep[e.g.][]{Liu06,Ghisellini10b},  whose energy density in the co-moving frame of the jet is about two orders of magnitude higher than that of the torus (equation (\ref{eq:z_BLR}), \ref{eq:z_DT}, \ref{eq:UBLR_boost} and \ref{eq:UDT_boost}). This results in a small emitting region located fairly close to the black hole; the short variability timescales observed at $\gamma$-ray energies are in agreement with this scenario \cite[e.g.][]{Tavecchio10}. A key prediction of the IC/BLR model is that the VHE flux should be strongly suppressed by $\gamma-\gamma$ absorption between the high energy photons produced in the jet, and the UV photons originating the broad line clouds \citep{Donea03,Liu06}. However, several FSRQs have been detected by ground-based Cherenkov telescopes at TeV energies \citep{Albert08b,Aleksic11,Aleksic14}, and none of these sources show evidence of a spectral cutoff in their $\gamma$-ray spectra. The same also holds for FSRQ detected with the \Fermi/LAT telescope, which do not show traces of $\gamma-\gamma$ absorption in their spectra \citep[e.g.][]{Pacciani14,Costamante18b}. These findings imply that the emitting region has to be located at pc scales, beyond the broad line region, in which case the dominant source of seed photons is likely to be the torus. However, it remains unclear whether this behaviour should apply to the entirety of the FSRQ population, or only a limited number of sources.

\begin{figure*}
    \centering
    \includegraphics[width=0.54\textwidth]{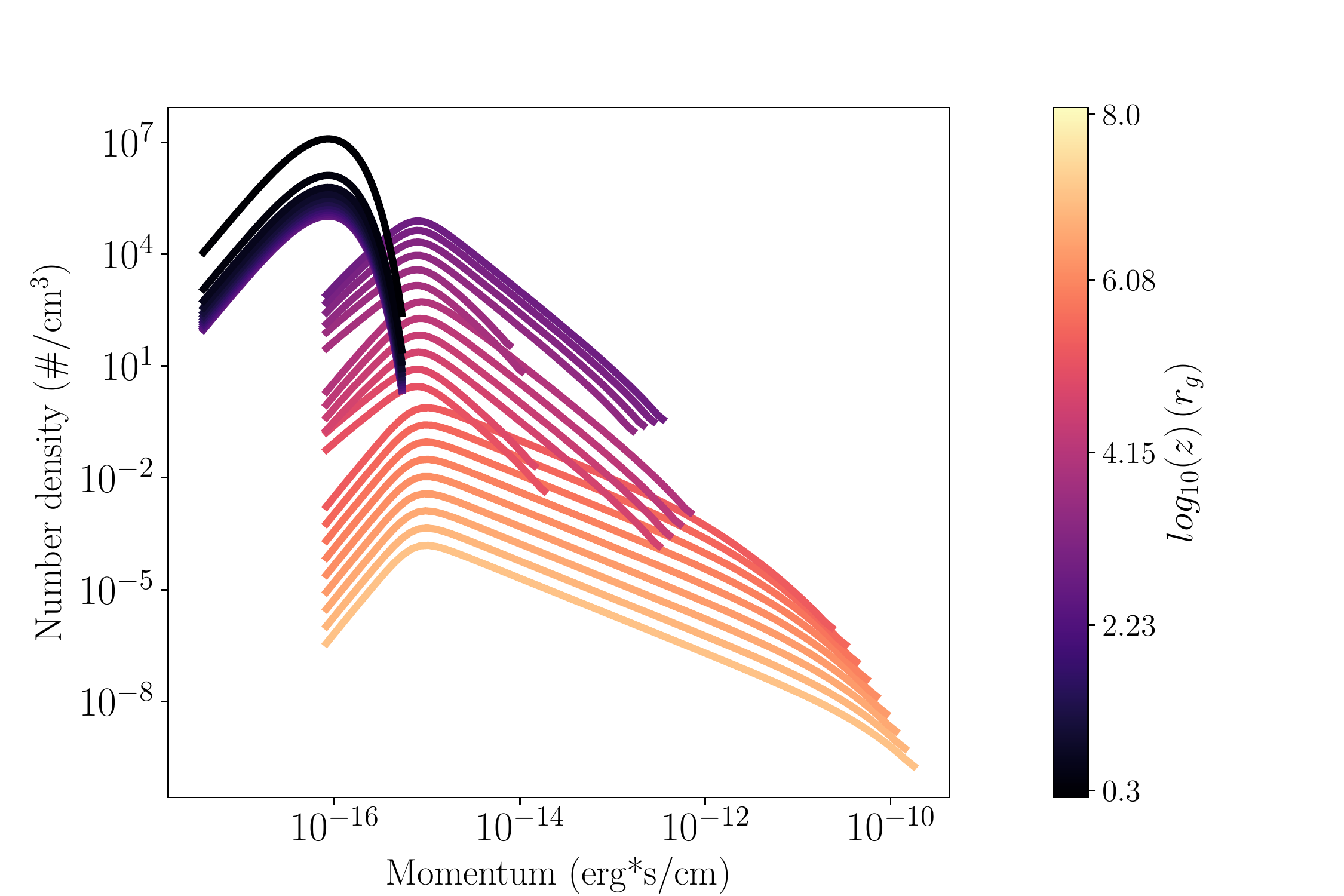}
    \includegraphics[width=0.45\textwidth]{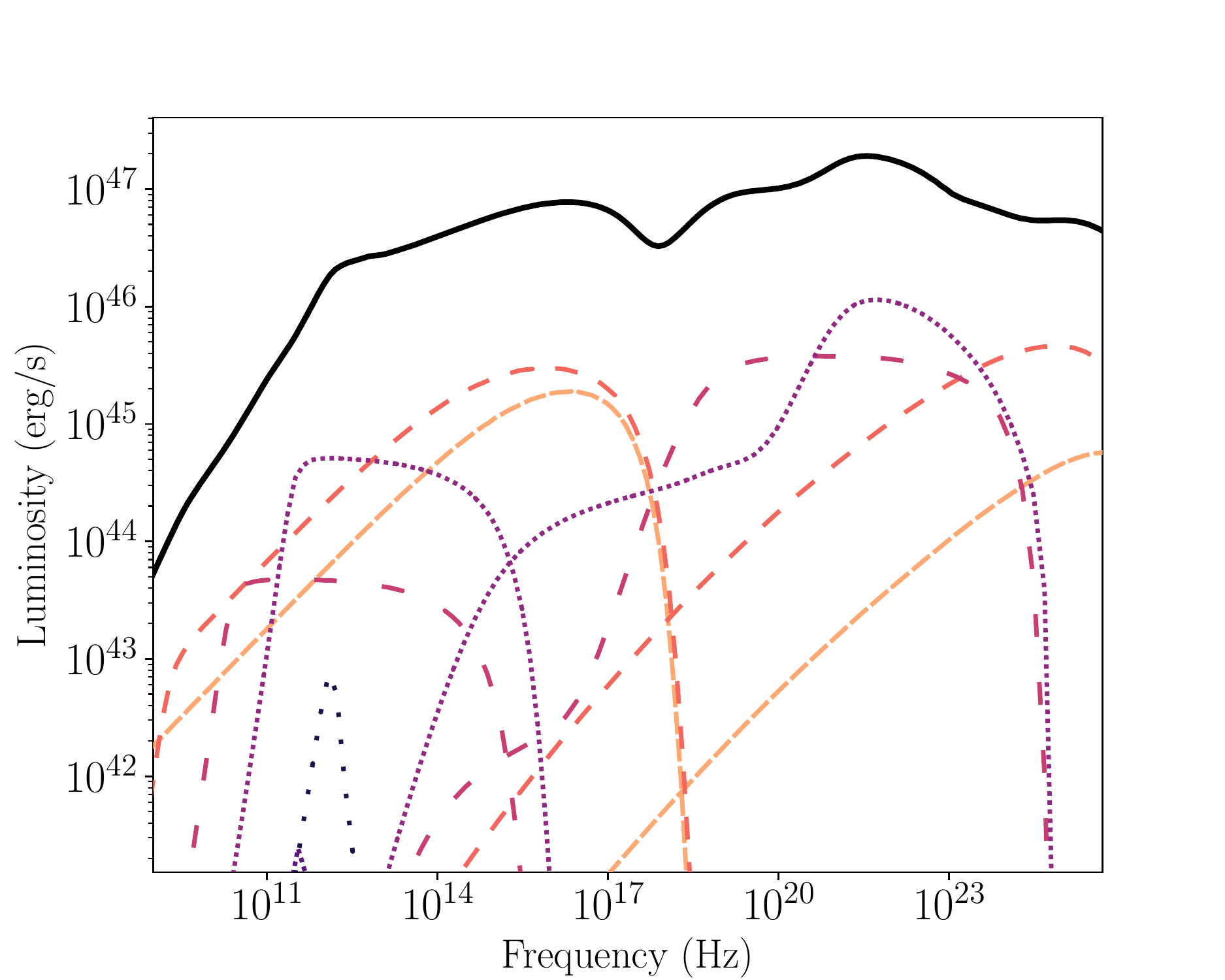}
    \includegraphics[width=0.54\textwidth]{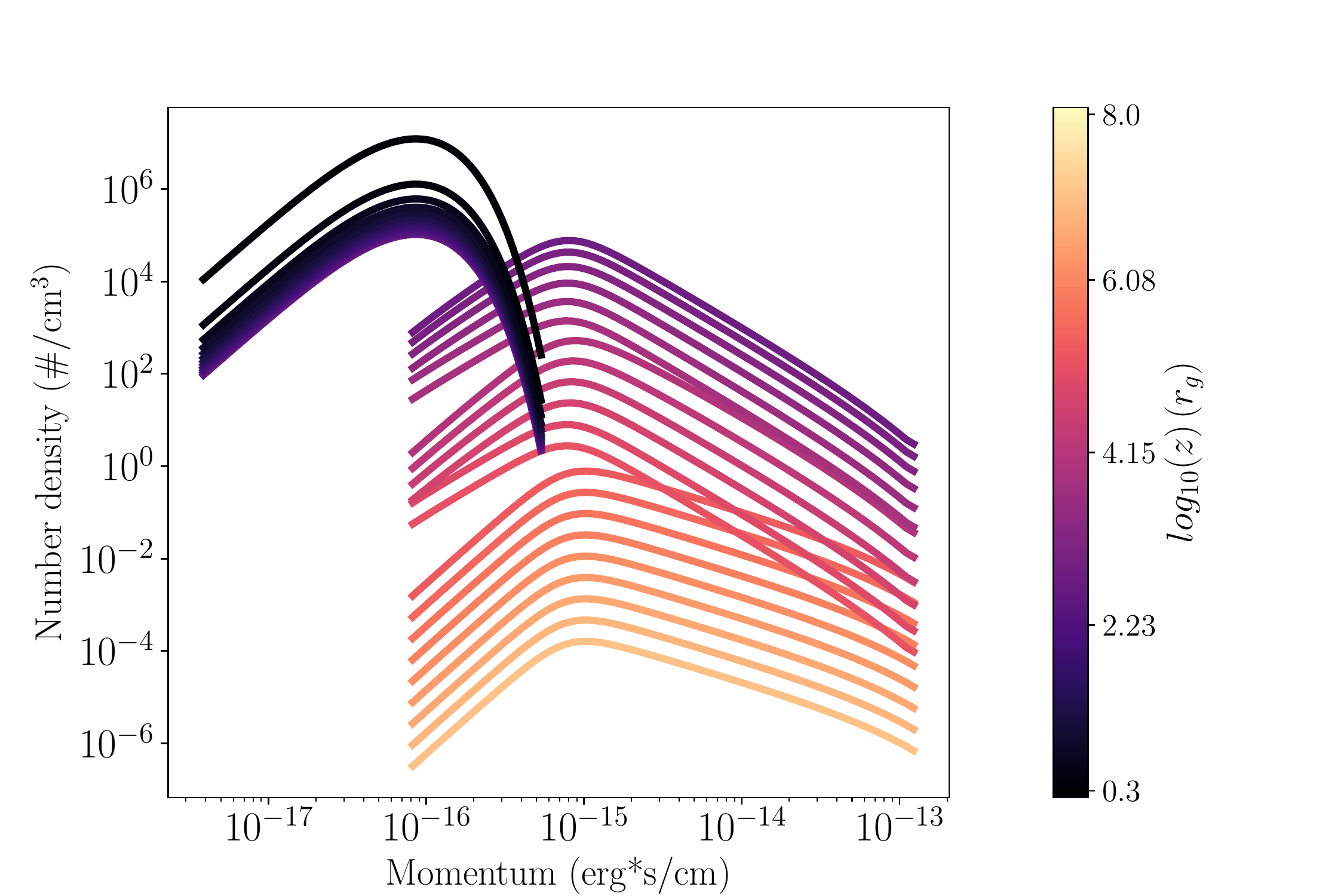}
    \includegraphics[width=0.45\textwidth]{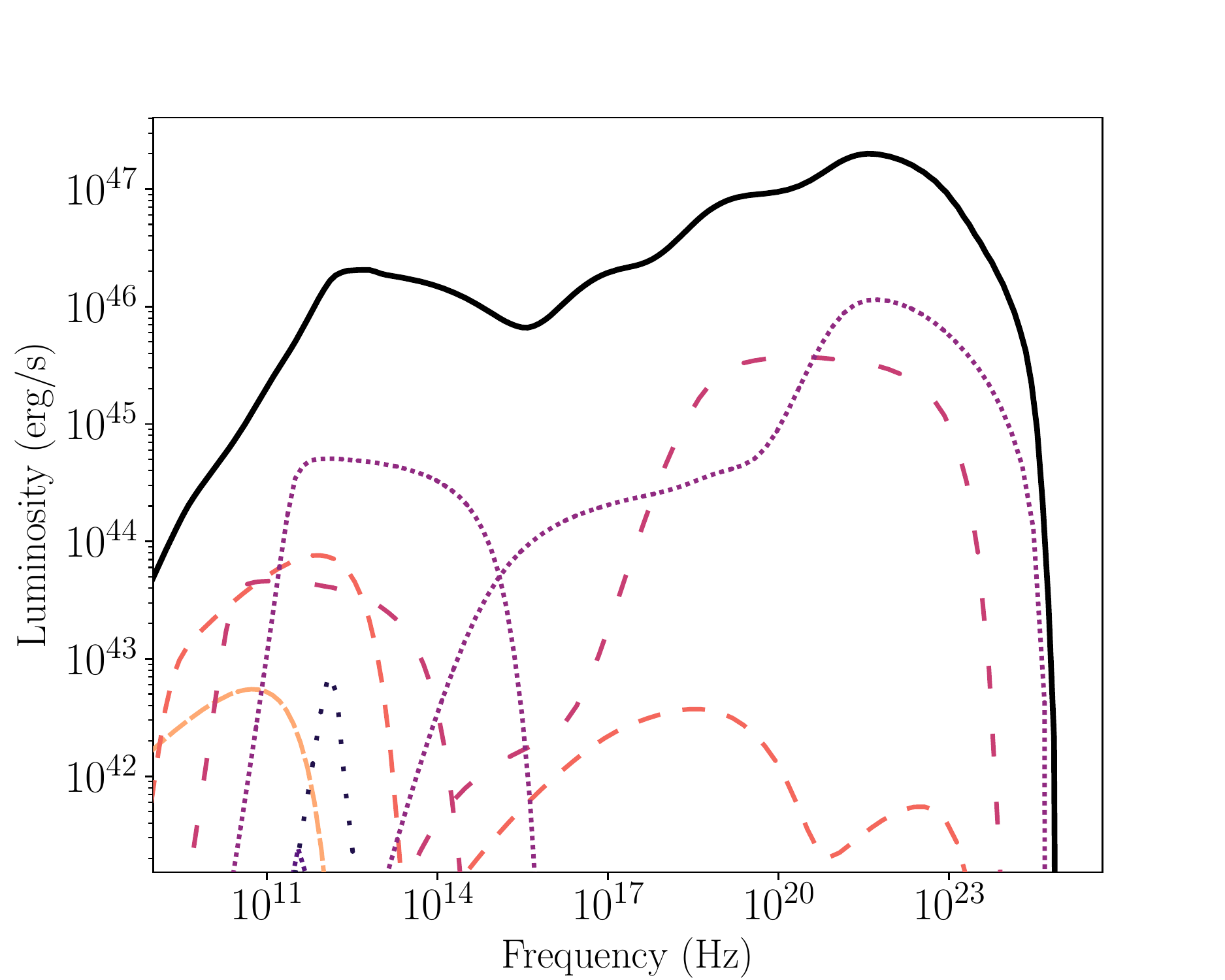}
    \caption{Top row: particle distribution (left panel) and SED of each region along the jet (right panel) computed by taking constant $f_{\rm sc}=10^{-6}$ along the jet. Bottom row: similar plots, but by taking constant $\gamma_{\rm max}=5\cdot10^{3}$ along the jet. The black line in the SED plot indicates the sum of all components, the color scale corresponds to distance along the BH axis, with lighter colours indicating increasing distance. In the plots on the right column, the difference in the contribution to the SED of the outermost jet regions is clearly visible. For clarity, the left panel shows the distributions in one out of three zones, and the right panel shows the SED of one out of fifteen zones.}
    \label{fig:FSRQ_fsc}
\end{figure*}

The goal of this section is to highlight how the inclusion of these additional photon fields has a large impact on shaping the lepton distribution in the jet, resulting in more complex behaviour of the high-energy spectra predicted by \texttt{BHJet}. This behaviour is mainly driven by the fact that, unlike single zone models \citep[e.g.][]{Ghisellini09,Boettcher13}, we aim to compute the emission from the entirety of the outflow; as a result, the dominant seed photon field changes as a function of distance along the jet axis. As we will show, the result is that some amount of fine tuning is required in describing the radiating particle distribution in the jet.

We leave the first thorough study of an FSRQ source with the model for a future paper, and instead simply attempt to produce an SED similar to that of a ``typical'' FSRQ. The full list of parameters for the models in this section is reported in table \ref{tab:FSRQs}. The only parameters which we vary are the electron temperature $T_{\rm e}$, the non-thermal fraction $f_{\rm nth}$, the acceleration timescale parameter $f_{\rm sc}$, and the the heating parameter $f_{\rm heat}$. We note that when modelling BHBs and LLAGN, recent works have used $f_{\rm nth} \approx 0.1$, $f_{\rm heat}\approx 1$, and a large range for $f_{\rm sc}$ \citep[e.g][]{Connors17,Connors19,Lucchini19a,Lucchini19b,Kantzas20,Lucchini21}, similarly to our results in subsec.~\ref{sec:GX} and \ref{sec:M81}.

Fig.~\ref{fig:FSRQ_plfrac} shows the effect of changing the non-thermal fraction parameter $f_{\rm nth}$ from the typical value of $f_{\rm nth}=0.1$ to $f_{\rm nth}=0.5$.  In the case of $f_{\rm nth} = 0.1$, two large bumps appear in the hard X-ray/soft $\gamma$-ray band. These bumps are caused by the inverse Compton scattering of black body photons from either the broad line region or torus, by the large number of thermal leptons in the jet. This behaviour is clearly un-physical, as blazars show smooth non-thermal bumps with no additional features (see e.g \citealt{Ballo02}, and references therein, for Beppo-SAX observations of 3C\,279 covering almost the entire inverse-Compton bump of the source). The case with $f_{\rm nth}=0.5$ is much more compatible with these observations. In this regime, as noted in sec.~\ref{sec:BHJet_additional}, the model automatically switches to using the \texttt{Bknpower} distribution, tuning the low energy part of the distribution to be quasi-Maxwellian. With this choice, the additional features are much less prominent, and are caused by the addition of the two external Comptonisation components on top of the jet SSC emission, rather than by the shape of the particle distribution.

\begin{figure*}
    \centering
    \includegraphics[width=0.49\textwidth]{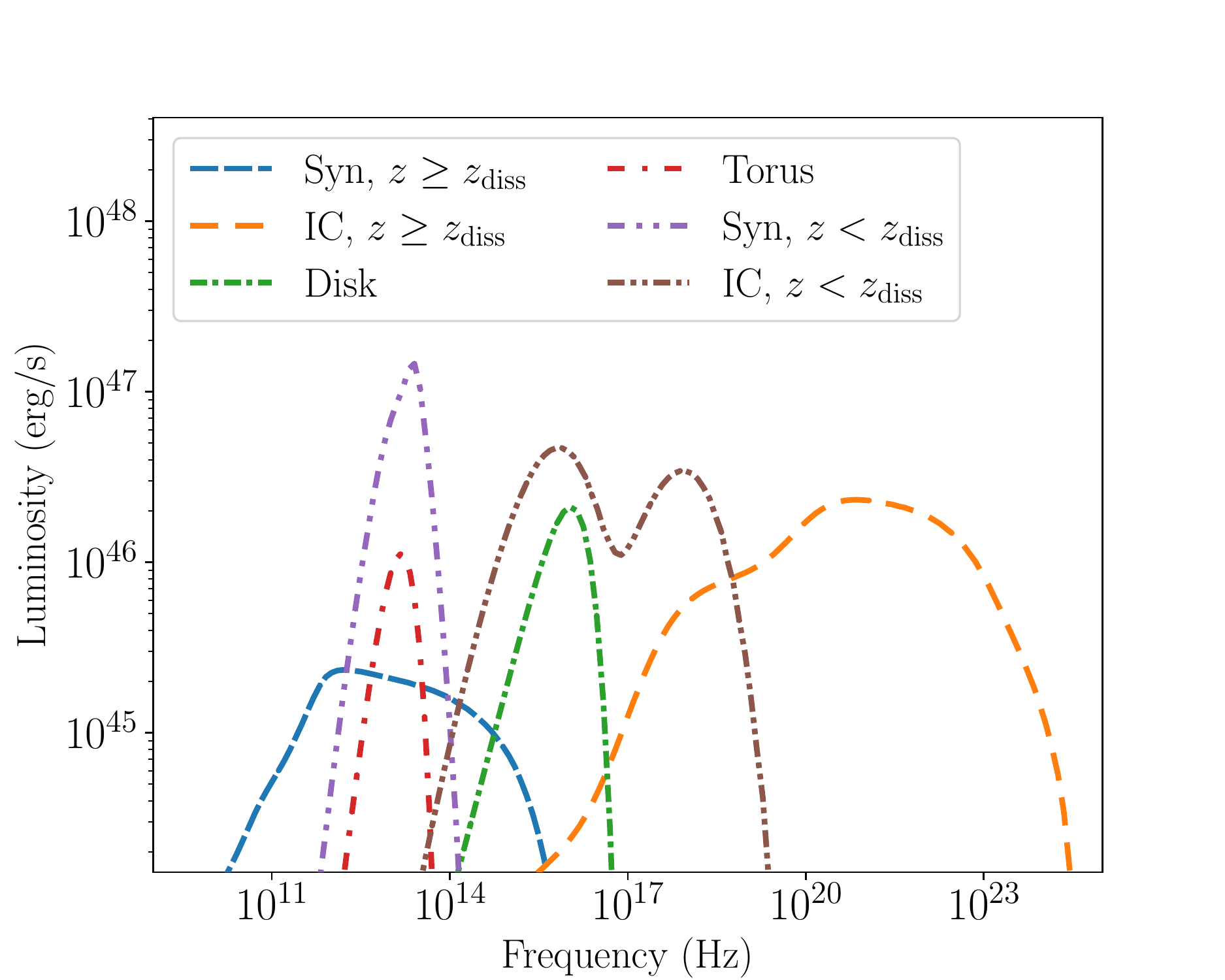}
    \includegraphics[width=0.49\textwidth]{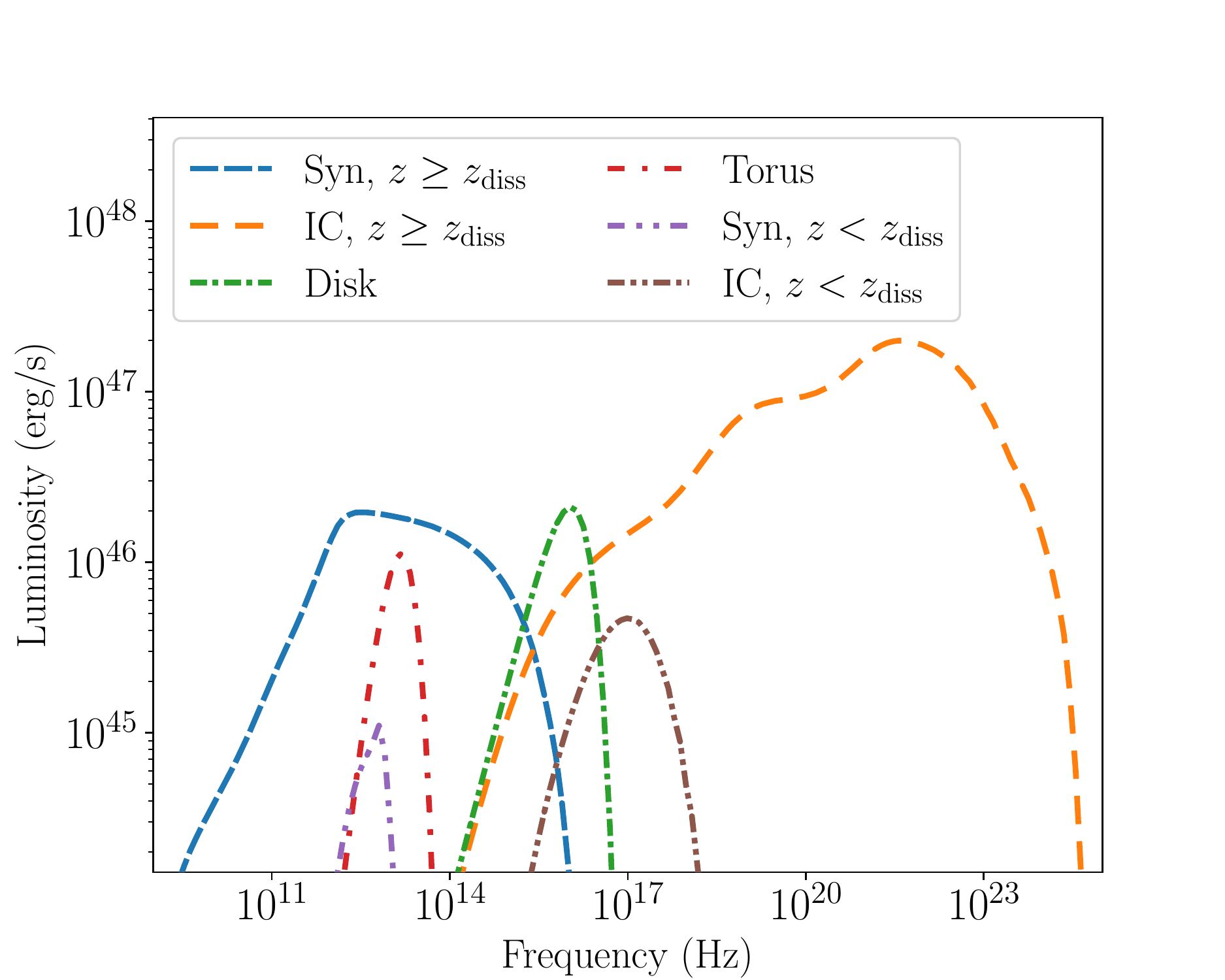}
    \caption{Left panel: SED computed with high initial electron temperature and no shock heating, resulting in prominent thermal bumps originating near the base of the jet. Right panel: SED computed with lower initial electron temperature, and substantial electron heating once particle acceleration begins. In this case, the thermal bumps are still present, but they do not dominate the SED.}
    \label{fig:FSRQ_fheat}
\end{figure*}

Regardless of the choice of $f_{\rm nth}$, both the SEDs in fig.~\ref{fig:FSRQ_plfrac} show a synchrotron peak in the UV/soft X-ray range, which is in disagreement with observations of powerful blazars, particularly the blazar sequence \citep{Fossati98,Ghisellini17}. The cause for this behaviour is highlighted in the top two panels of fig.~\ref{fig:FSRQ_fsc}. By taking a constant value for $f_{\rm sc} = 10^{-6}$, the maximum energy of the radiating particles increases along the jet as $U_{\rm rad}(z)$ changes (see equation (\ref{eq:max_gamma})); however, if external photons from the broad line region and/or torus are present, this increase is not smooth. This behaviour is caused because the main cooling channel along the jet changes abruptly. As long as $z\leq z_{\rm BLR}$, inverse Compton cooling off broad line region photons dominates, strongly reducing the break and maximum lepton energies. Then, for $z_{\rm BLR}\leq z \leq z_{\rm DT}$, the broad line region becomes de-boosted, and therefore the main photons driving IC cooling come from the torus. In this regime, due to Doppler boosting (equation (\ref{eq:UBLR_deboost}) and (\ref{eq:UDT_boost})) we have that $U_{\rm rad, DT} > U_{\rm rad, BLR}$; therefore, the break and maximum electron energies suddenly jump to intermediate values. Up to $z\approx z_{\rm DT}$, inverse Compton cooling is strong enough that for an injected particle spectrum with $s=2$, the steady-state distribution is fully cooled and reaches $p=3$ (equation (\ref{eq:FP_steadystate})). Finally, for $z\geq z_{\rm DT}$, all the external photon fields become de-boosted in the jet frame, further extending the maximum electron energy. Furthermore, in this regime the break energy becomes comparable to the maximum energy, which causes the slope of the steady-state particle distribution to be $p\approx 2$, rather than $p\approx 3$, as in the innermost regions of the jet. The result of these abrupt changes in the particle distributions on the SED are clearly shown in the top right panel of fig.~\ref{fig:FSRQ_fsc}: the outermost regions of the jet between $\approx 10^{6}-10^{8}\,\rm{R_g}$ contribute to a large fraction of the ultraviolet and soft X-ray luminosity through synchrotron emission, and $\gamma$-ray luminosity through SSC. 

The bottom panel of fig.~\ref{fig:FSRQ_fsc} shows how this behaviour can be mitigated in \texttt{BHJet}. In these SEDs, rather than take a fixed value of $f_{\rm sc}$, the non-thermal distribution is extended up to some constant Lorentz factor $\gamma_{\rm max}$ (we take $\gamma_{\rm  max} = 10^{3}$ here). While the particle distribution still switches abruptly from the fast-cooling to the slow-cooling regime for $z\geq z_{\rm DT}$, the emission from the outer jet regions is suppressed enough that the shape of the SED closely resembles a standard bright FSRQ. The contribution of each model component to the total SED is shown in the right panel of fig.\ref{fig:FSRQ_fheat}.

Finally, in fig.~\ref{fig:FSRQ_fheat} we highlight the effect of the $f_{\rm heat}$ heating parameter. This same issue has already been discussed in \cite{Lucchini19a}, for the case of a BL Lac object; we find that the same issue occurs in brighter FSRQs as well. The left SED is computed by taking a high initial electron temperature of $T_{\rm e}=1500\,\rm{keV}$ and $f_{\rm heat}=1$; in this case, the combination of high temperature and high magnetic fields at the base of the jet result in very bright thermal emission from the jet nozzle. Once again, these bright thermal bumps are in disagreement with observations. Furthermore, such high luminosity at the jet base result in extremely high compactness, possibly leading to runaway pair production \citep{Fabian15}. This un-physical regime can be suppressed by taking a lower temperature at the jet base (e.g. $T_{\rm e}= 511\,\rm{keV}$ here), together with a large value of the heating parameter ($f_{\rm heat}=20$ here). Combined with the large value of $f_{\rm nth}$ required, this indicates that the initial burst of particle acceleration should be very efficient in FSRQs. However, the necessity of suppressing the maximum energy of the particles along the jet suggests that beyond the initial region, particle re-acceleration should be fairly inefficient.

In summary, the newest version of \texttt{BHJet} can reproduce the SED of a canonical high power FSRQ without altering the prescription for the jet dynamics. In quasars, both broad line region and dusty torus photons are viable target photon fields for inverse Compton scattering; here, we have limited ourselves to a case in which the dissipation region is fairly close to the black hole ($z_{\rm acc} = z_{\rm diss} = 10^{3}\,\rm{R_g}$), in which case both contribute to the high-energy bump. However, the standard assumptions detailed in sec.\ref{sec:BHJet_additional} regarding the radiating particles (which are unique to \texttt{BHJet}), combined with the nature of the external photon fields in this class of sources, can cause un-physical features to appear in the SED. These features can be avoided with minor changes to the radiating particle distribution. This finding constitutes a large step forward for the model, which can now be applied to any class of jetted AGN or BHXBs. 

\section{Summary}
\label{sec:conclusion}

In this work we have presented the latest version of the \texttt{BHJet} code, which is designed to fit the steady-state multi-wavelength SED of any accreting black hole. \texttt{BHJet} is built on a \textsc{C++} library of classes called \texttt{Kariba}, which can treat common particle distributions and radiative mechanisms invoked in modelling accreting compact objects. We have also introduced several improvements to the inverse Compton calculation in the code, including a more physical treatment of radiative transfer (based on the \texttt{CompPS} Comptonisation code) and the ability to include external photon fields typical of bright AGN (which are necessary to extend the model to FSRQs).

A comparison of the scalings of number density, magnetic field and jet speed over distance between our code (using fiducial parameters) and the results of global GRMHD simulations, shows that our model captures effectively the physics inferred in theoretical works to first order, if one only considers the outer sheath region. However, we found that compared to simulations our model seems to require much more efficient jet acceleration in the inner jet spine (with the caveat that the results of simulations in this region should be interpreted with care).

Modelling a bright hard state of the BHXB GX\,339$-$4 shows that in these bright states, the bulk of the X-ray emission likely originates in the jet launching region, very close to the black hole, and is caused by inverse Compton scattering of both disc and cyclo-synchrotron photons. The data are equally well modelled with a compact, fairly optically thick corona ($r_{\rm0}\approx 6\,\rm{R_g}$, $\tau\approx 1$) in which the electrons have fairly low temperature ($T_{\rm e} \approx 100\,\rm{keV}$), or with a more extended, optically thin, hotter jet launching region ($r_{\rm0}\approx 140\,\rm{R_g}$, $\tau\approx 0.1$, $T_{\rm e} \approx 430\,\rm{keV}$). The former scenario is essentially a physical realisation for a lamp-post corona \citep[e.g.][]{Matt91,Matt92,Martocchia96,Beloborodov99}. 

Modelling the SED of the LLAGN M\,81 shows that in the very sub-Eddington regime the model behaves very differently. Invoking a low jet power, as one would expect for this type of source, causes the optical depth of the jet launching region to decrease significantly, resulting in synchrotron becoming the main channel for X-ray emission in the jet. Attempting an inverse-Compton dominated fit of the same data results in much worse statistical agreement with the data. This conclusion is valid regardless of black hole mass, as the optical depth at the base of the jet also does not depend on mass (see equation (\ref{eq:ne}) and (\ref{eq:tau_base})). This switch between thermal inverse Comptonisation and non-thermal synchrotron is a general prediction of \texttt{BHJet}. The exact jet power/accretion rate at which this switch should occur depends strongly on model parameters, and in principle can be tested by the upcoming X-ray polarimetric mission IXPE \citep{IXPE}, as the different radiative mechanisms predict very different polarisation fractions \citep[e.g.][]{McNamara09,Ingram15}. Additionally, our model suggests that in off-axis sources, a flat radio spectrum can only be produced if the jet does not accelerate to highly relativistic speeds, suggesting that in these objects we observe the outer jet sheath, rather than the inner spine.

Finally, we have shown that thanks to the improvements to the inverse Compton code discussed in this paper, \texttt{BHJet} can now be used to reproduce a Compton-dominant SED typical of a powerful FSRQ. However, doing so requires some amount of fine tuning in the details of the radiating particle distribution, unlike in every other object which has been studied with the model so far (including the low power BL Lac PKS\,2155$-$304). This finding, together with the results of our comparison with the results of GRMHD simulation, suggests that these objects (in which the spine likely dominates the observed emission) may challenge the standard theoretical paradigm of magnetically-driven jets.

\section*{Acknowledgements}

We thank the anonymous referee for their thorough comments which greatly clarified the manuscript. We thank Mike Nowak and J{\"o}rn Wilms for their contributions to the model over the years. M. L. and S. M. are thankful for support from an NWO (Netherlands Organisation for Scientific Research) VICI award, grant Nr. 639.043.513. C. C. acknowledges support from the Swedish Research Council (VR). This research has made use of \textsl{ISIS} functions (ISISscripts) provided by ECAP/Remeis observatory and MIT (http://www.sternwarte.uni-erlangen.de/isis/). 

\section*{Data Availability}
All data in this paper are publicly available. The code and scripts are available on github at https://github.com/matteolucchini1/BHJet/. 

\bibliographystyle{mnras}
\bibliography{references}

\end{document}